\documentclass[fleqn,usenatbib]{mnras}

\usepackage[T1]{fontenc}

\DeclareRobustCommand{\VAN}[3]{#2}
\let\VANthebibliography\thebibliography
\def\thebibliography{\DeclareRobustCommand{\VAN}[3]{##3}\VANthebibliography}

\usepackage{graphicx}	% Including figure files
\usepackage{amsmath}	% Advanced maths commands
\usepackage{amssymb}	% Extra maths symbols
\usepackage{float}
\usepackage{newtxtext,newtxmath}
\usepackage[dvipsnames]{xcolor}

\newcommand{\Msun}{{\rm M_\odot}}
\newcommand{\Mdyn}{M_{\rm dyn}}
\newcommand{\MdynL}{M_{\rm dyn}/L}
\newcommand{\re}{r_{\rm e}}
\newcommand{\retd}{r_{\rm e, 3D}}
\newcommand{\res}{r_{\rm e, *}}
\newcommand{\rer}{r_{\rm e, r}}
\newcommand{\Mtot}{M_{\rm tot}}
\newcommand{\stot}{\sigma_{\rm tot}}
\newcommand{\sstar}{\sigma_*}
\newcommand{\fdm}{f_{\rm DM}}
\newcommand{\fgas}{f_{\rm gas}}
\newcommand{\kms}{{\rm km\,s^{-1}}}
\newcommand{\vc}{v_{\rm c}}
\definecolor{purple}{RGB}{76, 0,153}

\title[Deconstructing the Fundamental Plane]{A common origin for the Fundamental Plane of quiescent and  star-forming galaxies in the EAGLE simulations}

\author[A. de Graaff et al.]{
Anna de Graaff,$^{1,2}$\thanks{E-mail: graaff@strw.leidenuniv.nl}
Marijn Franx$^{1}$,
Eric F. Bell$^3$,
Rachel Bezanson$^4$,
Matthieu Schaller$^{5,1}$,
Joop Schaye$^1$,
\newauthor\ Arjen van der Wel$^6$
\\
$^{1}$Leiden Observatory, Leiden University, P.O.Box 9513, NL-2300 AA Leiden, The Netherlands\\
$^{2}$Max-Planck-Institut f\"ur Astronomie, K\"onigstuhl 17, D-69117, Heidelberg, Germany \\
$^{3}$Department of Astronomy, University of Michigan, 1085 S. University Avenue, Ann Arbor, MI 48109, USA\\
$^4$Department of Physics and Astronomy, University of Pittsburgh, Pittsburgh, PA 15260, USA\\
$^{5}$Lorentz Institute for Theoretical Physics, Leiden University, PO Box 9506, NL-2300 RA Leiden, The Netherlands\\
$^{6}$Sterrenkundig Observatorium, Universiteit Gent, Krijgslaan 281 S9, B-9000 Gent, Belgium
}

\date{Accepted XXX. Received YYY; in original form ZZZ}

\pubyear{2022}

\begin{document}
\label{firstpage}
\pagerange{\pageref{firstpage}--\pageref{lastpage}}
\maketitle

% This is a simple template for authors to write new MNRAS papers.
% The abstract should briefly describe the aims, methods, and main results of the paper.
% It should be a single paragraph not more than 250 words (200 words for Letters).
% No references should appear in the abstract.

\begin{abstract}
We use the EAGLE cosmological simulations to perform a comprehensive and systematic analysis of the $z=0.1$ Fundamental Plane (FP), the tight relation between galaxy size, mass and velocity dispersion. We first measure the total mass and velocity dispersion {(including both random and rotational motions)} within the effective radius to show that simulated galaxies obey a total mass FP that is very close to the virial relation ($<10\%$ deviation), indicating that the effects of non-homology are weak. When we instead use the stellar mass, we find a strong deviation from the virial plane, which is driven by variations in the dark matter content. The dark matter fraction is a smooth function of the size and stellar mass, and thereby sets the coefficients of the stellar mass FP without substantially increasing the scatter. Hence, both star-forming and quiescent galaxies obey the same FP, with equally low scatter ($0.02\,$dex). We employ simulations with a variable stellar initial mass function (IMF) to show that IMF variations have a modest additional effect on this FP. Moreover, when we use luminosity-weighted mock observations of the size and spatially-integrated velocity dispersion, the inferred FP changes only slightly. However, the scatter increases significantly, due to the luminosity-weighting and line-of-sight projection of the velocity dispersions, and measurement uncertainties on the half-light radii. Importantly, we find significant differences between the simulated FP and observations, which likely reflects a systematic difference in the stellar mass distributions. Therefore, we suggest the stellar mass FP offers a simple test for cosmological simulations, requiring minimal post-processing of simulation data.

\end{abstract}


%%%%%%%%%%%%%%%%%%%%%%%%%%%%%%%%%%%%%%%%%%%%%%%%%%

%%%%%%%%%%%%%%%%% BODY OF PAPER %%%%%%%%%%%%%%%%%%

\section{Introduction}\label{sec:intro}

Quiescent galaxies have been found to obey a tight, planar scaling relation between the stellar velocity dispersion, size and surface brightness, which is known as the luminosity Fundamental Plane \citep[FP; e.g.,][]{Djorgovski1987,Dressler1987,Jorgensen1996}. On the other hand, the star-forming population has been shown to follow a linear relation between the luminosity and kinematics, referred to as the Tully-Fisher (TF) relation \citep{TullyFisher1977}. More recent work has demonstrated that the two galaxy populations can be reconciled within the framework of one scaling relation \citep[e.g.,][]{Cortese2014,Bezanson2015,Aquino2020,deGraaff2021}, by either modifying the TF relation \citep{Cortese2014,Aquino2018} or the FP \citep[e.g.,][]{Zaritsky2008,HydeBenardi2009,Aquino2020}. However, these studies are largely empirically driven, aiming to construct a dynamical scaling relation with minimal scatter. A firmer theoretical footing is imperative to gain a better understanding of why different types of galaxies may lie on a single dynamical scaling relation.

\subsection{Interpreting the luminosity FP}

The properties of the FP of quiescent galaxies have typically been interpreted in terms of the dynamical mass-to-light ratio 
\citep[$\MdynL$\,; as first suggested by][]{Faber1987}. 
The zero point of the FP is directly propertional to $\MdynL$, and the redshift evolution of the FP therefore directly traces the evolution in the $M_{\rm dyn}/L$ with cosmic time %, placing strong constraints on the formation epoch of quiescent galaxies while requiring few model assumptions
\citep[e.g.,][]{vanDokkum1996}.
Although at a fixed redshift the scatter about the FP is small ($\lesssim 0.1\,$dex), it cannot be explained by measurement uncertainties alone \citep[e.g.,][]{Jorgensen1996,Forbes1998}. Rather, the scatter in the zero point can also be linked to variations in the $\MdynL$, and correlations between the offsets of galaxies from the FP and various stellar population and structural properties hence have provided insight into the formation histories of early-type galaxies \citep[e.g.,][]{Gargiulo2009,Graves2009II}.

Furthermore, the tilt of the plane can also be interpreted by the $\MdynL$ \citep[e.g.,][]{Bender1992,Trujillo2004,Cappellari2006}. Fundamentally, a tight scaling relation between the velocity dispersion ($\sigma$), effective radius ($r_{\rm e}$) and surface brightness within $r_{\rm e}$ ($I_{\rm e}$) is to be expected for systems that are in virial equilibrium. Interestingly, however, the FP is tilted with respect to this simple virial prediction:
\begin{equation}
    r_{\rm e} \propto  \sigma^a I_{\rm e}^b\,
    \label{eq:lfp_powerlaw}
\end{equation}
where the coefficients $a$ and $b$ describe the tilt of the plane, which in the case of virial equilibrium would equal $a=2$ and $b=-1$ for a homologous set of galaxies. In practice, values have been found to be in the range $a\approx [0.7,1.5]$ and $b\approx[-0.9, -0.6]$, depending on the passband and fitting method used, as well as the redshift \citep[e.g.,][]{Jorgensen1996,LaBarbera2010,HydeBenardi2009,Joergensen2013}. By rewriting Eq.~\ref{eq:lfp_powerlaw} in terms of $\MdynL$ \citep[using $\Mdyn\propto r_{\rm e}\sigma^2$ and $L=2\pi r_{\rm e}^2 I_{\rm e}$;][]{Cappellari2006},
\begin{equation}
    \frac{\Mdyn}{L} \propto r_{\rm e}^{-2-(2+a)/(2b)} \Mdyn^{1+(a/2b)}\,,
\end{equation}
which is constant if $a=2$ and $b=-1$, it becomes apparent that the observed tilt of the FP reflects a correlation between $\MdynL$ and $\Mdyn$ or $r_{\rm e}$, the latter of which is often assumed to be subdominant. 

Although these various studies of the FP have led to a consensus on the existence of a rotation of the FP with respect to the virial plane, the origins for this rotation and precise values of the tilt have been debated extensively for the past decades, without reaching a consensus. In addition to being highly sensitive to measurement choices and uncertainties, the tilt depends strongly on the chosen fitting method and sample selection biases \citep[see, e.g.,][]{HydeBenardi2009,Magoulas2012}, which leads to large uncertainties especially toward higher redshifts \citep[e.g.,][]{Holden2010,Joergensen2013,deGraaff2021}. Nevertheless, multiple causes have been proposed to explain the observed deviation from the scalar virial theorem, which can be best understood by decomposing $\MdynL$ and assessing how the different components scale with $\Mdyn$ \citep{HydeBenardi2009}:
\begin{equation}
    \frac{\Mdyn}{L} = \frac{\Mdyn}{\Mtot} \frac{\Mtot}{M_*} \frac{M_*}{L}\,,
    \label{eq:mdynl_decomp}
\end{equation}
where $M_*$ and $\Mtot$ are the stellar and total (dark matter and baryonic) mass, respectively.

First, the departure from the expected virial plane may reflect the fact that the assumption of homology is inaccurate, captured by the ratio $\Mdyn/\Mtot \neq 1$. Quantified using the S\'ersic index \citep{Sersic1968}, the effects of non-homology were shown by some studies to play a key role \citep{Bender1992,Graham1997,Prugniel1997,Trujillo2004,Desmond2017}. However, others have found more modest or negligible contribution arising from variation in the galaxy structure, based on dynamical modelling or strong lensing results \citep[e.g.,][]{Cappellari2006,Bolton2007,Bolton2008,DEugenio2021}.

Second, broader agreement has been reached on the magnitude of the contribution to the tilt from the mass dependence of $\Mtot/M_*$, which depends on the `dark' mass within galaxies. Crucially, this ratio is not simply the dark matter fraction (assuming negligible gas mass), but also includes missing mass due to uncertainties in the stellar initial mass function (IMF), as stellar masses that are estimated from spectral energy distribution (SED) modelling often rely on the assumption of a universal IMF and therefore carry a systematic uncertainty \citep[for a review on SED modelling, see][]{Conroy2013}. Although this dark component is expected to contribute significantly to the tilt of the FP \citep[$\sim50\%$;][]{Renzini1993,HydeBenardi2009,Graves2010III}, distinguishing
between the effects of variations in the IMF versus the dark matter fraction is challenging. Recent observational work based on simple dynamical models has suggested that IMF variations can fully explain the observed relation between $\Mdyn/M_*$ and $\Mdyn$, by allowing for a non-universal IMF that can vary between galaxies as well as radially within galaxies \citep{Bernardi2018,Marsden2022}. However, others have shown that this would be difficult to reconcile with the observed correlations of stellar population properties throughout the FP \citep{Graves2010III}, or have found evidence for variations in both the IMF and dark matter content in galaxies \citep[e.g.,][]{Cappellari2013a,Cappellari2013b}.

The third component arises from variations in the stellar population properties across galaxies, i.e., variations in $M_*/L$. By evaluating the tilt of the FP in different passbands or by explicitly estimating $M_*/L$, the effects of $M_*/L$ variations have been shown to be insufficient to fully explain the observed tilt, but may account for up to half of this tilt \citep[e.g.,][]{LaBarbera2008,HydeBenardi2009,Graves2010III,Bernardi2020,DEugenio2021}.

\subsection{The stellar mass FP}\label{sec:intro_mfp}

The effects of $M_*/L$ variations across and along the FP can be addressed by explicitly estimating $M_*/L$ independently, by fitting the spectral or photometric SEDs with stellar population models.
We can then gain insight into $\Mdyn/M_*$ alone, a quantity that is of great interest, as it depends on the formation and structural evolution of galaxies, such as the effects of mergers \citep[e.g.,][]{Hopkins2008}.
\citet{Zaritsky2006,Zaritsky2008} first proposed the fundamental manifold, a 3D scaling relation within the 4D parameter space of the galaxy kinematics, size, surface brightness and $M_*/L$.  \citet{HydeBenardi2009} showed that similar results can be achieved by modifying the FP, replacing the surface brightness by the stellar mass surface density ($\Sigma_*$):
\begin{equation}
    r_{\rm e} \propto  \sigma^\alpha \Sigma_*^\beta\,,
    \label{eq:mfp_powerlaw}
\end{equation}
which is referred to as the stellar mass FP. Including $M_*/L$ also results in a lower intrinsic scatter about the scaling relation (i.e., the scatter after accounting for measurement uncertainties) than the standard luminosity FP.

Importantly, \citet{Zaritsky2008} showed that this framework, which up to then had focused on dynamically-hot spheroids, can be extended to disc-like structures as well, if the dynamical measurement ($\sigma$) explicitly includes galaxy rotation in addition to the random motions of stars. Later work demonstrated that both star-forming and quiescent galaxies follow the same stellar mass FP, with nearly identical tilt, zero point and scatter \citep{Bezanson2015,Aquino2020}, and that this result holds out to $z\sim 1$ with minimal evolution in the FP \citep{deGraaff2020,deGraaff2021}.

These results appear to be at odds with the observation that star-forming galaxies obey the TF relation, which is explicitly independent of surface brightness or another third parameter \citep[e.g.,][]{Zwaan1995,Courteau1999,Meyer2008,Lelli2019}. Furthermore, it casts doubt on earlier theoretical studies, which suggested that the dissipation of gas in galaxies plays a critical role in shaping the FP: using simulations of merging galaxies, dissipational mergers were shown to give rise to the observed tilt of the FP, with the tilt of the FP being preserved under further dissipationless mergers \citep{Boylan2006,Robertson2006a,Hopkins2008}. As a result, \citet{Hopkins2008} showed explicitly that discs and spheroids have a different dependence of $\Mdyn/M_*$ on mass.

\subsection{The FP in cosmological simulations}

Cosmological simulations may offer new insight into the origins of the FP, as, unlike the simulations of galaxy mergers, they do not a priori assume a formation channel for the FP. These large simulations have been shown to produce a wide diversity in galaxy morphologies and kinematic structures \citep[e.g.,][]{Snyder2015,Correa2017,Thob2019}, and also reproduce key observed relations such as the galaxy stellar mass function and stellar mass-size relation \citep[e.g.,][]{Schaye2015,Genel2018,Dave2019}, although simulations are typically calibrated to achieve these latter goals. 

However, the galaxy structure and dynamics are not `tuned' explicitly: the FP therefore poses both an interesting test of the realism of a simulation, and an opportunity to gain understanding of the drivers behind the relation itself. Focusing solely on early-type galaxies, different studies have shown that simulations such as Illustris, Illustris-TNG, Horizon-AGN and EAGLE form a FP that approximately resembles observations, but with significant variation in the measured tilt and scatter \citep{Rosito2019,Rosito2021,DOnofrio2020,Lu2020}. Additionally, \citet{DOnofrio2020} showed that galaxies follow a complex trajectory through the parameter space of the FP, and suggest that the low-redshift FP arises from a combination of galaxy mergers and the passive ageing of galaxies. On the other hand, \citet{Rosito2021} used the Horizon-AGN and Horizon-noAGN simulations to show that black hole feedback is a critical factor to reproduce the observed FP.

Taking a more holistic approach, \citet{Ferrero2021} evaluated the relation between the circular velocity, stellar mass and size for dispersion-dominated quiescent galaxies and rotation-dominated star-forming galaxies in the EAGLE and Illustris-TNG simulations. They suggest that, as a consequence of the stellar-halo mass relation, by which galaxies of fixed $M_*$ occupy a narrow range in halo mass, galaxy size becomes the only differentiating parameter. Star-forming discs are larger than quiescent spheroids at fixed $M_*$, and therefore encompass relatively more dark matter within the effective radius. The TF and FP relations are therefore suggested to arise solely from variations in the dark matter fraction, with the TF relation being independent of surface brightness due to the independence of the circular velocity on size at large enough radii.

Although an intriguing result, it omits the fact that the observed structures of star-forming and quiescent differ not only in size, but also in morphology. As a result, the shape of the gravitational potential may be expected to vary as a function of galaxy type, leading to the aforementioned effects of non-homology on the FP. Furthermore, observational biases, due to $M_*/L$ gradients in galaxies and differences in measurement methods, have been shown to have a significant effect on the obtained galaxy scaling relations and are important to take into account when comparing simulations and observations \citep[e.g.,][]{Price2017,Bottrell2017a,Bottrell2017b,vdSande2019,deGraaff2022}. 

In this paper, we aim to assess the different effects of non-homology, the dark matter content and observational uncertainties on the tilt and scatter of the stellar mass FP for both quiescent and star-forming galaxies. By using the EAGLE cosmological hydrodynamical simulations \citep{Schaye2015}, we systematically introduce one of these components at a time, and evaluate whether these results differ for quiescent and star-forming galaxies. We build on the mock observations and measurements presented in \citet{deGraaff2022} to arrive at a FP that is as close as possible to the observed FP, and show how selection biases affect the measurement and interpretation of the FP.

The simulations used and the different definitions of galaxy size, mass and velocity dispersion are described in Section~\ref{sec:data}. In Section~\ref{sec:FP_3D} we present the simulated FP, and discuss the effects of non-homology and variations in the dark matter fractions. We introduce observational effects, measurement and selection biases in Section~\ref{sec:FP_obs}, where we demonstrate how the tilt of the FP is sensitive to these different effects. Moreover, we explore the possible additional complication of a non-universal IMF. We discuss these results in Section~\ref{sec:discussion} and show how the FP and TF relation may be reconciled. Our main results are summarised in Section~\ref{sec:conclusion}.

\section{Data and methods}\label{sec:data}

\subsection{EAGLE simulations}\label{sec:eagle}

The EAGLE simulations are a set of cosmological smoothed particle hydrodynamics (SPH) simulations  \citep{Schaye2015,Crain2015}. These simulations all assume a flat $\Lambda$CDM cosmology with cosmological parameters from the \citet{Planck2014} ($\Omega_{\rm m}=0.307$, $\Omega_{\rm b}=0.0482$ and $H_0 = 67.77\,\rm km\,s^{-1}\,Mpc^{-1}$), but vary in the volume, resolution and subgrid model used. In this work, we will focus mainly on the reference model with a volume of $100^3\,$ comoving Mpc$^3$ (cMpc; L0100N1504), which has a mass resolution of $m_{\rm DM}=9.7\times10^6\,{\rm M_\odot}$ and $m_{\rm b}=1.81\times10^6\,{\rm M_\odot}$ for the dark matter particles and initial mass of the gas particles, respectively. With the Plummer-equivalent gravitational softening scale of $\epsilon=0.70$\,proper kpc at $z<2.8$, this amounts to an effective spatial resolution of $\approx2\,$proper kpc. The reference model assumes a Chabrier IMF \citep{Chabrier2003} for the star formation prescription, which together with the other subgrid prescriptions was calibrated to reproduce the $z=0$ stellar mass function and stellar mass-size relation.

To assess the numerical convergence of our results, we use the smaller simulation of $25^3\,$cMpc$^3$ for the recalibrated model (L0025N0752), which has a resolution that is 8 times higher. Furthermore, to examine the effects of a non-universal IMF, we use the simulations by \citet{Barber2018}. These simulations implement a variable IMF into the reference EAGLE model, by allowing either the low- or high-mass end of the Kroupa double power law IMF \citep[i.e., above or below $0.5\rm\,M_\odot$;][]{Kroupa2001} to vary according to the pressure of the local interstellar medium. The models were calibrated to reproduce the scaling relation between the excess stellar mass-to-light ratio and stellar velocity dispersion that has been observed for early-type galaxies in the local Universe \citep{Cappellari2013b}, while simultaneously matching other key observables such as the $K$-band luminosity function and the relation between the half-light radius and luminosity. The resulting bottom-heavy (LoML0050N0752) and top-heavy (HiML0050N0752) models were run in $50^3\,$cMpc$^3$ volumes, which can be readily compared with the reference run of the same volume.

We will focus on galaxies at $z=0.1$, for which mock images of the light distributions that include realistic dust attenuation, noise and seeing are available from \citet{Trayford2017} and \citet{deGraaff2022}. Throughout, galaxies are defined in the usual way, as the self-bound substructures that are identified within haloes by the \textsc{subfind} algorithm \citep{Springel2001,Dolag2009}. This mechanism also allows for a separation of central and satellite galaxies, which is used in Section~\ref{sec:FP_3D}. Moreover, we distinguish between star-forming and quiescent galaxies based on the specific star formation rate (sSFR) measured within a spherical aperture of radius $30$\,proper kpc centred around the potential minimum \citep[obtained from the online public database;][]{McAlpine2016}: quiescence is defined as $\rm sSFR<10^{-11}\,yr^{-1}$. Lastly, in what follows all length units will be quoted as proper lengths unless explicitly noted otherwise.

\subsection{Galaxy sizes and masses}\label{sec:size_mass}

As discussed in \citet{deGraaff2022}, the sizes of galaxies depend strongly on whether these are measured from the stellar mass or optical light distributions. A secondary effect is the measurement technique used, i.e., whether quantities are measured with a growth curve method or by using parametric models. 
The mass that is enclosed within the effective radius then changes correspondingly. 

Scaling relations, such as the FP, may be expected to be sensitive to these differences. To examine to what extent this makes a difference on the obtained FP, we will use multiple definitions of galaxy size and mass throughout the paper:

\begin{itemize}
    \item $r_{\rm e, 3D}$\,: the radius that encloses half of the stellar mass within a spherical aperture of radius 100\,kpc centered around the potential minimum \citep[see also][]{Furlong2017}. We consequently define the stellar mass within a spherical aperture of this radius as $M_*(<r_{\rm e,3D})$, and the total mass within the same aperture $M_{\rm tot}(<r_{\rm e,3D})$. The total mass is the sum of the dark matter, stellar, gas and black hole particle masses.
    \item $r_{\rm e, *}$\,: half-mass semi-major axis obtained from S\'ersic profile fitting to projected images (along the $z$-axis of the simulation box) of the stellar mass distributions from \citet{deGraaff2022}. $M_*(<r_{\rm e, *})$ is half of the stellar mass of the integrated, best-fit S\'ersic model.
    \item $r_{\rm e, r}$\,: half-light semi-major axis obtained from S\'ersic profile fitting to images of the optical light distributions in the $r$-band \citep[again, using the random projection along the $z$-axis of the simulation box;][]{Trayford2017,deGraaff2022}. We obtain stellar masses $M_*(<r_{\rm e, r})$ by multiplying the mass-to-light ratio within a spherical aperture of radius 30\,kpc ($M_*/L_r$) by half of the luminosity of the best-fit S\'ersic profile.    
\end{itemize}

\subsection{Sample selection}\label{sec:sample}

\begin{figure}
    \centering
    \includegraphics[width=\linewidth]{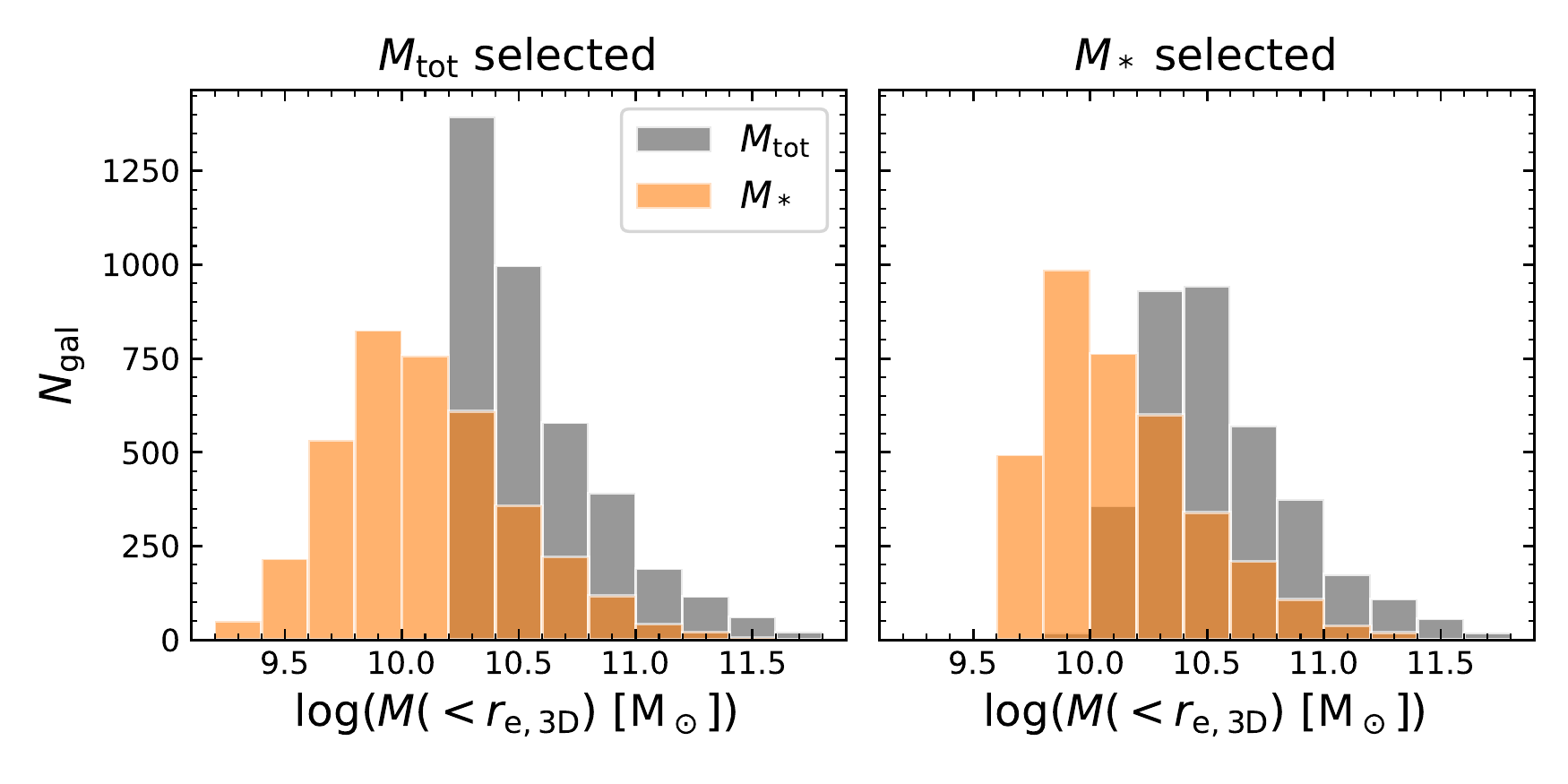}
    \caption{Distributions of the total mass and stellar mass enclosed within the 3D effective radius, for the samples selected by the total mass (left) and stellar mass (right). Total masses include all particle masses, i.e., dark matter, stellar, gas and black hole particles.}
    \label{fig:mass_hists}
\end{figure}

A key goal of this work is to quantify the effects of sample selection on the obtained FP. In large imaging surveys, the selection of galaxies is limited by total flux and/or surface brightness, depending on the apparent size and point spread function. For a chosen maximum distance, the flux-limited samples can then be used to construct sub-samples that form an accurate representation of the galaxy stellar mass function, down to a specified stellar mass limit.

Although cosmological simulations are by construction complete in mass, low-mass galaxies in the simulation are affected by the limited resolution of the simulation \citep[resulting in, e.g., unreliable sizes;][]{Ludlow2019,Ludlow2021}.
We therefore impose a selection on the galaxy mass, and construct two samples that are complete in (i) total mass and (ii) stellar mass. First, we calculate the total mass enclosed within $r_{\rm e,3D}$, and select galaxies for which $M_{\rm tot}(<r_{\rm e,3D})>10^{10.2}\,\rm M_\odot$ and that also contain $>10^3$ stellar particles (96\% of the sample contain $>1\times10^4$ stellar particles). For the $100^3$\,cMpc$^3$ box (Section~\ref{sec:eagle}), this results in a sample of 3758 galaxies.

The second sample follows the selection of \citet{deGraaff2022}: this sample is selected by requiring the aperture stellar mass $M_*>10^{10}\,\rm M_\odot$, and thus effectively selected by $M_*(<r_{\rm e,3D})>10^{9.7}\,\rm M_\odot$, with additional criteria imposed on the quality of the S\'ersic profile fits. {Namely, as discussed in detail in \citet{deGraaff2022}, we require that the fit has converged within the parameter boundaries (removing 35 objects) and pass our visual inspection (by not showing strong residual features; removing 29 objects). We refer the reader to this previous work for further examples of the fitting procedure and discussion of the obtained sizes and morphologies.} The stellar mass-selected sample consists of 3624 galaxies, of which 3560 are flagged as having good S\'ersic profile fits.
The mass distributions for these two different selections are shown in Fig.~\ref{fig:mass_hists}. The two samples were constructed to contain an approximately equal number of objects, and to have a significant overlap, with 3183 galaxies appearing in both samples ($\approx 90\%$).

\subsection{Velocity dispersion measurements}\label{sec:vdisp}

The third critical component that enters the FP is the velocity dispersion, which reflects the depth and shape of the gravitational potential. Observationally, this quantity is traced by the light emitted by stars, as their motion along the line of sight leads to a broadening of stellar absorption lines. {Importantly, this motion can come from both the disordered motion and the ordered rotation of the stars (see also Section~\ref{sec:intro_mfp}).}

\begin{figure}
    \centering
    \includegraphics[width=\linewidth]{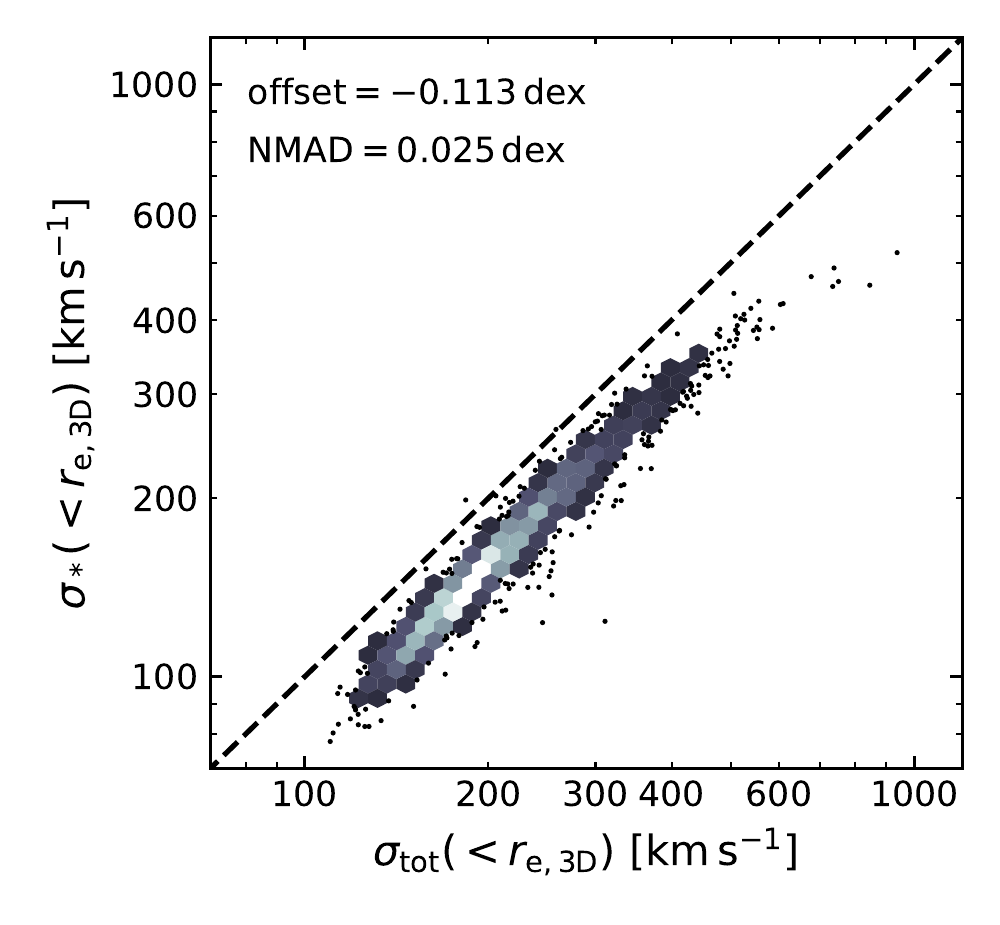}
    \caption{Comparison between the velocity dispersion of the stellar particles within $r_{\rm e,3D}$ and the velocity dispersion of the stellar, dark matter, and gas particles within the same aperture (Eq.~\ref{eq:kinetic_energy} and \ref{eq:sigma_3d}). Individual data points are shown for sparsely sampled areas of the figure  \citep[created using \textsc{densityplot};][]{Krawczyk:13361}. The total velocity dispersion is greater than the stellar velocity dispersion, suggesting that the dark matter particles are dynamically hotter. }
    \label{fig:sigma_tot}
\end{figure}

To systematically assess the impact of these different observational effects, we begin by measuring the velocity dispersion within spherical apertures of radius $r_{\rm e, 3D}$. Following \citet{McAlpine2016}, the kinetic energy of a collection of particles is calculated as
\begin{equation}
    K = \frac{1}{2} \sum_i m_i (\mathbf{v} - \mathbf{v_{\rm pec}})^2\,,
    \label{eq:kinetic_energy}
\end{equation}
where $m$ and $\mathbf{v}$ are the mass and velocity of the particle, respectively, and $\mathbf{v_{\rm pec}}$ is the peculiar velocity of the galaxy, which we calculate as the mass-weighted average velocity of the stellar particles within an aperture of 30\,kpc centred around the potential minimum. We calculate the velocity dispersion within a radius $r$ as the mean-square speed {\citep[thereby including both the random motion and rotation of the particles, see][Chapter 4.8.3]{BinneyTremaine1987}}, which depends on the kinetic energy and mass of the particles enclosed within the same radius:
\begin{equation}
    \sigma(<r) = \sqrt{\frac{2 K(<r)}{ M(<r) }}\,.
    \label{eq:sigma_3d}
\end{equation}
We calculate two versions of this velocity dispersion: the stellar velocity dispersion $\sigma_*(< r_{\rm e,3D})$ that is based on the kinetic energy and mass of the stellar particles within $r_{\rm e,3D}$, and the total velocity dispersion $\sigma_{\rm tot}(< r_{\rm e,3D})$, which includes the dark matter, stellar and gas particles in Eq.~\ref{eq:kinetic_energy} and \ref{eq:sigma_3d}. These two different velocity dispersions are compared in Fig.~\ref{fig:sigma_tot}, which shows that the total velocity dispersion is systematically larger than the stellar velocity dispersion and with small scatter {(calculated as the normalised median absolute deviation; NMAD)}.

Next, we apply a measurement that is in better agreement with observational methods. We use the size, axis ratio and position angle from the best-fit S\'ersic profile to construct an elliptic cylindrical aperture, which is centred around the potential minimum and has a length along the $z$-axis of the simulation box of $\pm 50\,$kpc. Selecting all particles within the aperture, we obtain the spatially-integrated line-of-sight velocity dispersion ($\sigma_{\rm los}$; i.e., along the $z$-axis) by first calculating the weighted mean
\begin{equation}
    \langle v_{\rm los} \rangle = \frac{\sum_i w_i v_{{\rm z},i}}{\sum_i w_i}\,, 
\end{equation}
followed by
\begin{equation}
    \sigma^2_{\rm los} = \frac{\sum_i w_i (v_{{\rm z},i} - \langle v_{\rm los} \rangle )^2}{\sum_i w_i}\,,
\end{equation}
where $v_{\rm z}$ is the velocity of the particle along the line of sight, and $w_i$ is the weight. {This spatially-integrated measurement of the velocity dispersion therefore also includes both the rotational and random motions of the particles along the line-of-sight direction.} Using the S\'ersic profile fits to the stellar mass images (see Section~\ref{sec:size_mass}) for the apertures and the current mass of the stellar particles as weights, we obtain $\sigma_*(<r_{\rm e, *})$. Similarly, using the S\'ersic profile fits to the optical light and weighting by the luminosities of the particles, we obtain $\sigma_*(<r_{\rm e, r})$. Here, we have chosen to use the $r$-band S\'ersic profile fits to construct the apertures, but $g$-band luminosities for the weighting of the velocity dispersions, to mimic observations where the more prominent absorption lines are around $\sim5000\,$\AA. We show in Appendix~\ref{apdx:nodust} that the mismatch between the waveband chosen for these two different measurements has only a small effect.

\begin{figure}
    \centering
    \includegraphics[width=0.95\linewidth]{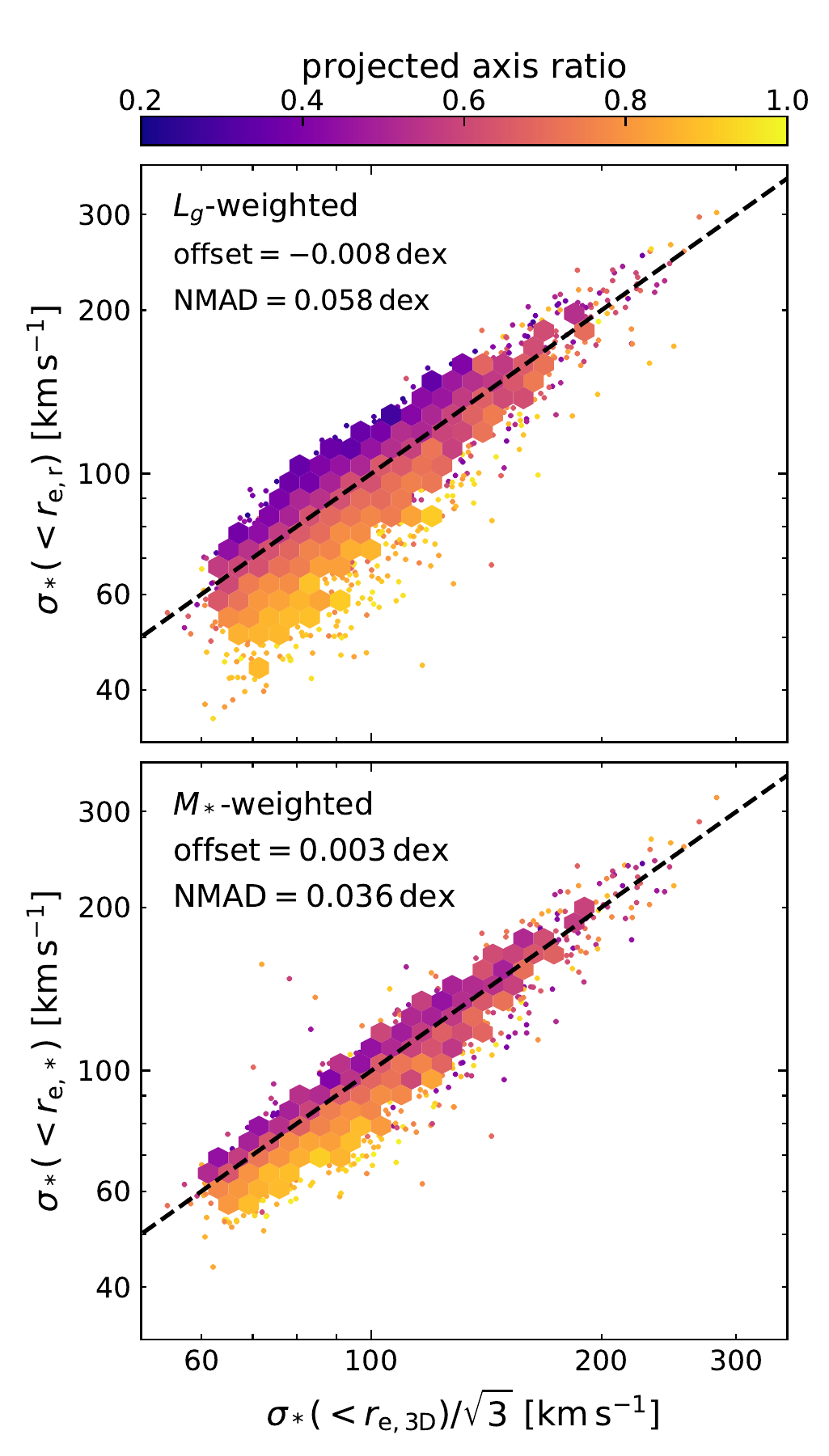}
    \caption{The {spatially-integrated} line-of-sight stellar velocity dispersions measured in elliptic cylindrical apertures versus the 3D stellar velocity dispersion, for the luminosity-weighted (top) and stellar mass-weighted (bottom) measures. The colour coding reflects the median projected axis ratio in each bin, demonstrating that the scatter about the unit slope can be attributed mainly to projection effects {and the degree to which the rotational motion along the line of sight contributes to the measured velocity dispersion.}  }
    \label{fig:sigma_obs}
\end{figure}

We compare the three different stellar velocity dispersions in Fig.~\ref{fig:sigma_obs}, which shows the luminosity-weighted and stellar mass-weighted velocity dispersion along the line of sight as a function of the 3D velocity dispersion calculated with Eq.~\ref{eq:sigma_3d}. Although there is good agreement between the different measures, there is a large scatter that is particularly strong toward low $\sigma_*$. By colour coding the data with the projected axis ratios of the S\'ersic models, it becomes apparent that this is due to projection effects: galaxies that are near face-on are observed to have a significantly lower velocity dispersion than edge-on galaxies, which is to be expected for oblate systems that are strongly rotating.

\section{The simulated Fundamental Plane}\label{sec:FP_3D}

As discussed in Section~\ref{sec:intro}, the tilt of the FP reflects the deviation from the simple prediction of virial equilibrium for homologous systems: $M(r) \propto r\,\sigma^2$. In this Section, we present the theoretical perspective on the cause of these deviations, by evaluating the effects of structural non-homology and variations in the mass compositions of galaxies on the FP. Observational effects and selection biases will then be discussed in Section~\ref{sec:FP_obs}.

\subsection{Dynamical tracers of the total galaxy mass}\label{sec:mtot_fp}

In Section~\ref{sec:vdisp} we defined two different tracers of the galaxy dynamics: a stellar velocity dispersion, and the total velocity dispersion. Clearly, $\stot$ is a quantity that cannot be measured observationally, however, it may seem a natural choice when the aim is to recover $\Mtot$. We begin by examining the planar relation
\begin{equation}
    \retd \propto \stot^\alpha(<\retd) \, \Sigma_{\rm tot}^\beta\,,
\end{equation}
where $\Sigma_{\rm tot} = \Mtot(<\retd)/(\pi\retd^2)$, and the coefficients $\alpha=2$ and $\beta=-1$ for homologous systems in virial equilibrium.

\begin{figure*}
    \centering
    \includegraphics[width=0.94\linewidth]{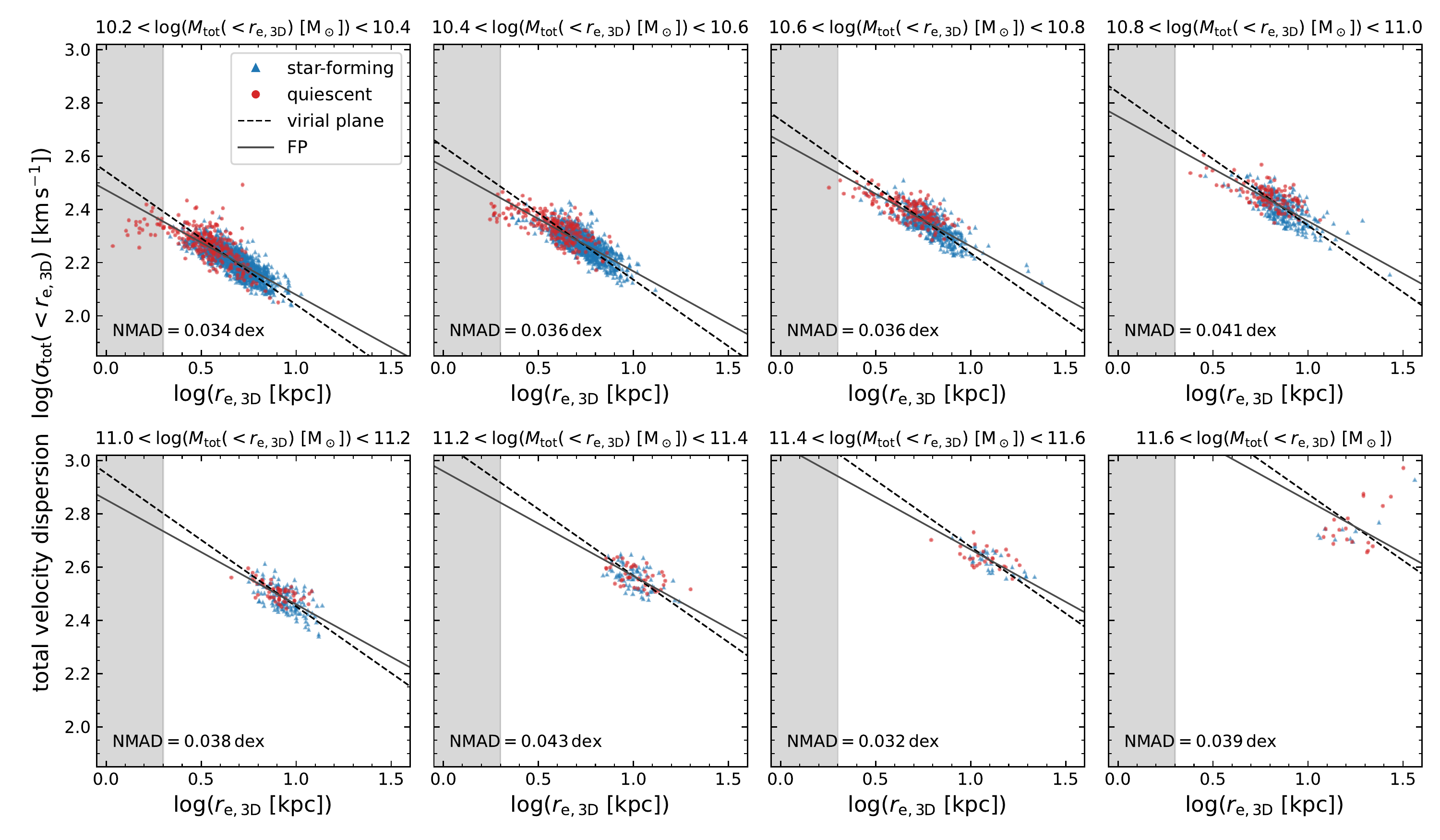}
    \caption{Velocity dispersion of all particles (dark matter, stellar, gas{; see Section~\ref{sec:vdisp}}) as a function of the 3D half-mass radius, in bins of the total mass $\Mtot(<\retd)$. The sample is divided into star-forming (blue) and quiescent (red) galaxies by the sSFR; the grey shaded area indicates $\retd<2\,$kpc, where measurements are likely to be affected by the resolution of the simulation. Dashed lines, with the intercept fit to the data in each panel, show {the slope of} the virial plane that galaxies are expected to follow if they are both in virial equilibrium and homologous. The best-fit total mass FP (solid lines; Table~\ref{tab:FP_3D}) deviates slightly from the virial plane due to the effects of non-homology; the orthogonal scatter about this line is printed in each panel.  }
    \label{fig:stot_fixedmtot}
\end{figure*}

To aid in the visualisation of this 3D relation, we form narrow bins in $\log(\Mtot(<\retd))$, and show the relation between $\stot(<\retd)$ and $\retd$ for galaxies in the $\Mtot$-selected sample in Fig.~\ref{fig:stot_fixedmtot}. Here, the sample is divided into star-forming (blue) and quiescent (red) galaxies, and shaded areas show the region in parameter space that is likely to be affected by the limited resolution of the simulation. We fit linear relations with a fixed slope ($m=-0.5$) to the data in each panel, which represent lines of constant $\Mdyn$ (dashed lines) and thus the {tilt of the} virial plane.

\begin{figure*}
    \centering
    \includegraphics[width=0.94\linewidth]{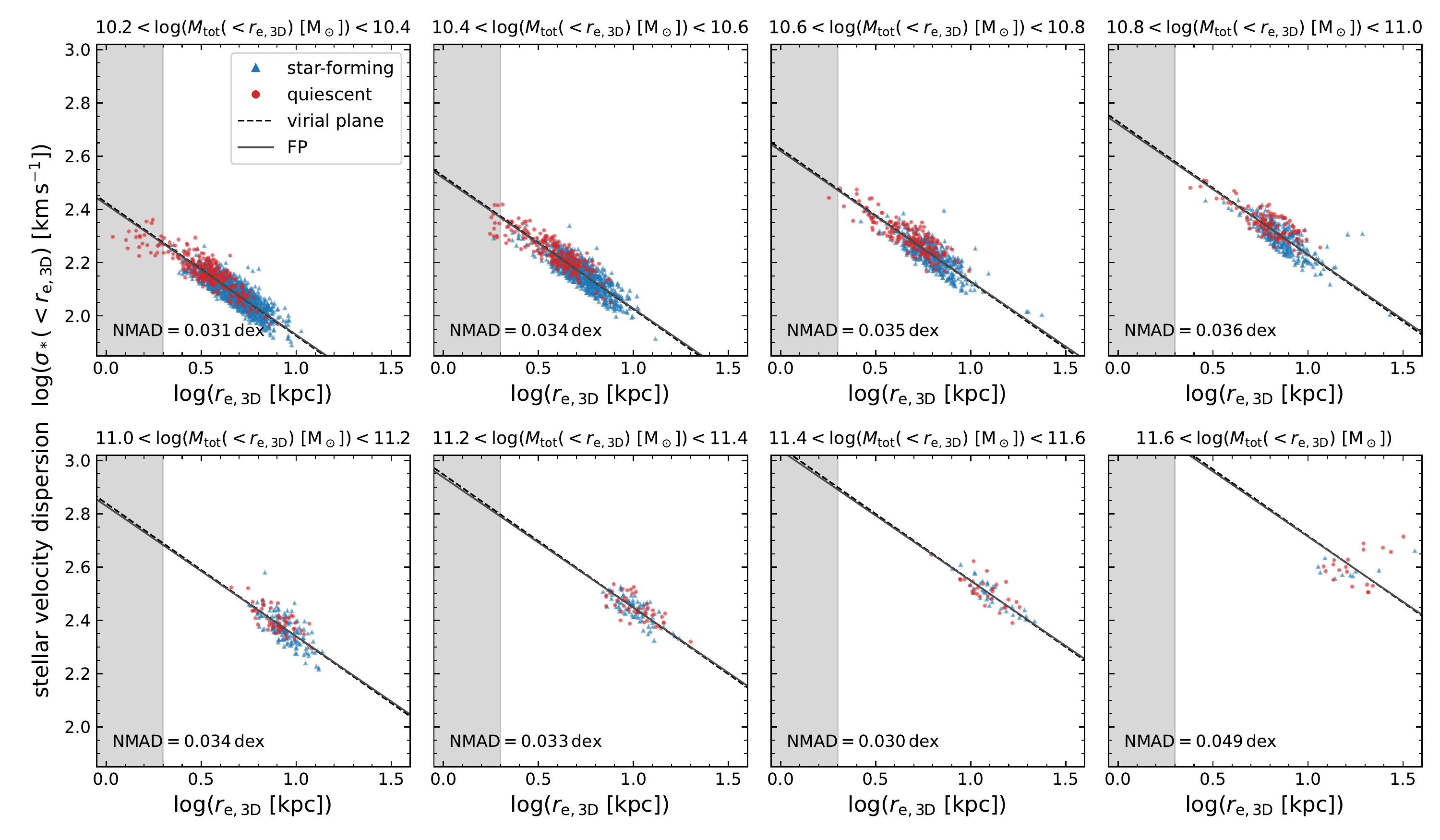}
    \caption{As Fig.~\ref{fig:stot_fixedmtot}, but showing the velocity dispersion of only the stellar particles. The scatter about the dashed lines is marginally lower, whereas the number of strong outliers (caused by recent mergers) is slightly increased. The {tilt of the} FP is closer to the virial plane when using $\sigma_*\,$, likely due to the decreased dependence on the galaxy structure and the fact that $\stot$ is strongly influenced by the dynamically-hot dark matter particles (see main text). }
    \label{fig:sstar_fixedmtot}
\end{figure*}

There is a tight sequence around these relations in all mass bins, except for the very highest mass bin that spans a broad range in mass (up to $10^{12.25}\,\rm M_\odot$). {Moreover, toward higher total mass, the sequence itself shifts toward larger sizes and higher velocity dispersions.} The assumption of virial equilibrium and homology, which would imply $\Mdyn/\Mtot =1$, therefore seems to be reasonable. However, toward smaller sizes, there appears to be a systematic offset with respect to the dashed lines. 

We fit the coefficients $\alpha$ and $\beta$ by minimising the sum of the absolute orthogonal distances to the plane:
\begin{equation}
    \Delta_{\rm FP} = \frac{\left | \log(\retd) -  \alpha\, \log(\sigma_{\rm tot}(<\retd)) -  \beta\,\log(\Sigma_{\rm tot}) - \gamma \right|}{\sqrt{1+\alpha^2 + \beta^2}}\,,
    \label{eq:delta_fp}
\end{equation}
where $\gamma$ is the zero point of the plane. We obtain errors on the fit parameters by bootstrapping the data with 1000 subsamples. Excluding galaxies for which $\retd<2\,$kpc, we find that the coefficients deviate significantly from the virial plane (Table~\ref{tab:FP_3D}), with a stronger deviation for quiescent galaxies than for star-forming galaxies, and with very low scatter about the plane ($0.018\,$dex). Solid lines in Fig.~\ref{fig:stot_fixedmtot} show the best-fit total mass FP of the combined sample, and the scatter about these lines, calculated using the NMAD, is indicated in each panel. We note that the scatter measured in Fig.~\ref{fig:stot_fixedmtot} is larger than presented in Table~\ref{tab:FP_3D} due to the finite bin widths used.

Next, in Fig.~\ref{fig:sstar_fixedmtot} we replace the total velocity dispersion by $\sstar$, which is expected to be a good tracer of the galaxy dynamics because of the collisionless nature of stellar orbits. The results are qualitatively similar to those of Fig.~\ref{fig:stot_fixedmtot}, except with slightly lower scatter. The overall scaling is also lower, as the stellar velocity dispersion is systematically lower than the total dispersion (Fig.~\ref{fig:sigma_tot}). By inspecting the merger trees and images of the strong outliers that are visible in the figure, we find that these few systems are either currently merging with another galaxy or did so in their recent history, and have therefore likely not yet reached equilibrium. 

Fitting the planar relation with $\sstar(<\retd)$ instead of $\stot(<\retd)$ results in coefficients that are even closer, although still not equal, to the virial plane. Interestingly, the separate fits to the quiescent and star-forming subsamples are also in better agreement than before, and the scatter about these different planes is reduced even further ($0.013\,$dex).

\begin{figure}
    \centering
    \includegraphics[width=\linewidth]{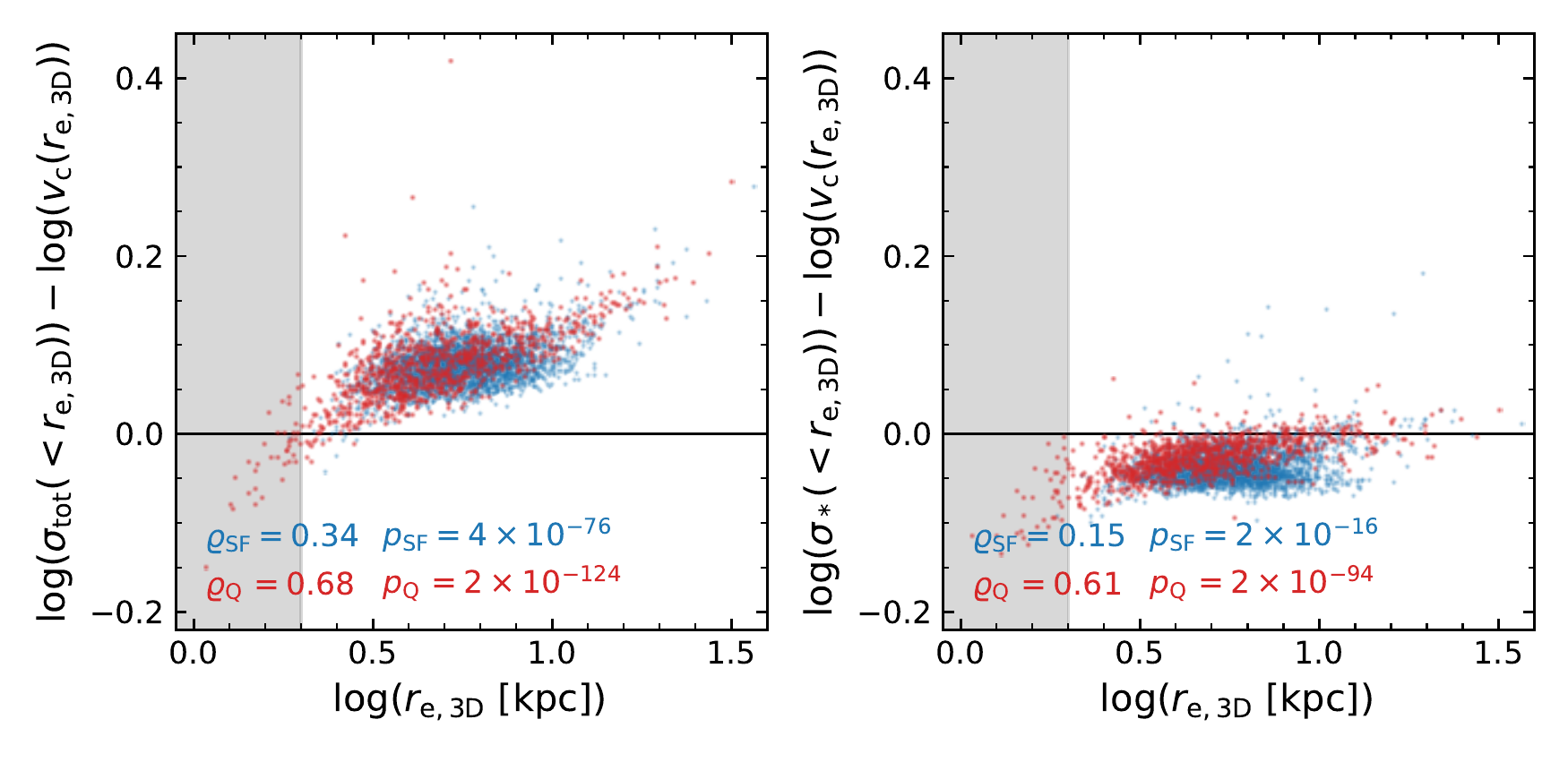}
    \caption{{The offset between the FP and the virial plane, measured as the difference between the total (left) or stellar (right) velocity dispersion and the dispersion predicted by Eq.~\ref{eq:vcirc}, as a function of the half-mass radius. There is not only a difference in the zero-point of the two planes, but also in the tilt: the positive Spearman rank correlation coefficients (and corresponding p-values) indicate that the FP is tilted with respect to the virial plane. This tilt is significant, and is stronger for quiescent galaxies than for star-forming galaxies.} }
    \label{fig:delta_sigma_size}
\end{figure}

{Whereas Figs.~\ref{fig:stot_fixedmtot} and \ref{fig:sstar_fixedmtot} focus on the qualitative differences between tilt of the FP and virial plane, we quantify the differences between the best-fit FP and the virial plane in Fig.~\ref{fig:delta_sigma_size}. We compute the velocity dispersion predicted from the virial plane, which is equivalent to the circular velocity at the effective radius for a spherically symmetric mass distribution,
\begin{equation}
    v_{\rm c}(\retd)= \sqrt{\frac{G \Mtot(<\retd)}{\retd}}\,,
    \label{eq:vcirc}
\end{equation}
where $G$ is the gravitational constant, and hence evaluate how the deviation between the measured velocity dispersion and this circular velocity depends on the half-mass radius.
Fig.~\ref{fig:delta_sigma_size} shows that there is clearly a systematic offset between the zero-points of the total mass FP and virial plane, for both $\stot$ (left) and $\sstar$ (right), which we discuss in the following section (\ref{sec:nonhomology}). By measuring the Spearman rank correlation coefficients, we also find that there is a positive correlation between the deviation from the virial plane and the half-mass radius. This implies that the total mass FP is tilted with respect to the virial plane, and that this tilt is stronger for quiescent galaxies than for star-forming galaxies, and is consistent with the results of our planar fits (Table~\ref{tab:FP_3D}).  }

\begin{table*}
 \caption{Best-fit coefficients of the total mass FP and stellar mass FP, and the orthogonal scatter about the planes, for different dynamical tracers and selection criteria. Galaxies for which $\retd<2\,$kpc are excluded from the fits. }
 \label{tab:FP_3D}
 \begin{tabular}{llcccc}
  \hline
  Relation & Sample selection & $\alpha$ & $\beta$ & $\gamma$ & NMAD  \\
  \hline
  $\log(\retd) = \alpha \log(\stot(<\retd)) + \beta \log(\Sigma_{\rm tot}) + \gamma$  & $\Mtot$  & $1.700 \pm 0.010$ & $-0.834 \pm0.004$ & $4.40\pm0.03$ & $0.0176\pm0.0003$ \\

    & $\Mtot$ \& quiescent & $1.612 \pm 0.018$ & $-0.747 \pm0.011$ & $3.79\pm0.10$ & $0.0184\pm0.0009$ \\
   
   & $\Mtot$ \& star-forming & $1.776 \pm 0.015$ & $-0.865 \pm0.006$ & $4.49\pm0.04$ & $0.0158\pm0.0005$ \\   
 
  $\log(\retd) = \alpha \log(\sstar(<\retd)) + \beta \log(\Sigma_{\rm tot}) + \gamma$  & $\Mtot$  & $1.799 \pm 0.008$ & $-0.940 \pm0.005$ & $5.29\pm0.03$ & $0.0134\pm0.0002$ \\

   & $\Mtot$ \& quiescent & $1.769 \pm 0.012$ & $-0.885 \pm0.007$ & $4.86\pm0.06$ & $0.0122\pm0.0004$ \\
   
  & $\Mtot$ \& star-forming & $1.848 \pm 0.012$ & $-0.949 \pm0.006$ & $5.28\pm0.03$ & $0.0121\pm0.0003$ \\  
  
  $\log(\retd) = \alpha \log(\sstar(<\retd)) + \beta \log(\Sigma_{\rm tot}) + \gamma$  & $M_*$  & $1.829 \pm 0.009$ & $-0.953 \pm0.005$ & $5.35\pm0.04$ & $0.0137\pm0.0003$ \\

   & $M_*$ \& quiescent & $1.790 \pm 0.011$ & $-0.882 \pm0.006$ & $4.80\pm0.05$ & $0.0124\pm0.0004$ \\
   
  & $M_*$ \& star-forming & $1.871 \pm 0.009$ & $-0.963 \pm0.006$ & $5.36\pm0.04$ & $0.0123\pm0.0003$ \\   

  $\log(\retd) = \alpha \log(\sstar(<\retd)) + \beta \log(\Sigma_*) + \gamma$  & $M_*$  & $1.559 \pm 0.008$ & $-0.562 \pm0.003$ & $2.26\pm0.02$ & $0.0186\pm0.0004$ \\

   & $M_*$ \& quiescent & $1.514 \pm 0.010$ & $-0.564 \pm0.005$ & $2.37\pm0.04$ & $0.0191\pm0.0007$ \\
   
  & $M_*$ \& star-forming & $1.591 \pm 0.011$ & $-0.558 \pm0.004$ & $2.17\pm0.02$ & $0.0180\pm0.0005$ \\

  \hline
 \end{tabular}
\end{table*}

\subsection{Effects of non-homology}\label{sec:nonhomology}

The deviation of the different fits for the `total mass' FP from the virial plane raises several questions. Most importantly, we may ask why the dynamical and total masses are different. Secondly, it is unclear why the use of the stellar velocity dispersion results in a FP that is closer to virial than is the case for the total velocity dispersion, given that the orbits of the cold dark matter are also collisionless and the gas fractions are small (see Section~\ref{sec:DM_frac}). 

A difference between $\Mdyn$ and $\Mtot$ within the same spherical aperture of $\retd$ indicates that the measured velocity dispersion differs from the expected dispersion. Either the assumption of virial equilibrium does not hold, or the systems are not homologous. The first is unlikely, as the age of the galaxies at $z=0.1$ is $\sim10^{10}\,$yr, and thus significantly larger than the crossing time ($\sim 10^8\,$yr). Therefore, only for systems that have very recently merged with a significantly large neighbour, might we expect virial equilibrium to not have yet been established, which explains some of the apparent outliers in Figs.~\ref{fig:stot_fixedmtot} and \ref{fig:sstar_fixedmtot}. 

{To examine the effects of non-homology, we again use the difference between the measured velocity dispersion and the velocity dispersion predicted from the virial plane with Eq.~\ref{eq:vcirc} ($\Delta\log\sigma$; equivalent to the offset between the measured FP and the virial plane). We then evaluate how this calculated deviation depends on different galaxy properties.}
% {the velocity dispersion predicted from the virial plane}, which is equivalent to the circular velocity at the effective radius for a spherically symmetric mass distribution,
% \begin{equation}
%     v_{\rm c}(\retd)= \sqrt{\frac{G \Mtot(<\retd)}{\retd}}\,,
%     \label{eq:vcirc}
% \end{equation}
% where $G$ is the gravitational constant, 
%We then evaluate how the deviation between the measured velocity dispersion and this circular velocity ($\Delta\log\sigma$; equivalent to the offset between the measured FP and the virial plane) depends on different galaxy properties.}

\begin{figure}
    \centering
    \includegraphics[width=\linewidth]{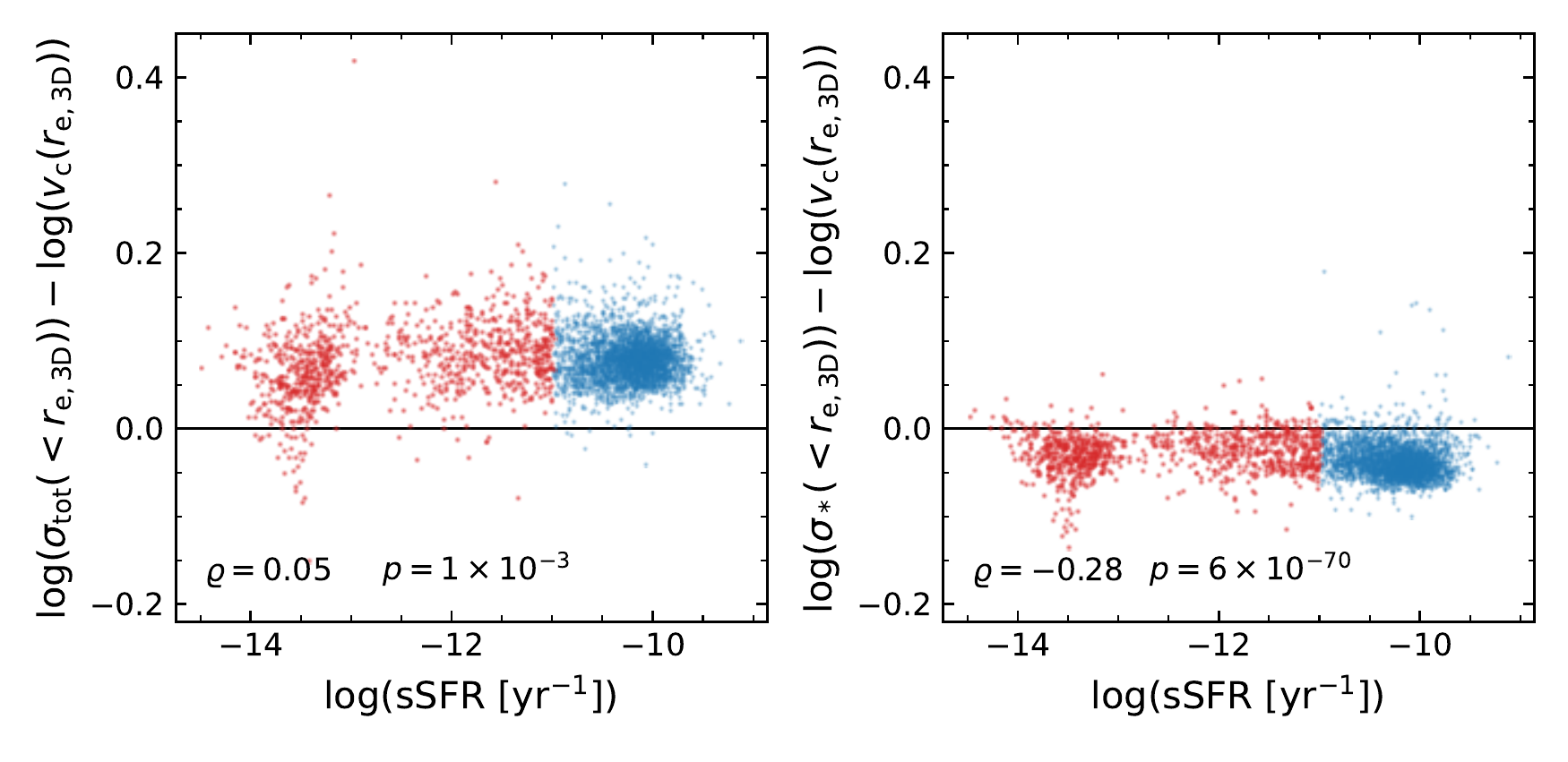}
    \caption{Difference between the measured total (left) or stellar (right) velocity dispersion and the dispersion predicted from the virial theorem (Eq.~\ref{eq:vcirc}), as a function of the instantaneous sSFR. The Spearman rank correlation coefficients indicate that the sSFR has negligible impact on the deviation in $\stot$, and only a weak effect on $\sigma_*$. }
    \label{fig:sigma_ssfr}
\end{figure}

Fig.~\ref{fig:sigma_ssfr} shows $\Delta\log\sigma = \log(\sigma(\retd))-\log(v_{\rm c}(\retd))$ for both the total and stellar velocity dispersion as a function of the instantaneous sSFR. This sSFR of course cannot be expected to drive the effects of structural non-homology, but may correlate with the galaxy structure and therefore lead to a correlation with $\Delta\log\sigma$. {Indeed, \citet{Correa2017}, \citet[][Figs. 2 and 3]{Thob2019} and \citet[][Figs. 11, 12 and 14]{deGraaff2022} show that the 3D shape and dynamical properties as well as the inferred projected structural parameters depend on the colour, sSFR and stellar mass.} We note that 419 galaxies have $\rm SFR=0\,M_\odot\,yr^{-1}$, which for visualisation purposes only have been given an offset of $\rm 0.001\,M_\odot\,yr^{-1}$ (corresponding to the cloud of points below ${\rm sSFR}\lesssim -13\,{\rm yr}^{-1}$). The Spearman rank correlation coefficients ($\varrho$) indicate that the dependence on sSFR is, at most, weak. This is not unexpected, given that the individual fits to the quiescent and star-forming populations (Table~\ref{tab:FP_3D}) both deviate from the virial plane in an approximately equal way.

Next, we evaluate explicitly whether $\Delta\log\sigma$ correlates with differences in the galaxy structure. To do so, we use the S\'ersic indices measured from the projected stellar mass distributions (Section~\ref{sec:size_mass}), as well as the 3D structural parameters measured by \citet{Thob2019}. The 3D stellar mass distributions were modelled with ellipsoids and quantified by the parameters $\epsilon_* = 1 - C/A$, which describes the flattening of the short axis ($C$) relative to the longest axis ($A$), and the triaxiality $T = (A^2-B^2)/(A^2-C^2)$, which also depends on the intermediate axis ($B$). A value of $T\approx0$ thus corresponds to an oblate system, whereas $T\approx 1$ implies a prolate shape. The shape of the dark matter (within an approximately equal aperture as the stellar mass distribution) was measured in a similar way, and is quantified by the flattening parameter $\epsilon_{\rm DM}$.
Furthermore, the kinematic structural parameters from \citet{Thob2019} provide information on the mean orbital properties of the stars, and are measured from cylindrical apertures that are aligned along the long axis with the total angular momentum of the stellar particles. The anisotropy in the velocity dispersion was then calculated as $\beta = 1- (\sigma_z/\sigma_0)^2$, where $\sigma_z$ is the stellar velocity dispersion along the long axis of the cylinder (the rotation axis of the galaxy) and $\sigma_0$ is the dispersion in the plane perpendicular to this axis, and thus reflects the degree of disordered motion along the radial or tangential direction. Finally, the quantity $\kappa_{\rm co} = K_{\rm co}^{\rm rot} / K$ measures the fraction of the total kinetic energy that is due to the co-rotation of stars along the axis defined by the total angular momentum. A value of $\kappa_{\rm co}=1$ therefore corresponds to a dynamically-cold disc in which all stars follow circular orbits.

\begin{figure*}
    \centering
    \includegraphics[width=0.76\linewidth]{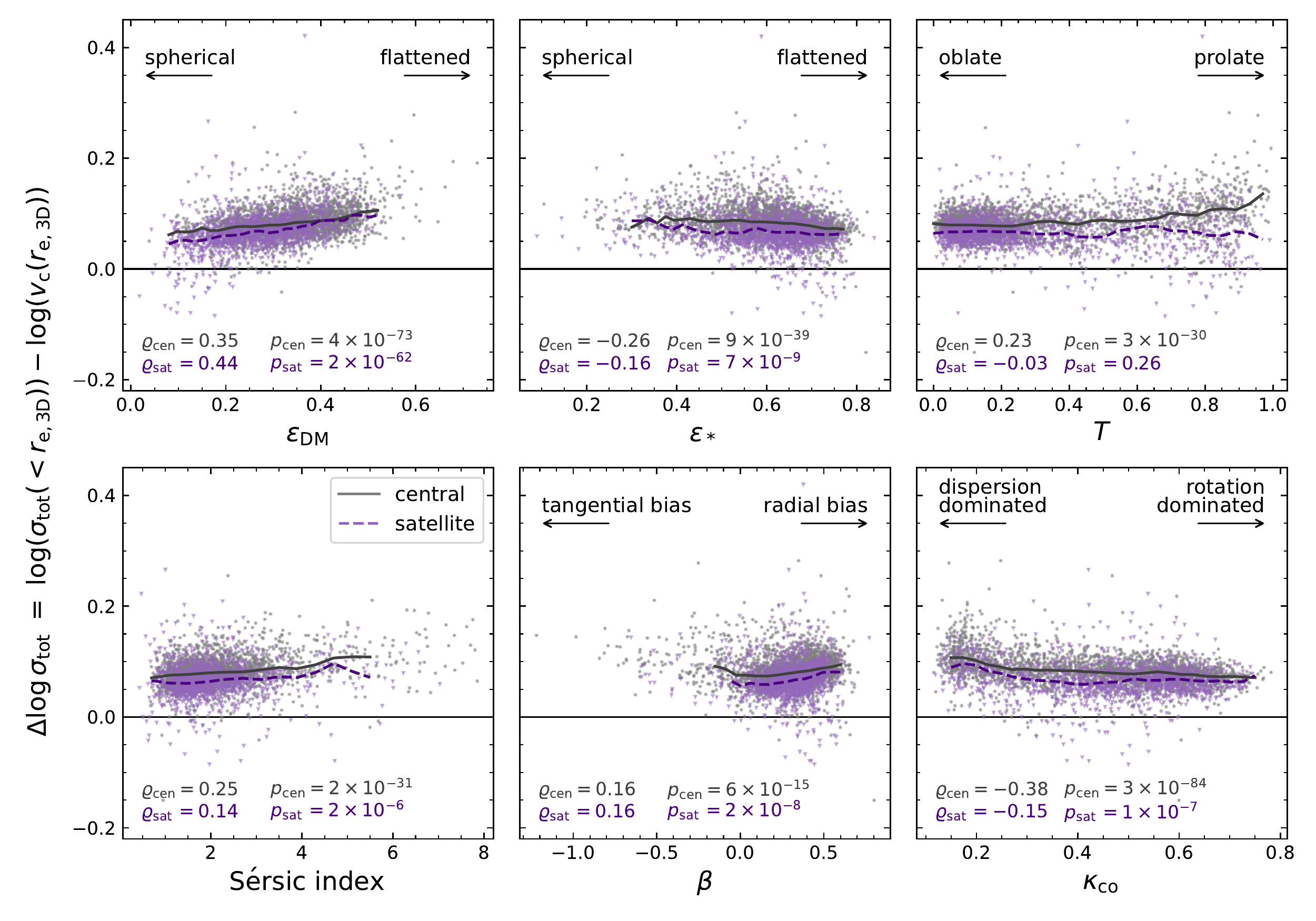}
    \caption{Deviation of the total velocity dispersion from {the prediction of the virial plane} (Eq.~\ref{eq:vcirc}) as a function of various structural properties: the flattening of the 3D dark matter ($\epsilon_{\rm DM}$) and stellar mass ($\epsilon_*$) distributions, triaxiality of the stellar mass ($T$), S\'ersic index ($n$), anisotropy ($\beta$), and the co-rotational kinetic energy fraction ($\kappa_{\rm co}$). Central (grey) and satellite (purple) galaxies are indicated separately, with solid and dashed lines showing the respective running medians in each panel. The correlations between $\Delta\log\stot$ and the galaxy structure and dynamics demonstrate the effects of non-homology: galaxies have highly diverse morphologies and kinematic structures, which affects the measured value of $\stot$ and hence causes a tilt in the FP with respect to the viral plane.}
    \label{fig:stot_nonhomology}
\end{figure*}

\begin{figure*}
    \centering
    \includegraphics[width=0.76\linewidth]{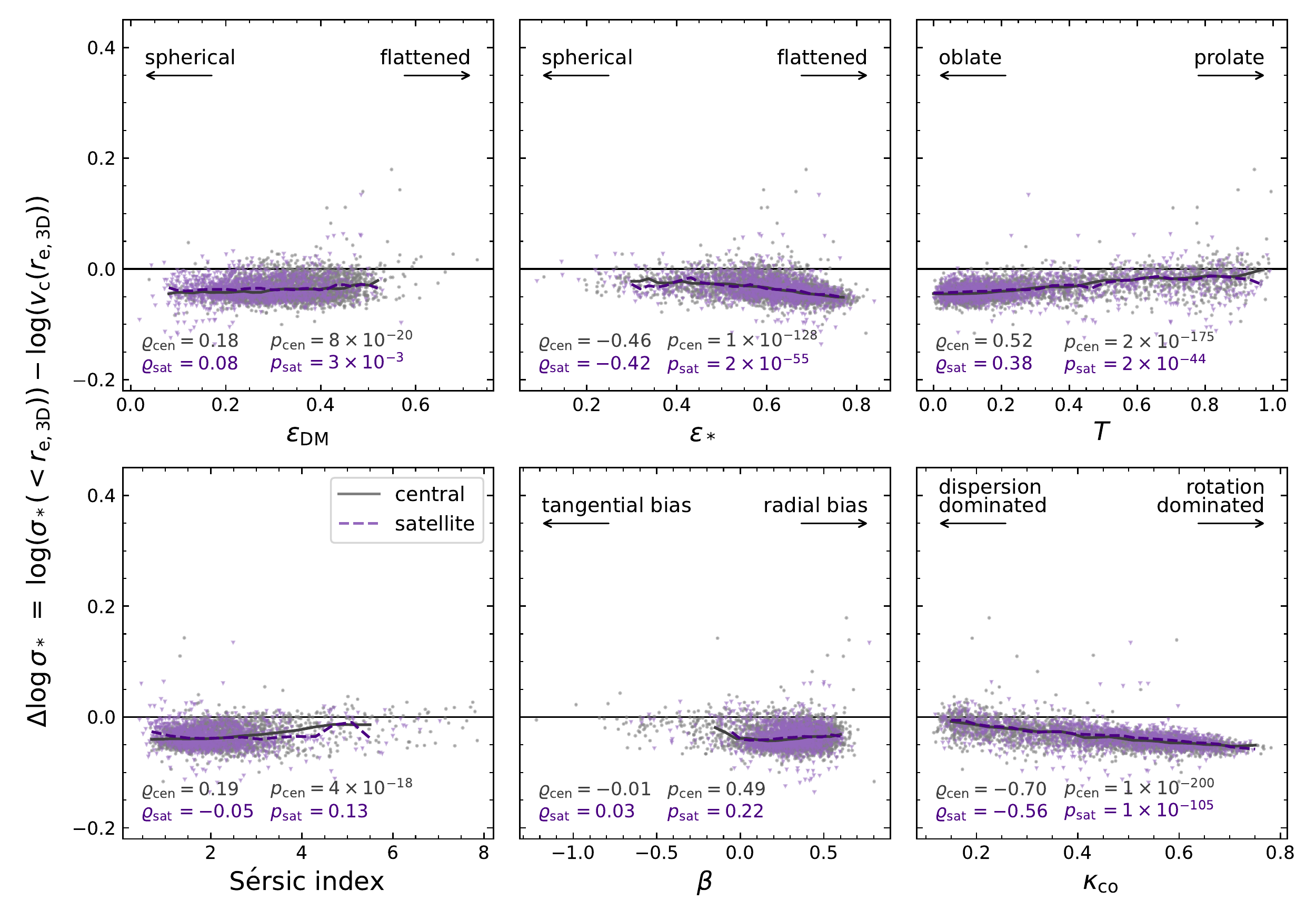}
    \caption{Deviation of the stellar velocity dispersion from {the prediction of the virial plane (Eq.~\ref{eq:vcirc})} as a function of galaxy structure. Compared to Fig.~\ref{fig:stot_nonhomology}, the stellar velocity dispersion is less dependent on the shape of the dark matter distribution ($\epsilon_{\rm DM}$), but instead depends more strongly on the structure of the stellar mass. {For spherical, dispersion-supported systems $\sigma_*$ approximates the virial plane prediction, but galaxies that are more flattened and strongly rotating diverge from this, due to the difference in the shape of the gravitational potential.} Correlations are slightly weaker for satellite galaxies than centrals, which reflects the additional effect of the local environment on these systems.  }
    \label{fig:sstar_nonhomology}
\end{figure*}

Figs.~\ref{fig:stot_nonhomology} and \ref{fig:sstar_nonhomology} show how $\Delta\log\sigma$ varies with these different structural properties, for both the total and stellar velocity dispersion, respectively. As the sSFR has minimal impact on the measured deviation, we omit the colour coding by sSFR in these figures. Instead, however, we distinguish between central (grey) and satellite (purple) galaxies, as the structural properties of satellite galaxies may be expected to be influenced by their local environment (e.g., through tidal stripping). The running median is plotted in each panel for the central galaxies (solid lines) and satellites (dashed lines).

Starting with $\stot$\,, we find correlations with all structural properties shown, particularly with the $\epsilon_{\rm DM}$ and $\kappa_{\rm co}$. These trends are generally stronger for central galaxies than the satellites, except for the $\epsilon_{\rm DM}$\,, and shows that the local environment has a small effect on the measured $\stot$ and the other structural parameters. On the other hand, $\sigma_*$ is less dependent on the structure of the dark matter, and instead depends strongly on the morphology and dynamics of the stellar mass. 

The wide variety in shapes and structures among galaxies \citep[discussed more extensively by][]{Thob2019} clearly shows that the assumption of homology is incorrect. The measurement of $\sigma(<\retd)$ reflects these variations in the density profiles: for instance, for oblate, rotating systems $\sigma_*(<\retd)$ underestimates the total mass, whereas it is a good approximation of the total mass for more {spherical} systems with greater dispersion support. These differences in the structure therefore also lead to a deviation of the total mass FP from the virial plane. {This is to be expected, as the virial plane (Eq.~\ref{eq:vcirc}) assumes a spherically symmetric mass distribution, but the true circular velocities of galaxies depend on the shape of the mass distribution \cite[see][Chapter 2]{BinneyTremaine1987}.} The fact that the FP with $\stot$ deviates more strongly from the virial plane than $\sstar$ can then be attributed to the fact that $\stot$ is sensitive to not only the stellar mass distribution, but also the dark matter.

However, as shown by \citet{Trayford2019}, \citet{vdSande2019} and \citet{deGraaff2022}, the galaxy morphologies in the $100^3\,$cMpc$^3$ EAGLE simulation are different from observed galaxies: the simulated galaxies tend to be thicker, and with significantly lower S\'ersic indices. This is possibly the result of the pressure floor in the simulation, or due to the limited resolution, as \citet{Ludlow2019,Ludlow2021} showed that the 2-body scattering of the relatively massive dark matter particles with the baryonic particles in the simulation affects the resulting stellar mass density profiles. In Appendix~\ref{sec:apdx_highres} we show results from the higher-resolution EAGLE simulations (described in Section~\ref{sec:eagle}), and demonstrate that the morphology (particularly the S\'ersic index) is strongly dependent on the resolution, but that our conclusions on the effects of non-homology on the FP are robust to changes in the resolution.

Lastly, although we have explained the relative differences in $\Delta\log\sigma$, we have thus far neglected the fact that there is also a systematic offset in $\Delta\log\stot$ visible in Fig.~\ref{fig:sigma_ssfr} and \ref{fig:stot_nonhomology}: $\stot$ systematically overpredicts the total mass within $\retd$. This suggests that there must be a factor missing in Eq.~\ref{eq:vcirc}, which can most plausibly be attributed to the assumptions made in obtaining the virial theorem. To arrive at the scalar virial theorem of $2K + W=0$, where $W$ is the gravitational potential energy, one has to assume that the mass density $\rho(r\rightarrow \infty) = 0$ \citep{BinneyTremaine1987}. Albeit a reasonable assumption for the stellar mass distribution, the distribution of the dark matter is more complex. The dark matter particles are more likely to be on highly eccentric orbits with semi-major axes that are significantly larger than the stellar half-mass radius of the galaxy. This can also be interpreted as a surface pressure term in the virial theorem, such that $2K + W + S_{\rm p} = 0$\,, with $S_{\rm p}/|W| < 0$ and hence $2K/|W| > 1$ \citep[see also][]{Shapiro2004}. Therefore, we would expect to find $\stot>v_{\rm c}$ (from Eq.~\ref{eq:vcirc}), which is exactly what Fig.~\ref{fig:sigma_ssfr} and \ref{fig:stot_nonhomology} show.

\subsection{Variations in the dark matter fraction} \label{sec:DM_frac}

Having quantified the $\Mdyn/\Mtot$ contribution to the tilt of the FP, we now add in the effects caused by the different mass compositions of galaxies, i.e. the contribution from stellar, gas and dark matter mass (we neglect the black hole mass, as this typically comprises $<1\%$ of $\Mtot(<\retd)$). This is also coupled with a change in the sample selection, as instead of using the $\Mtot$ complete sample, we from hereon focus on the $M_*$-selected sample and stellar velocity dispersions only.

First, we examine the effect of this change in the sample on the total mass FP. Because of the strong overlap between the two samples the coefficients are changed only weakly, although this is statistically significant. The total mass FP spanned by the $M_*$-selected sample is slightly closer to the virial plane than before, but the effects of non-homology discussed in the previous section still apply.

\begin{figure*}
    \centering
    \includegraphics[width=0.94\linewidth]{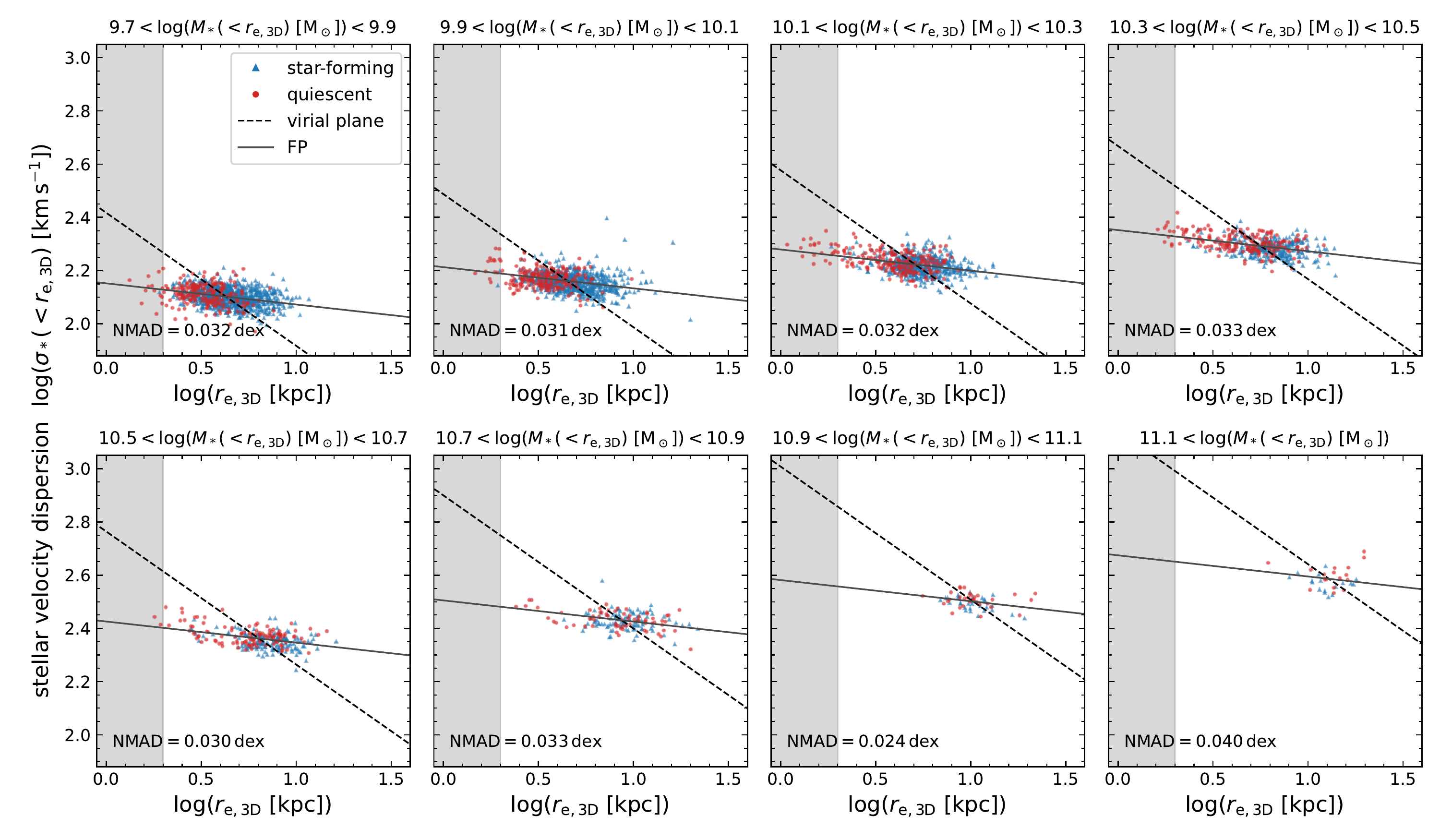}
    \caption{Stellar velocity dispersion {(see Section~\ref{sec:vdisp})} as a function of the half-mass radius, binned by the stellar mass $M_*(<\retd)$ rather than the total mass (Fig.~\ref{fig:sstar_fixedmtot}). Symbols indicate the same as in Fig.~\ref{fig:sstar_fixedmtot}. Galaxies clearly deviate strongly from the relations that they are expected to follow {if $\Mdyn \propto M_*$ (dashed lines show the slope corresponding to the virial plane}). In addition to the effects of non-homology established in Section~\ref{sec:nonhomology}, variations in the dark matter and gas fractions within $\retd$ must contribute to these deviations, and hence set the tilt of the stellar mass FP (solid lines).  }
    \label{fig:sstar_fixedmstar}
\end{figure*}

Second, we change from the total mass FP to the stellar mass FP (Eq.~\ref{eq:mfp_powerlaw}), and investigate the relation between $\sstar(<\retd)$ and $\retd$ in narrow bins of $M_*(<\retd)$. Fig.~\ref{fig:sstar_fixedmstar} differs from Fig.~\ref{fig:sstar_fixedmtot} only by the choice of the mass used to bin the data, with an additional small effect due to differences in the samples used. Dashed lines show the predicted slope in each panel for homologous galaxies in virial equilibrium, which is clearly a poor prediction. 
By fitting the tilt of the stellar mass FP, we find a much stronger deviation from the virial plane than before, particularly for the $\beta$ coefficient, and with increased scatter (solid lines; Table~\ref{tab:FP_3D}).   

We again define a predicted velocity dispersion, by replacing the total mass in Eq.~\ref{eq:vcirc} by the stellar mass:
\begin{equation}
    \sigma_{\rm pred} = \sqrt{\frac{G M_*(<r)}{r}}\,,
    \label{eq:sigma_pred}
\end{equation}
using $r=\retd$ and evaluate how the dynamical and stellar mass differ from each other. The effects from non-homology discussed in the previous section ($\ref{sec:nonhomology}$) still hold here. However, there are now two new factors to consider: the dark matter ($\fdm$) and gas fraction ($\fgas$) within $\retd$.

\begin{figure}
    \centering
    \includegraphics[width=\linewidth]{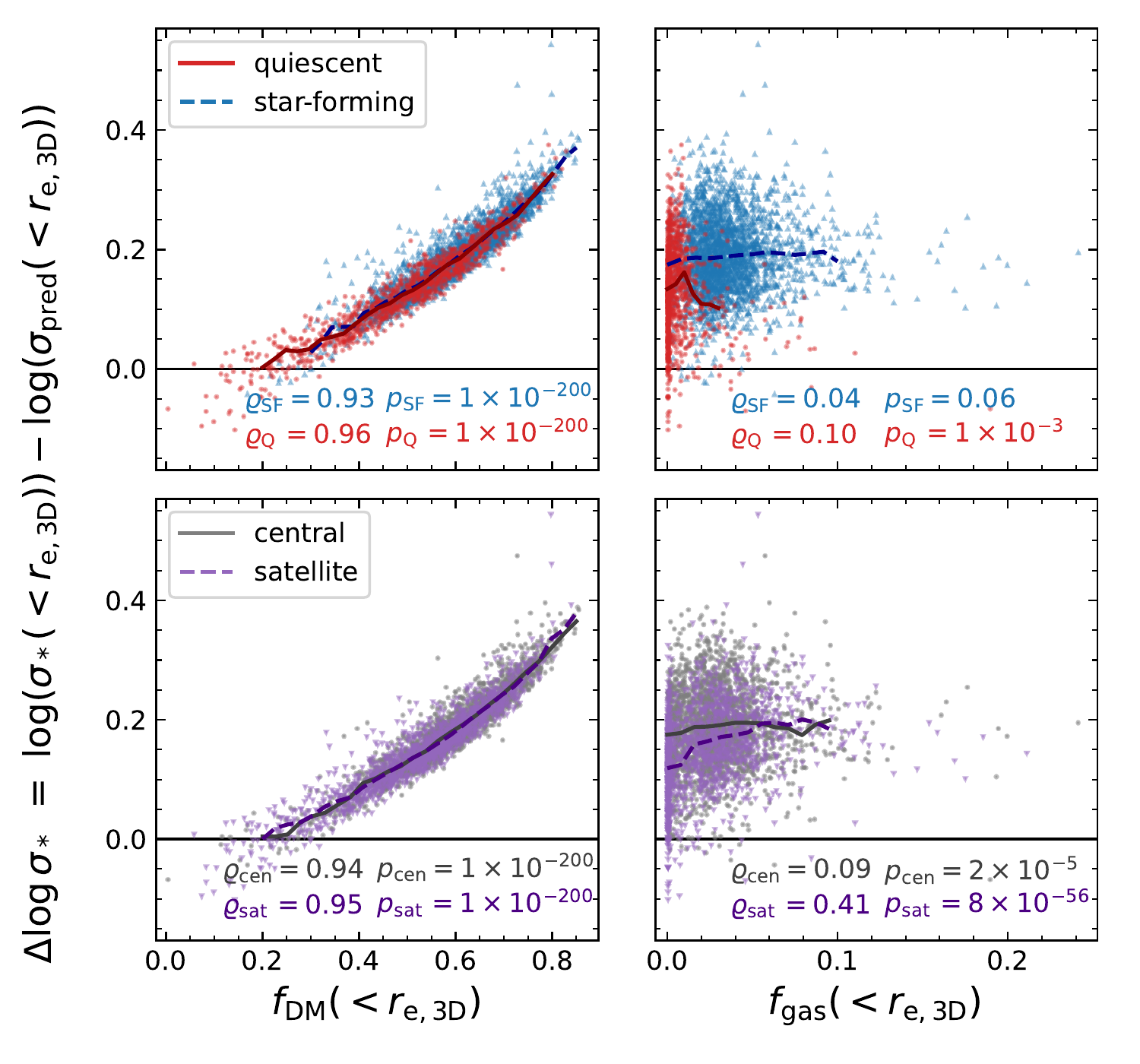}
    \caption{Deviation between the measured stellar velocity dispersion and the dispersion predicted from the stellar mass and stellar half-mass radius (Eq.~\ref{eq:sigma_pred}; {corresponding to the offset from the virial plane}) versus the dark matter (left) and gas (right) fractions within the half-mass radius. Top panels distinguish between star-forming (blue) and quiescent (red) galaxies; bottom panels separate central (grey) and satellite (purple) galaxies. Solid and dashed lines show the running medians. The correlation with the dark matter fraction is stronger than for any other parameter (Fig.~\ref{fig:sstar_nonhomology}), indicating that it is the primary driver of the tilt of the stellar mass FP, for both star-forming and quiescent galaxies.}
    \label{fig:sstar_mass_fraction}
\end{figure}

In Fig.~\ref{fig:sstar_mass_fraction} we show the difference between the stellar velocity dispersion and the predicted dispersion ($\Delta\log\sstar$) as a function of $\fdm(<\retd)$ and $\fgas(<\retd)$. The upper panels are colour coded using the division into star-forming (blue) and quiescent (red) galaxies; the lower panels distinguish between central (grey) and satellite (purple) galaxies, for comparison with Fig.~\ref{fig:sstar_nonhomology}. There is a systematic offset in the obtained $\Delta\log\sigma_*$, as Eq.~\ref{eq:sigma_pred} misses a significant fraction of the total galaxy mass and therefore leads to a systematically lower value of $\sigma_{\rm pred}$.

Most importantly, we find a very strong correlation between $\Delta\log\sigma_*$ and $\fdm$, for all four categories of galaxies. These trends are stronger than any of the correlations with galaxy structure found in Fig.~\ref{fig:sstar_nonhomology}, and therefore demonstrate that systematic variations in the dark matter fractions are the main driver of the tilt of the stellar mass FP. 
{We find no correlation with the gas fraction for star-forming galaxies, which may be contrary to expectations: the gas fractions are generally very low ($\lesssim 5\%$), which is likely due to the relatively small aperture considered ($1\,r_{\rm e}$), as \citet{Bahe2016} showed that the overall gas content in EAGLE galaxies agrees well with observations within our selected stellar mass range. } On the other hand, there is a correlation among the satellite galaxies, which suggests an additional, weak effect of the local environment on the stellar mass FP.

\begin{figure}
    \centering
    \includegraphics[width=0.98\linewidth]{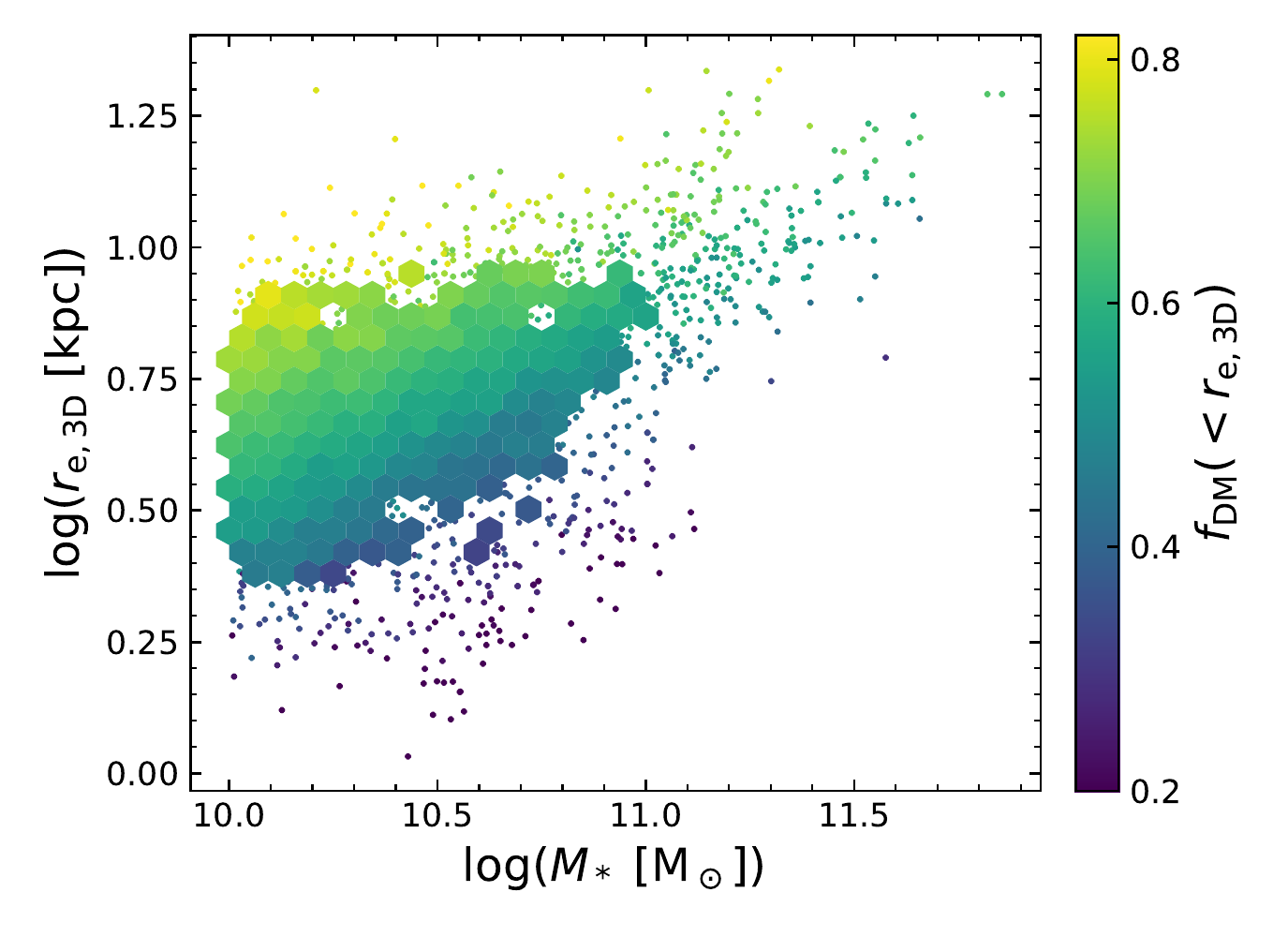}
    \caption{Total stellar mass versus the stellar half-mass radius for all (quiescent and star-forming) galaxies, colour-coded by the dark matter fraction within the half-mass radius. The variation in $\fdm$ is a smooth function of both size and stellar mass, which leads to the observed correlation between $\Delta\log\sstar$ and $\fdm$ in Fig.~\ref{fig:sstar_mass_fraction}. As a result, quiescent and star-forming galaxies lie on a common stellar mass FP that deviates strongly from the virial plane. }
    \label{fig:fdm_var}
\end{figure}

Fig.~\ref{fig:fdm_var} further examines the variation in $\fdm$ across the stellar mass-size plane. This indicates that there is not simply a large variation in $\fdm$, but that the variation in $\fdm$ is a smooth power-law function of both $M_*$ and $\retd$. In turn, these variations result in the observed strong correlation between $\Delta\log\sstar$ and $\fdm$ found in Fig.~\ref{fig:sstar_mass_fraction}, and hence the tilt of the simulated stellar mass FP. The common FP for star-forming and quiescent galaxies can then be interpreted as the power-law relation $\fdm(<\retd) \propto M_*^a \retd^b$ having similar coefficients $a$ and $b$ for both galaxy populations. We estimate the coefficients by minimising the sum of the offsets orthogonal to the planar relation:
\begin{equation}
    \Delta_{\rm DM} = \frac{\left | \log(\fdm(<\retd)) -  a \log(M_*) -  b\log(\retd) - c \right|}{\sqrt{1+a^2 + b^2}}\,,
    \label{eq:delta_fdm}
\end{equation}
where $c$ is the zero-point of the relation and $M_* = 2 M_*(<\retd)$. We present the results of these fits in Table~\ref{tab:fdm}.

\begin{table*}
 \caption{Best-fit coefficients for the relation $\log(\fdm(<\retd)) = a\log(M_*) + b\log(\retd)+c$. Galaxies for which $\retd<2\,$kpc are excluded from the fits. }
 \label{tab:fdm}
 \begin{tabular}{lcccc}
  \hline
  Sample selection & $a$ & $b$ & $c$ & NMAD  \\
  \hline
  $M_*$ & $-0.214\pm0.004$ & $0.573\pm0.009$ & $1.58\pm0.04$ & $0.0312\pm0.0006$ \\
  $M_*$ \& quiescent & $-0.317\pm0.008$ & $0.781\pm0.015$ & $2.54\pm0.07$ & $0.0339\pm0.0014$ \\
  $M_*$ \& star-forming & $-0.194\pm0.004$ & $0.518\pm0.008$ & $1.41\pm0.04$ & $0.0280\pm0.0007$ \\
  \hline
 \end{tabular}
\end{table*}

\begin{table*}
 \caption{Best-fit coefficients of the stellar mass FP and orthogonal scatter, for different measures of galaxy size and velocity dispersion. Galaxies for which $r_{\rm e}<2\,$kpc are excluded from the fits. }
 \label{tab:FP_obs}
 \begin{tabular}{llcccc}
  \hline
  Relation & Sample selection & $\alpha$ & $\beta$ & $\gamma$ & NMAD  \\
  \hline
  $\log(\res) = \alpha \log(\sstar(<\res)) + \beta \log(\Sigma_*) + \gamma$  & $M_*$  & $1.760 \pm 0.018$ & $-0.610 \pm0.006$ & $2.40\pm0.04$ & $0.0353\pm0.0008$ \\

    & $M_*$ \& quiescent & $1.62 \pm 0.03$ & $-0.550 \pm0.013$ & $2.14\pm0.09$ & $0.043\pm0.002$ \\
   
   & $M_*$ \& star-forming & $1.82 \pm 0.02$ & $-0.624 \pm0.006$ & $2.42\pm0.04$ & $0.0326\pm0.0009$ \\   
 
  $\log(r_{\rm circ, *}) = \alpha \log(\sstar(<\res)) + \beta \log(\Sigma_*) + \gamma$  & $M_*$  & $1.578 \pm 0.015$ & $-0.617 \pm0.005$ & $2.71\pm0.03$ & $0.0311\pm0.0006$ \\

   & $M_*$ \& quiescent & $1.49 \pm 0.02$ & $-0.584 \pm0.008$ & $2.60\pm0.07$ & $0.0345\pm0.0016$ \\
   
  & $M_*$ \& star-forming & $1.618 \pm 0.016$ & $-0.634 \pm0.006$ & $2.77\pm0.04$ & $0.0296\pm0.0007$ \\   
  
  $\log(\rer) = \alpha \log(\sstar(<\rer)) + \beta \log(\Sigma_*) + \gamma$  & $M_*$  & $1.407 \pm 0.014$ & $-0.578 \pm0.005$ & $2.81\pm0.04$ & $0.0436\pm0.0009$ \\

   & $M_*$ \& quiescent & $1.45 \pm 0.03$ & $-0.532 \pm0.015$ & $2.33\pm0.12$ & $0.0478 \pm0.0017$ \\
   
  & $M_*$ \& star-forming & $1.38 \pm 0.015$ & $-0.576 \pm0.006$ & $2.84\pm0.05$ & $0.0388\pm0.0011$ \\   

  $\log(r_{\rm circ, r}) = \alpha \log(\sstar(<\rer)) + \beta \log(\Sigma_*) + \gamma$  & $M_*$  & $1.266 \pm 0.017$ & $-0.545 \pm0.005$ & $2.70\pm0.04$ & $0.0485\pm0.0010$ \\

   & $M_*$ \& quiescent & $1.33 \pm 0.02$ & $-0.560 \pm0.010$ & $2.69\pm0.08$ & $0.0378\pm0.0014$ \\
   
  & $M_*$ \& star-forming & $1.19 \pm 0.02$ & $-0.546 \pm0.008$ & $2.84\pm0.06$ & $0.0530\pm0.0012$ \\

  \hline
 \end{tabular}
\end{table*}

\section{Observing the Fundamental Plane}\label{sec:FP_obs}

Measurements of the FP in the previous section relied entirely on 3D measurements of the size, mass and velocity dispersion. To be able to compare with observations, we need to take into account the different observational effects that may bias the observed FP with respect to the intrinsic `3D FP'. Broadly, these are the effects of projection along a random viewing angle, differences in the measurement methods and associated measurement uncertainties, gradients in the $M_*/L$ ratio and associated systematic uncertainties in the assumed IMF, and selection biases. In this section, all these effects are added in, to arrive at a realistic measurement of the FP.

\subsection{Impact of projection effects and measurement biases}\label{sec:obs_bias}

\begin{figure*}
    \centering
    \includegraphics[width=0.92\linewidth]{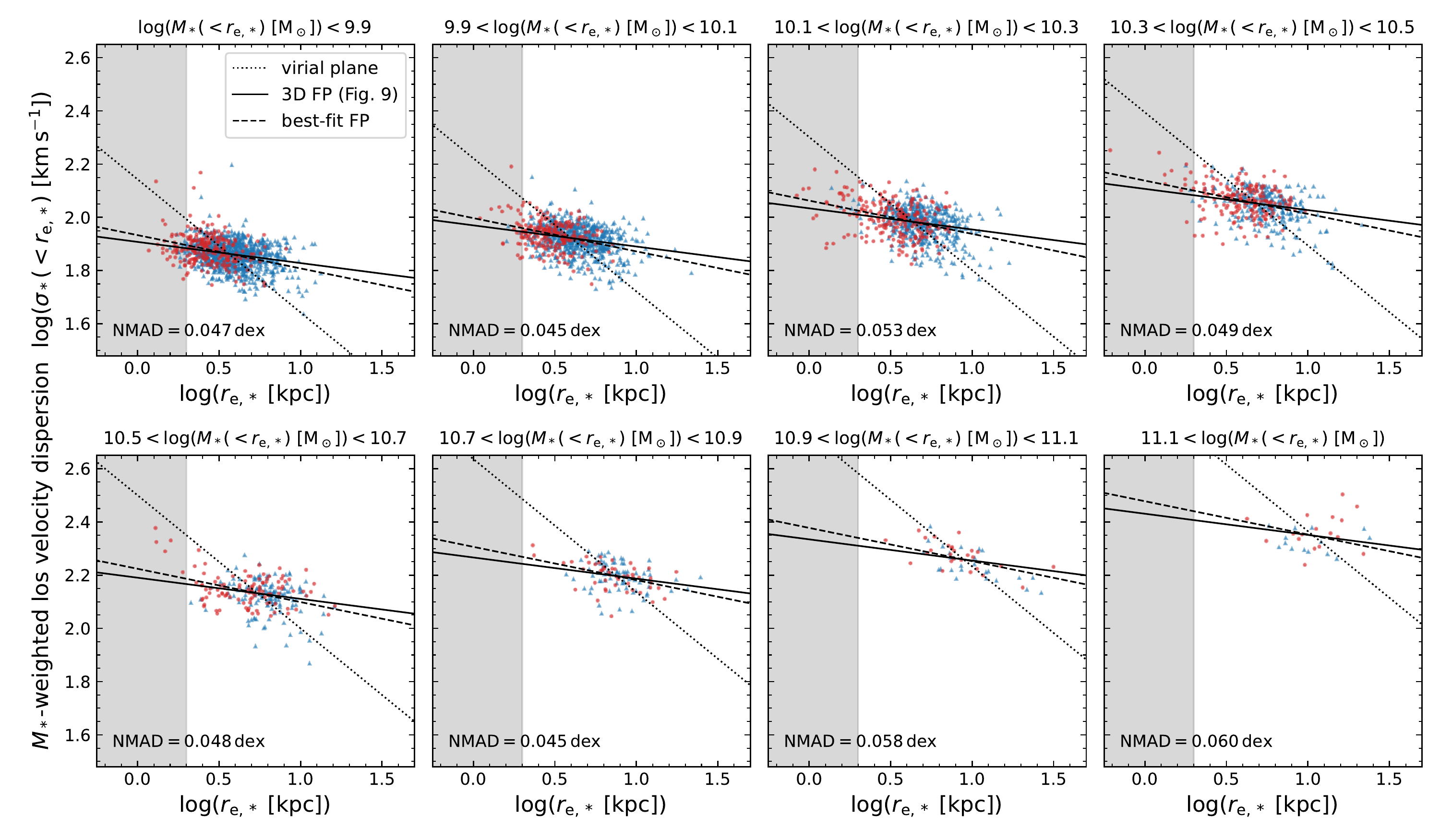}
    \caption{Stellar mass-weighted, line-of-sight velocity dispersion versus the half-mass radius from the best-fit S\'ersic profile, in bins of the stellar mass $M_*(<\res)$. Star-forming and quiescent galaxies are indicated by blue and red symbols, respectively. The virial plane is shown as the dotted line, and solid lines show the stellar mass FP from fitting to the 3D aperture measurements (Table~\ref{tab:FP_3D}). The orthogonal scatter about this `3D FP' is printed in each panel for comparison with Fig.~\ref{fig:sstar_fixedmstar}. Projection effects and differences in the applied measurement methods result in an increased scatter, and slightly alter the inferred stellar mass FP (dashed lines; Table~\ref{tab:FP_obs}).}
    \label{fig:fp_re_star}
\end{figure*}

We use the mock observations described in Section~\ref{sec:data}, and begin with the measurements of the projected stellar mass distributions. As before, we create bins in stellar mass, which are now changed to the mass inferred from the best-fit S\'ersic profile, and show the line-of-sight velocity dispersion as a function of the half-mass radius in Fig.~\ref{fig:fp_re_star}. Star-forming and quiescent galaxies are again indicated separately, using blue and red symbols, respectively. The virial plane is shown as the dotted line in each panel for easy comparison with previous figures.

The main differences with respect to Fig.~\ref{fig:sstar_fixedmstar} are in the scatter: the offset in the velocity dispersion $\Delta\log\sstar =-0.007\,$dex with a scatter of 0.05\,dex, and the offset in the size $\Delta\log r_{\rm e} =-0.06\,$dex, with a scatter of $0.07\,$dex. The mass bins are also changed slightly, although this effect is small (offset of $\Delta\log M_* = -0.02\,$dex with a scatter of 0.04\,dex). These systematic offsets and scatter arise from projection effects, which particularly affect the velocity dispersions, and, for the sizes and masses, differences in the measurement methods and measurement uncertainties. We note that the projected velocity dispersions as measured in Section~\ref{sec:vdisp} are noise-free (nor include PSF smoothing), and the scatter therefore is purely from projection along the line of sight. However, measurement errors are expected to be subdominant, as we find that the typical uncertainty on the velocity dispersion for galaxies of $M*>10^{10}\,\Msun$ at $z\sim0$ in the Sloan Digital Sky Survey is $\approx 0.02\,$dex \citep[SDSS; using the sample described in][]{deGraaff2021}.

To estimate the effect on the inferred stellar mass FP, we first calculate the stellar mass surface density within the elliptical aperture described by the $\res$ and the axis ratio $q_*$: $\Sigma_* = M_*(<\res)/ (\pi q_* \res^2)$, with $M_*$ being half of the stellar mass of the integrated S\'ersic profile. We then measure the tilt of the stellar mass FP with the same orthogonal fitting used previously (dashed lines in Fig.~\ref{fig:fp_re_star}; Table~\ref{tab:FP_obs}), and compare with the stellar mass FP measured from the 3D measurements (solid lines; Table~\ref{tab:FP_3D}). The scatter about this solid line is printed in each panel for comparison with Fig.~\ref{fig:sstar_fixedmstar}.

The increased scatter affects mainly the $\alpha$ parameter of the tilt, likely due to the asymmetric scatter toward low $\sigma_*$ from galaxies that are close to face-on. Moreover, the scatter about the FP itself is nearly doubled. The $\beta$ parameter is largely unchanged, however, despite an offset and significant scatter in the size, which can be understood from the fact that changes in the size correlate in a direction that is near-parallel to the FP itself \citep[see also Appendix B of][]{deGraaff2021}.

\begin{figure*}
    \centering
    \includegraphics[width=0.92\linewidth]{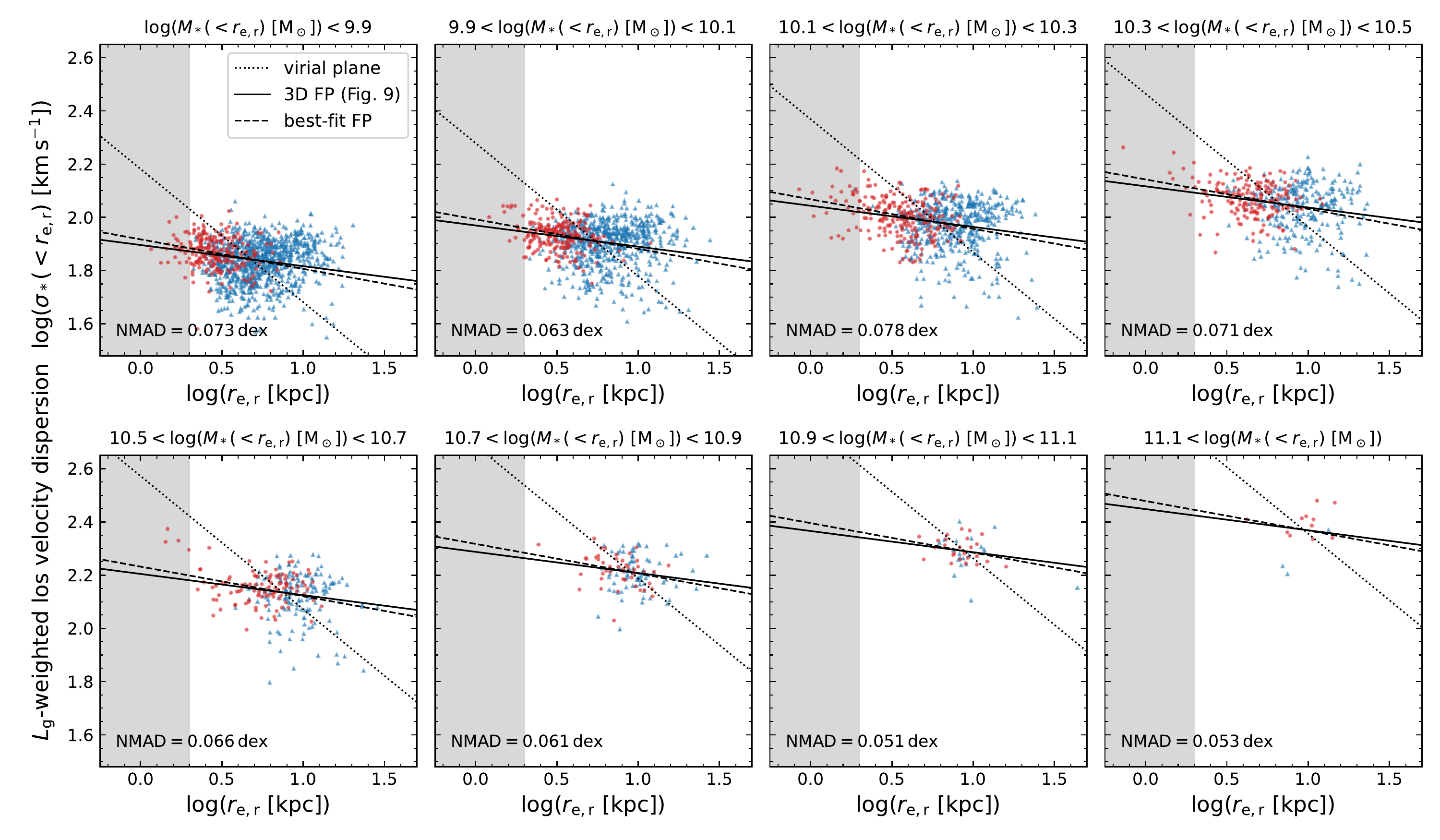}
    \caption{Projected, luminosity-weighted velocity dispersion versus the half-light radius from the best-fit S\'ersic profile, in bins of the stellar mass $M_*(<\rer)$. Symbols indicate the same as in Fig.~\ref{fig:fp_re_star}. Gradients in the $M_*/L$ ratio have a strong effect on the scatter, which is increased by $\approx 50\%$ in the lower mass bins, and make it more difficult to precisely retrieve the intrinsic (3D) stellar mass FP. }
    \label{fig:fp_re_rband}
\end{figure*}

However, observational studies of the FP rarely use the effective radius as we have here, i.e., the semi-major axis size. Rather, the effective radius is often circularised, such that $r_{\rm circ} = \sqrt{q}\, r_{\rm e}\,$, as this may be a better approximation of the galaxy size for systems that are not oblate in shape (the fraction of low-redshift early-types that have prolate shapes or are triaxial). It also serves as a crude correction for the projection effects on $\sigma_*$, mitigating residual correlations with $q$ throughout the FP. We therefore repeat our fits using the circularised size instead of the major axis size, and present the results in Table~\ref{tab:FP_obs} and Appendix~\ref{sec:apdx_rcirc}. 
Again, the changes to the $\beta$ parameter are small, because of the covariance between $r_{\rm circ}$ and $\Sigma_*$. On the other hand, the $\alpha$ parameter depends strongly on the measure of size that is used, due to the corrective effect on the velocity dispersion. The FP resulting from the circularised sizes is closer to the intrinsic, 3D FP than is the case for the major axis sizes, despite the fact that the circularised size is a poor approximation of the galaxy size for oblate systems (i.e., most of the galaxy population). This also suggests that a properly-calibrated correction for the projected velocity dispersions, as derived empirically by van der Wel et al. (2022, submitted) using dynamical Jeans models, may provide even better results than the ad hoc correction from the circularised sizes.

Lastly, we examine the effects of variations in $M_*/L$ (for a universal IMF) within galaxies, which can lead to significantly larger sizes, depending on the star formation activity and dust attenuation. The velocity dispersions as calculated in Section~\ref{sec:vdisp} do not include the effects of dust, however, the measurements will be biased toward the location of the younger stellar populations. As a result, the scatter in the velocity dispersion (Fig.~\ref{fig:sigma_obs}) is increased, as the younger stellar populations tend to lie in dynamically-cold discs \citep{Trayford2019} and the projected velocity dispersions therefore will be more strongly dependent on the inclination angle.

Fig.~\ref{fig:fp_re_rband} shows the equivalent of Fig.~\ref{fig:fp_re_star}, but for the luminosity-weighted measurements. Except for the two highest mass bins, the scatter is significantly increased (by $\approx 50\%$), due to the greater scatter in the size (0.14\,dex in comparison with the 3D half-mass radii) and the velocity dispersion (scatter of 0.07\,dex in comparison with the 3D dispersions). In Appendix~\ref{apdx:nodust}, we investigate whether the strong increase in the scatter is caused by the inconsistency in the tracer used for the size and velocity dispersion, i.e., the use of half-light radii measured from $r$-band imaging that include the effects of dust, while the velocity dispersions are measured using the unattenuated $g$-band luminosities. We show that the strong increase in the scatter between the top panels of Figs.~\ref{fig:fp_re_star} and \ref{fig:fp_re_rband} is driven primarily by the change from a $M_*$-weighting to a $L$-weighting for the velocity dispersions, with the (in)consistency between the tracers being of lesser importance. 

To measure the coefficients of the mock observed FP, we again calculate $\Sigma_* = M_*(<\rer)/ (\pi q_{\rm r}\rer^2)$, now using the stellar mass and axis ratio corresponding to the luminosity-weighted S\'ersic model (see Section~\ref{sec:size_mass}). As a result, the best-fit stellar mass FP has systematically different values for the $\alpha$ parameter (Table~\ref{tab:FP_obs}) than is the case for the mass-weighted measurements, and the scatter is further increased. 
We also perform the fits with the circularised half-light radii, and show the corresponding figure in Appendix~\ref{sec:apdx_rcirc}. These again lead to a difference in $\alpha$ alone, as the result of the circularised size effectively compensating for the projection effects on $\sstar$.

\subsection{Selection bias}\label{sec:selection_bias}

All fits of the stellar mass FP thus far have been based on the stellar mass-selected sample. In observational studies of the FP, however, these galaxies would likely not all be selected: imaging and spectroscopic surveys have lower completeness at low luminosities, as well as toward low velocity dispersions (due to the limitation in the spectral resolution of the instrument, or a selection against velocity dispersions with large measurement uncertainties). As also shown by, e.g. \citet{HydeBenardi2009}, these selection effects lead to a bias in the measured tilt of the FP.

Given the relatively high stellar mass (and therefore high luminosity) of our sample, we may expect all these galaxies to be identified in large imaging surveys of the $z\sim0$ Universe, except for possibly very rare, very low surface brightness objects. Moreover, luminosity biases are relatively easily corrected for using standard Vmax corrections. We therefore only examine the effects of selection cuts in $\log\sstar$, as the dispersions do not scale trivially with luminosity or stellar mass and are susceptible to strong variation from the random projection on the sky.

\begin{figure}
    \centering
    \includegraphics[width=\linewidth]{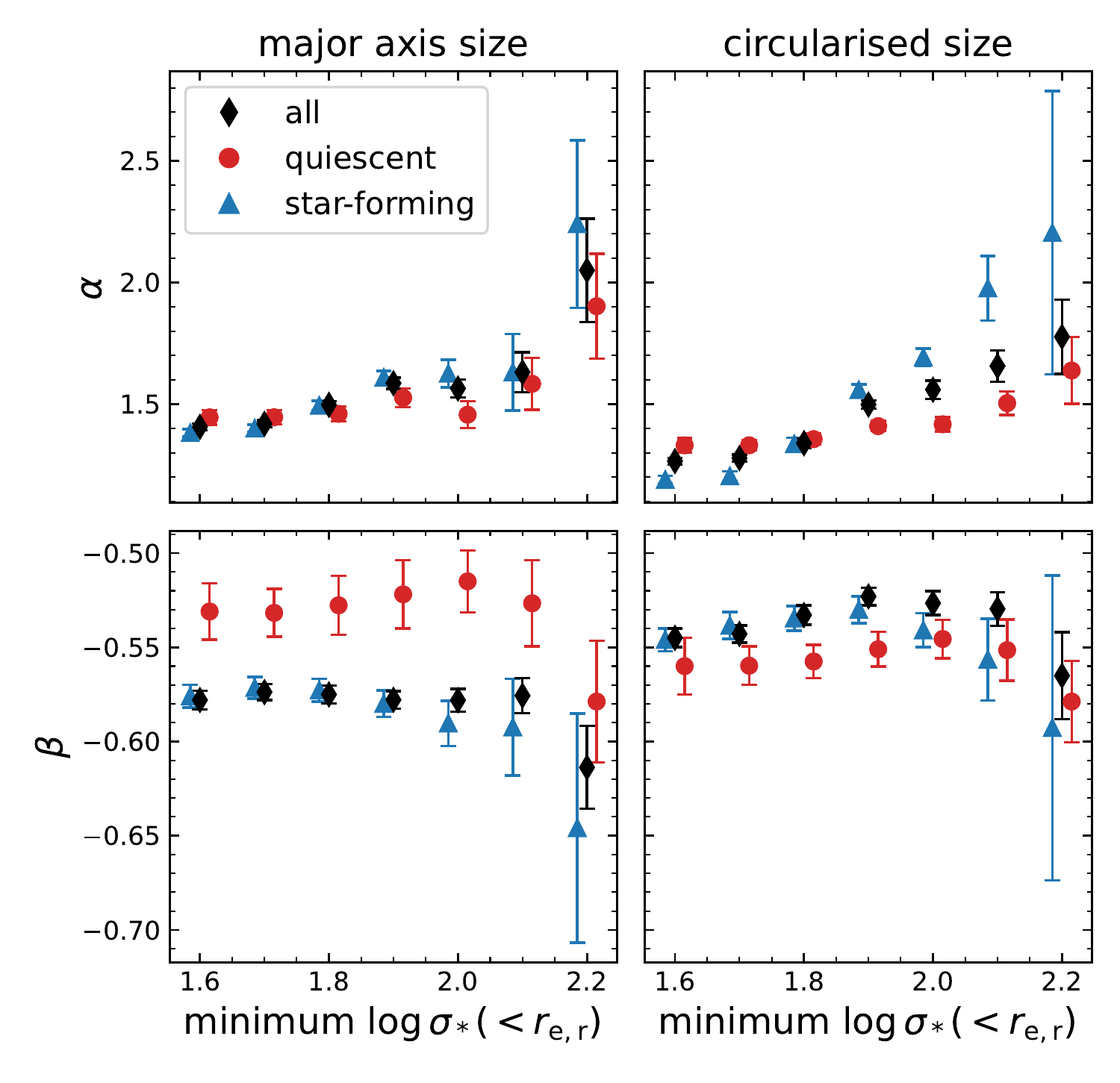}
    \caption{Effect of a selection bias in $\log\sstar$ on the measured tilt of the stellar mass FP for the luminosity-weighted measurements, using the major axis effective radius (left) and circularised effective radius (right) as the measure of size. Blue and red data points have been given a slight offset in $\log\sstar$ for visualisation purposes. {Although the $\beta$ parameter is quite stable (varying by $\lesssim 10\%$), the $\alpha$ parameter can vary strongly depending on the selection (differing by up to $\approx 80\%$ with respect to the complete sample).}}
    \label{fig:sel_bias}
\end{figure}

We use the luminosity-weighted measurements, and measure the stellar mass FP after imposing different selections on $\log\sstar$ (i.e., horizontal cuts in Fig.~\ref{fig:fp_re_rband}). Fig.~\ref{fig:sel_bias} shows the dependence of the parameters $\alpha$ and $\beta$ on the different selections in $\log\sstar$, for the full sample (black), and the quiescent (red) and star-forming (blue) sub-samples. Both fits using the semi-major axis half-light radii (left) and circularised sizes (right) are shown.

We find results are qualitatively similar to the effects found by \citet{HydeBenardi2009}: the $\beta$ parameter varies only weakly with differences in the sample selection, particularly for the fits using the major axis sizes. On the other hand, the $\alpha$ parameter increases toward higher cuts in $\log\sstar$, which is a particularly strong effect for the fits using the circularised sizes. The star-forming population even reaches super-virial values, albeit with large uncertainties.  

Although low-redshift studies of the FP will focus on samples of quiescent galaxies with high completeness down to $\log(\sstar/\kms)\approx 1.8$, these selection effects will likely become much more important toward higher redshift. Given the difficulty in measuring stellar absorption lines in high-redshift galaxies, the completeness boundary for the velocity dispersion shifts to $\log(\sstar/\kms)\approx 2.1-2.2$ \citep[e.g.,][]{Holden2010,deGraaff2021}. At the same time, the structural properties of quiescent galaxies also change, becoming more disc-like and with stronger rotational support \citep{Chang2013,Bezanson2018a}, which will likely lead to selection effects that are more similar to that of the star-forming population in Fig.~\ref{fig:sel_bias}. To measure the evolution in the FP, then requires not just a fair comparison sample (i.e., a low-redshift sample for which the velocity dispersions are greater than the high-redshift completeness limit in $\log\sstar$), but also a correction factor to account for evolution in the (dynamical) structures of the galaxy population.

\subsection{Contribution of IMF variations}\label{sec:imf}

Although we have measured in detail the stellar mass FP and its dependence on different observational effects, one potentially significant systematic uncertainty remains due to the assumption of a universal IMF in the simulations. We use the simulations with a pressure-dependent bottom-heavy (`LoM') and top-heavy (`HiM') IMF, which varies both between and within galaxies (further described in Section~\ref{sec:eagle}), to assess the magnitude of this uncertainty on the FP within the EAGLE simulations.

We compare the variable IMF runs with the $50^3\,$cMpc$^3$ simulation that uses the reference model (`Ref'), which assumes a fixed, Chabrier IMF. Because observational studies typically assume a universal IMF, we can no longer use a sample selection based on the summed stellar particle masses for a fair comparison between the different simulations. Instead, we therefore select by the $r$-band luminosity: luminosities for the stellar particles were computed using the \textsc{FSPS} software, and based on the age, metallicity and IMF of each particle \citep[for details, see][]{Barber2018}. By comparing the total luminosity within a spherical aperture of radius 30\,kpc to the stellar mass within the same aperture for galaxies in the Ref simulation, we find that a minimum (rest-frame) $r$-band luminosity of $10^{9.85}\,L_{\odot, r}$ ($M_r\approx-20.0$) results in a selection completeness of $\gtrsim50\%$ down to a stellar mass of $10^{10}\,\Msun$. Applying this limit provides a sample of 527, 528, and 415 galaxies in the Ref, LoM and HiM simulations, respectively. As before, we divide these samples into star-forming and quiescent subsamples using the boundary of $\rm sSFR=10^{-11}\,yr^{-1}$. We note that the SFR and stellar mass are both dependent on the IMF, but the effect on the sSFR and thus the definition of quiescence is negligible due to the approximately equal change in the SFR and the stellar mass \citep[see also][]{Clauwens2016}.

\begin{figure}
    \centering
    \includegraphics[width=\linewidth]{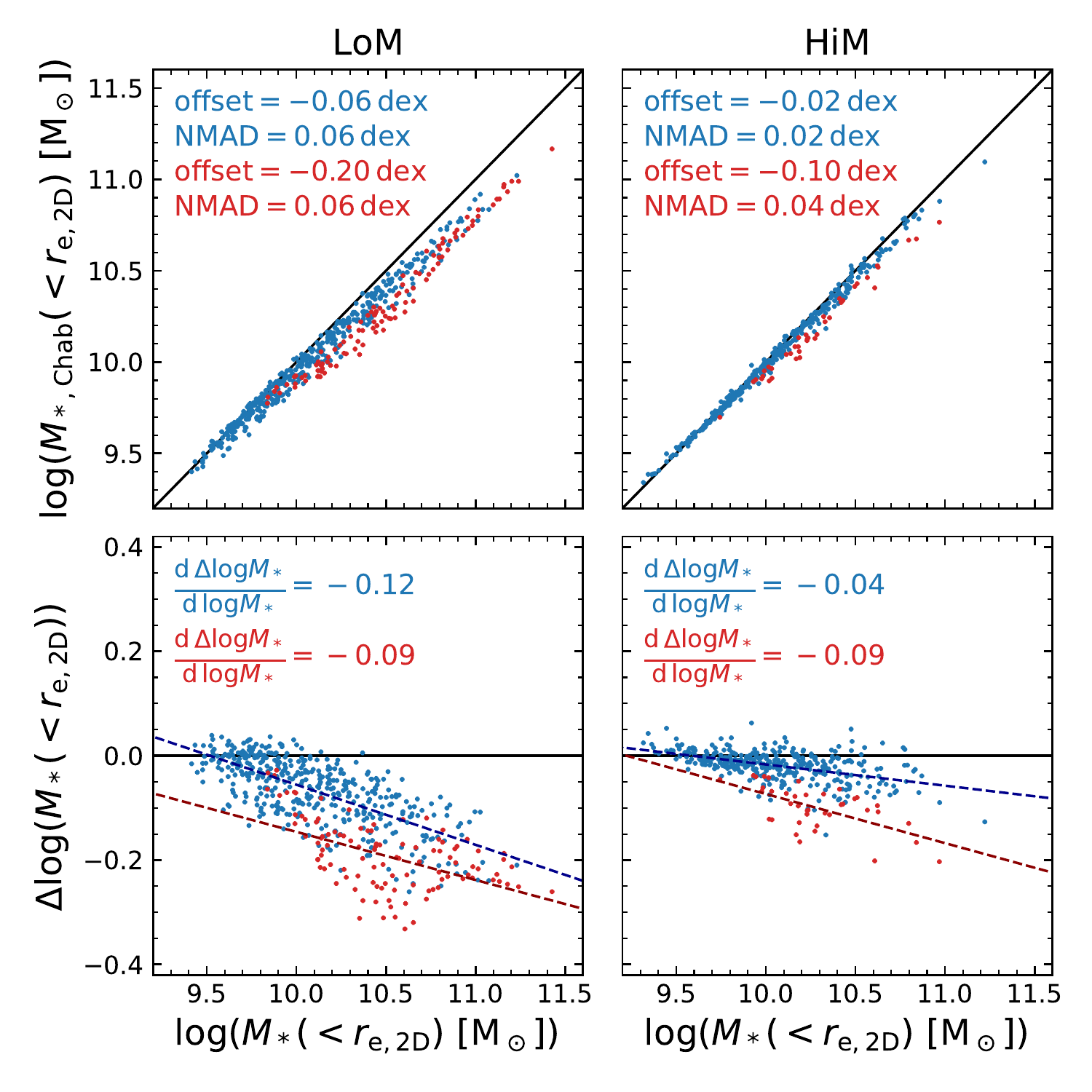}
    \caption{{Comparison between the true stellar mass within the effective radius and the Chabrier-reinterpreted stellar mass, for both the LoM (left) and HiM (right) variable IMF simulations. Top panels show a direct comparison, with star-forming and quiescent galaxies indicated in blue and red, respectively. The median offset and scatter about the one-to-one relation (solid line) are printed in each panel. Bottom panels show that the difference between the two stellar masses is mass-dependent, further quantified by a linear least squares fit (dashed lines).}}
    \label{fig:imf_mstar}
\end{figure}

We subsequently extract 3D aperture measurements as used in Section~\ref{sec:FP_3D}, measuring the particle masses and stellar velocity dispersions within the 3D half-mass radii. To estimate the effect of an incorrect assumption for the IMF requires luminosity-based measurements. We use measurements from the online catalogues \citep{McAlpine2016,Barber2018}, as individual particle luminosities are not available. 
The size used is the circular half-light radius in the $r$-band, $r_{\rm e, 2D}$, based on the total luminosity within a 30\,kpc spherical aperture\footnote{We note that these sizes are free from measurement uncertainties and do not include the effects of dust, therefore leading to considerably lower scatter than seen in Section~\ref{sec:obs_bias}. {The effects of dust are discussed further in Appendix~\ref{apdx:nodust}. } Moreover, these circular half-light radii are smaller than the 3D half-mass radii by $\approx 25\%$ for all galaxies, which differs strongly from the S\'ersic model half-light radii that are larger by $\approx 25\%$ and smaller by $\approx 10\%$ in comparison with the 3D half-mass radii for star-forming and quiescent galaxies, respectively.}; the velocity dispersion ($\sigma_*(<r_{\rm e, 2D})$) is measured within a circular aperture of radius $r_{\rm e, 2D}$ in projection along the $z$-axis of the simulation box, and weighted by the $r$-band luminosities of the particles. For the stellar mass, we use (i) the true stellar mass within the circular aperture (i.e., based on the varying IMF; $M_*(<r_{\rm e,2D})$) and (ii) the stellar mass within the same aperture that is reinterpreted under the assumption of a Chabrier IMF \citep[$M_{\rm *,Chab}(<r_{\rm e,2D})$;][]{Chabrier2003}. The latter quantity is calculated by multiplying the $r$-band luminosity with the $M_*/L_r$ ratio that is obtained for the particles when these are evolved with a Chabrier IMF (using \textsc{FSPS}), and therefore allows for a comparison with observations \citep{Barber2018}. 
{Fig.~\ref{fig:imf_mstar} shows the comparison between the true and Chabrier-reinterpreted stellar masses. By assuming an incorrect Chabrier IMF for the variable IMF simulations, the obtained stellar mass is systematically underpredicted with respect to the true stellar mass. The magnitude of this deviation depends on the stellar mass, as demonstrated by linear least squares fits (dashed lines).}

\begin{figure}
    \centering
    \includegraphics[width=\linewidth]{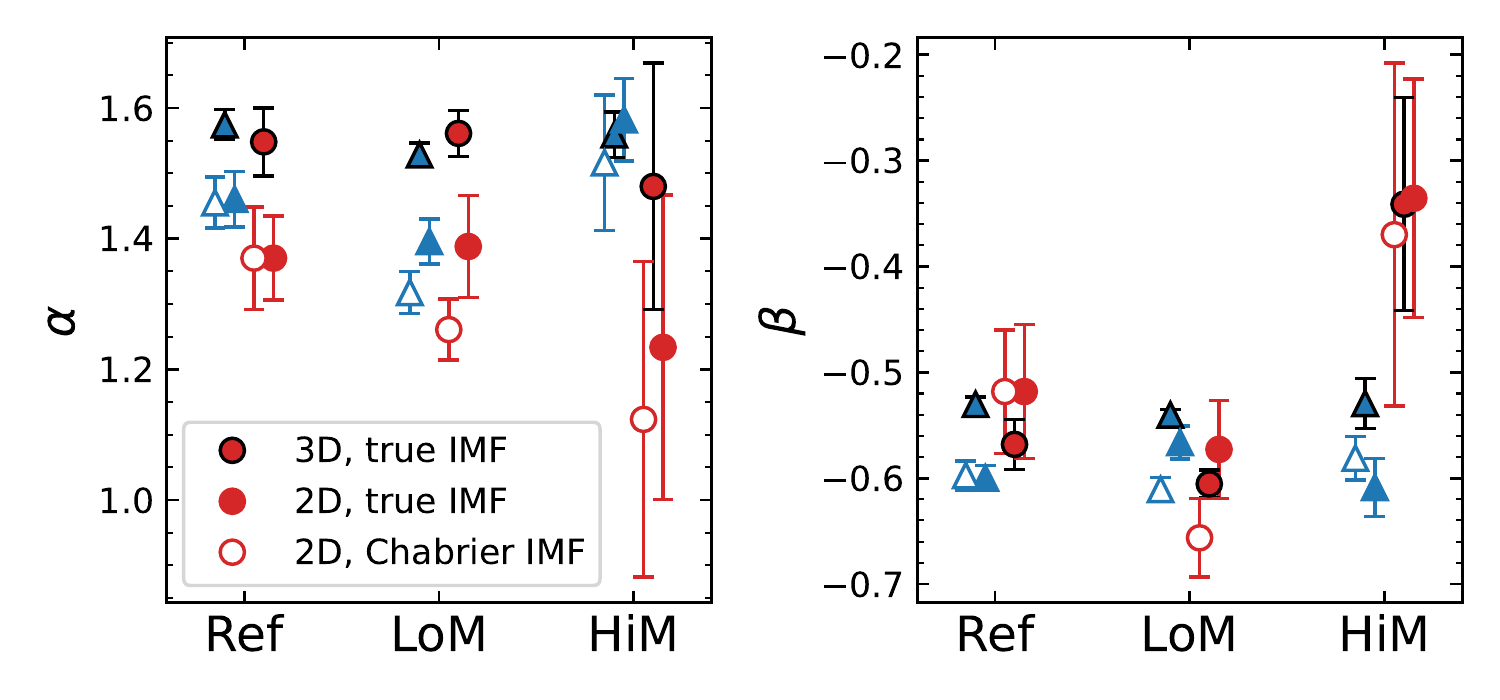}
    \caption{Measured tilt of the stellar mass FP for galaxies in the 50$^3$\,cMpc$^3$ EAGLE simulations for the reference, bottom-heavy IMF (LoM) and top-heavy IMF (HiM) models. Circles and triangles show results for the quiescent and star-forming populations, respectively. Black outlined symbols indicate the fits to the 3D aperture measurements (as in Table~\ref{tab:FP_3D}); coloured symbols show the fits to the projected measurements using the true stellar mass (filled symbols) and Chabrier-reinterpreted stellar mass (open symbols), respectively. The effects of a variable IMF are small for the intrinsic (3D) stellar mass FP. However, by assuming an incorrect IMF, the inferred tilt of the stellar mass FP can be changed by $\approx10\%$ {for star-forming galaxies}, and up to $\approx 20\%$ for quiescent galaxies.}
    \label{fig:imf_tilt}
\end{figure}

The tilt of the stellar mass FP is measured in the same way as before, with the results shown in Fig.~\ref{fig:imf_tilt} for the three simulations. The 3D aperture measurements are shown as the black outlined circles (triangles) for the quiescent (star-forming) populations. The results for the Ref simulation are consistent with the measurements in Table~\ref{tab:FP_3D}. For the LoM model, the fits to these 3D aperture measurements are very similar to those of the Ref model, with only a marginally lower value of $\beta$. The star-forming population in the HiM simulation is also consistent with the other models, and only the quiescent galaxies diverge, particularly in the measurement of $\beta$, although with a large measurement uncertainty.

Because of the differences in the projected measurements with respect to the previous sections of this paper, measurements of the tilt based on the 2D quantities cannot easily be compared to the results of Table~\ref{tab:FP_obs}. However, comparison of the 2D and 3D measurements using the true stellar masses gives insight into the effects of measurement biases: the filled coloured symbols show that there is indeed a small difference in the measured tilt due to a combination of effects from the projection, aperture definition and $M_*/L$ gradients.

\begin{figure}
    \centering
    \includegraphics[width=\linewidth]{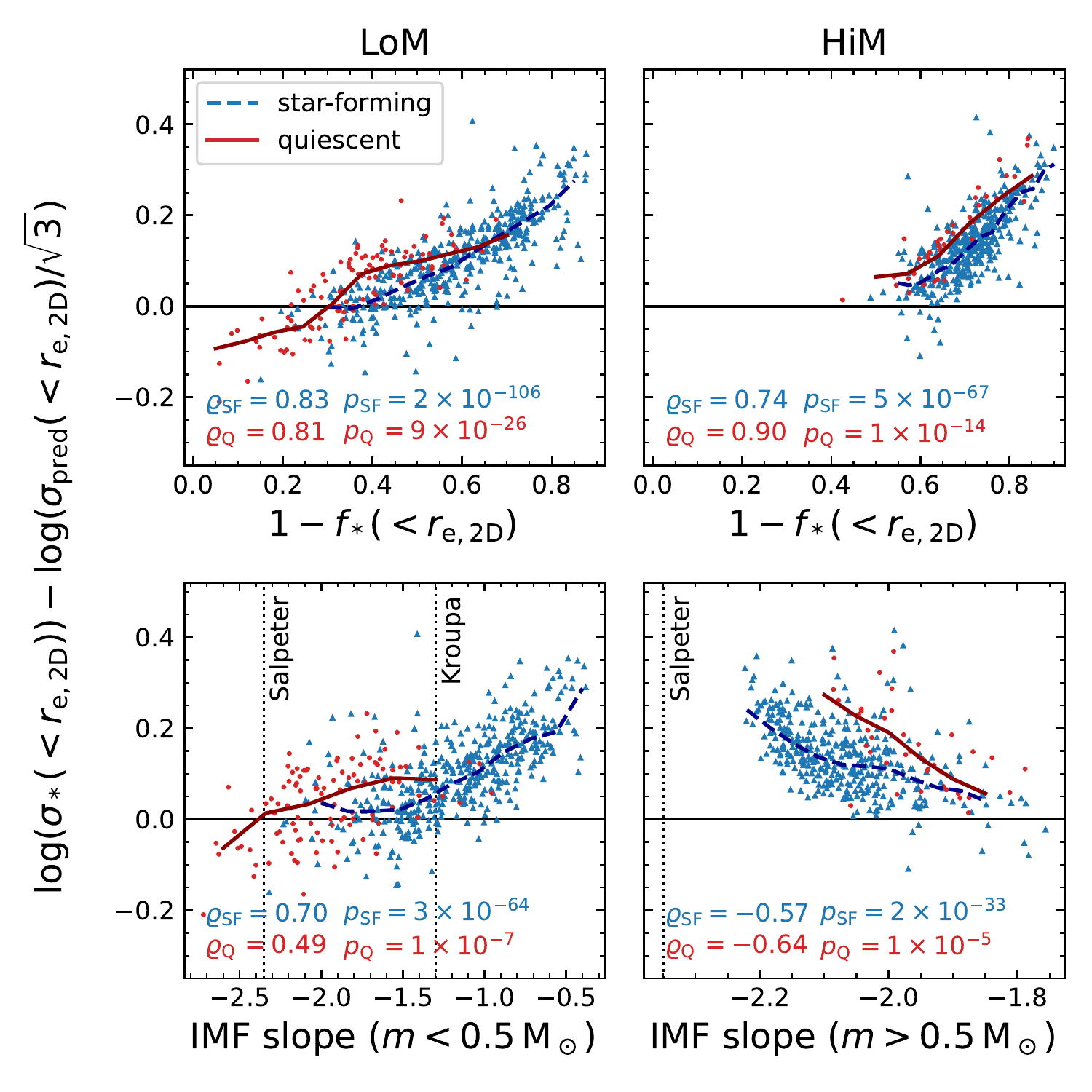}
    \caption{Deviation between the measured stellar velocity dispersion and the predicted velocity dispersion for galaxies in the LoM and HiM variable IMF simulations, as a function of the dark mass fraction and luminosity-weighted IMF slope within the circular aperture of $r_{\rm e,2D}$. Galaxies in the LoM simulation are slightly less dark matter-dominated than in the reference model (in comparison with Fig.~\ref{fig:sstar_mass_fraction}), whereas the HiM model produces strongly dark matter-dominated galaxies. In both cases, there is a somewhat stronger correlation between $\Delta\log\sigma_*$ and $1-f_*$ than with the IMF slope, indicating that fluctuations in the dark matter content are the primary driver of the stellar mass FP, and likely explains the weak variation in the inferred tilt between the different simulations (Fig.~\ref{fig:imf_tilt}). }
    \label{fig:imf_dsigma}
\end{figure}

Comparing the three simulations, we find that the measurements of the tilt for the LoM model are close to those of the Ref model ($<10\%$ difference), whereas the HiM model deviates more strongly. We calculate the difference in the measured (projected) stellar velocity dispersion and the predicted velocity dispersion (using the true stellar mass; Eq.~\ref{eq:sigma_pred}) and examine the drivers of these FPs in Fig.~\ref{fig:imf_dsigma}. Instead of the dark matter fraction, we compute the `dark mass' fraction as $1-f_*(<r_{\rm e,2D})$, where $f_* \equiv M_*/M_{\rm tot}$ and is calculated using 3D apertures of radius $r_{\rm e,2D}$. If the gas fraction is negligible, $f_*$ simply measures the dark matter fraction. The LoM model leads to slightly lower dark fractions than the Ref model, particularly for the quiescent galaxies. On the other hand, galaxies in the HiM simulation are strongly dark matter-dominated within $r_{\rm e,2D}$. Despite this difference, there is a strong correlation between $\Delta\log\sigma_*$ and $1-f_*$ for both models. We also show the correlations with the luminosity-weighted average IMF slope within the circular aperture of $r_{\rm e,2D}$. Although these correlations are strong, the fluctuations in $1-f_*$ still dominate, indicating that variations in the dark matter fractions are still the primary driver of the simulated stellar mass FP. This likely also explains why we find little variation in the tilt between the different simulations.

Finally, we can quantify the effect of IMF variations on the observed tilt of the stellar mass FP, by using the 2D measurements and comparing the fits obtained for the true stellar masses and the reinterpreted stellar masses, shown as the open symbols in Fig.~\ref{fig:imf_tilt}. As expected for the Ref model, the Chabrier and Chabrier-reinterpreted IMF measurements result in identical fits. For the LoM model both $\alpha$ and $\beta$ are slightly lower in value for the Chabrier-reinterpreted measurements, and lower than is measured for the Ref model (by $\approx10-25\%$), an effect that is stronger for the quiescent galaxy population. Star-forming galaxies in the HiM model are not affected significantly by a change in the assumed IMF, and agree well with the fit to the Ref model. On the other hand, the quiescent population does show a very different tilt from the Ref model. Although the formal statistical uncertainty on the fit is large (due to a small sample size and likely a small number of outliers affecting the fitting), the result itself is of significance, as the comparison is between three simulations with equal initial conditions.

\begin{figure}
    \centering
    \includegraphics[width=\linewidth]{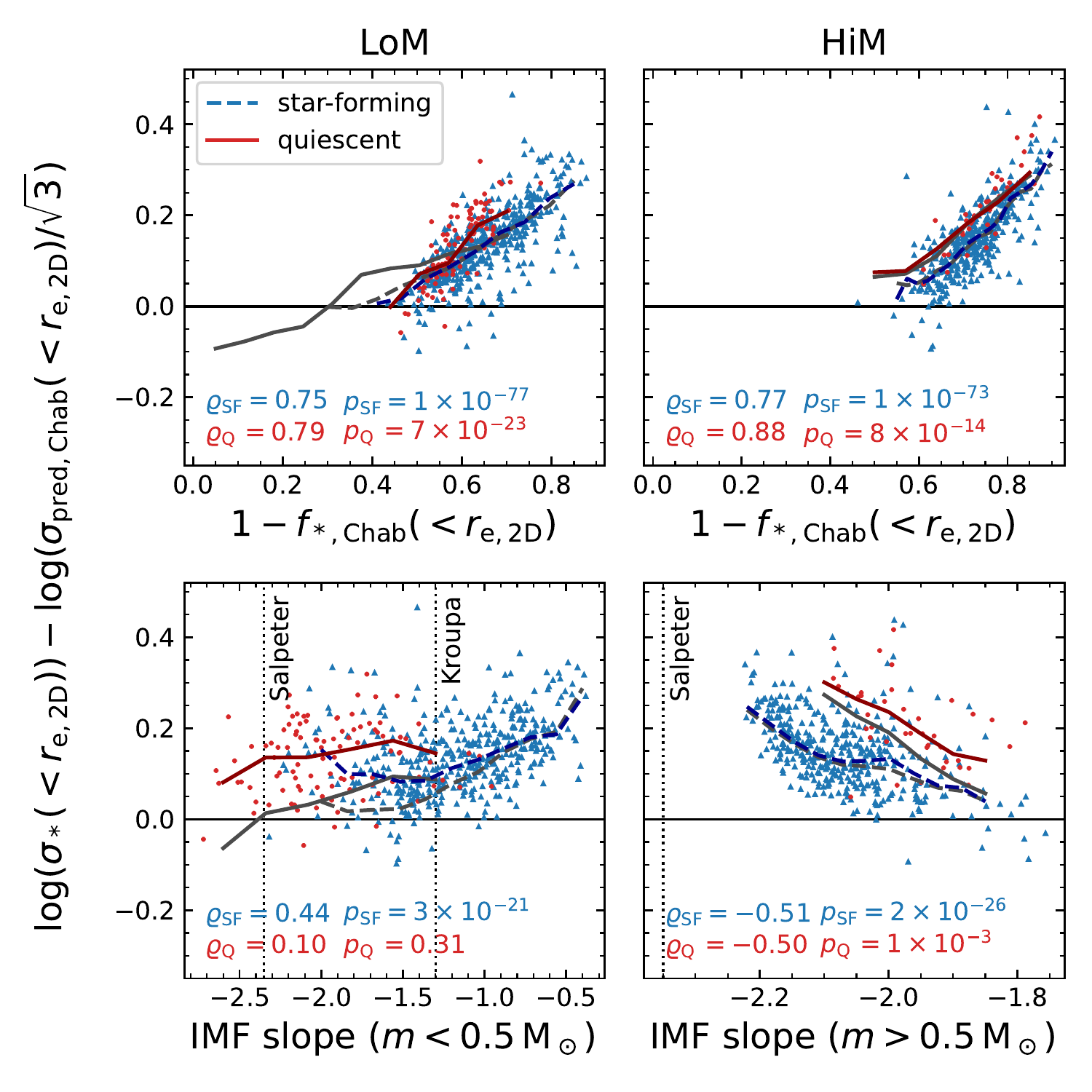}
    \caption{Deviation between the measured stellar velocity dispersion and the velocity dispersion predicted from the Chabrier-reinterpreted stellar mass for galaxies in the LoM and HiM variable IMF simulations, as a function of the dark mass fraction and luminosity-weighted IMF slope within the circular aperture of $r_{\rm e,2D}$. The Chabrier IMF underestimates the true stellar mass of galaxies in the LoM simulation, leading to increased dark mass fractions compared with Fig.~\ref{fig:imf_dsigma} (grey lines show the medians from Fig.~\ref{fig:imf_dsigma}), and weakened correlations with the IMF slope. Galaxies in the HiM simulation are largely unaffected by a difference in the assumed IMF, which may reflect strong structural differences between galaxies in the HiM simulation and the other models.}
    \label{fig:imf_dsigma_chab}
\end{figure}

In Fig.~\ref{fig:imf_dsigma_chab} we show the corresponding change to the results of Fig.~\ref{fig:imf_dsigma}, obtained by calculating the velocity dispersion predicted from the Chabrier-reinterpreted stellar mass, and similarly the reinterpreted stellar mass fraction ($1-f_{\rm *,Chab}$). For the LoM model, the effect of an incorrectly assumed IMF largely removes the correlation with the IMF slope for the quiescent galaxies, and strongly reduces the effect for the star-forming galaxies. Instead, the missing stellar mass is interpreted as extra dark matter, resulting in high values of and a strong correlation with $1-f_{\rm *,Chab}$. The difference in the tilt between the `true IMF' and Chabrier IMF measurements then must stem from the mismatch between the density profile that is traced by the measured velocity dispersion and the (incorrectly) estimated stellar mass surface density. Interestingly, the correlations for galaxies in the HiM simulation are largely unchanged with respect to Fig.~\ref{fig:imf_dsigma}, which is consistent with the very small changes found in the tilt between the `true IMF' and Chabrier IMF fits. Possibly, the top-heavy IMF strongly affects the structural properties of galaxies in the HiM model, such that the profile of the dark matter and corresponding variations in the dark matter content dominate the stellar mass FP (see also Section~\ref{sec:discussion_TF}), and a reinterpretation of the stellar mass with a different IMF therefore has a comparatively small impact. 

Overall, in comparison with the Ref model, the parameters of the observed stellar mass FPs using the reinterpreted stellar masses differ from each other by at most $\approx9\%$ for star-forming galaxies and at most $\approx29\%$ for quiescent galaxies. For star-forming galaxies, this effect is thus of similar magnitude to the effects of non-homology (Section~\ref{sec:nonhomology}); for quiescent galaxies, the uncertainty on the IMF has a larger effect, although this is still subdominant to the effects of variations in the dark matter content.

\section{Discussion}\label{sec:discussion}

\subsection{Interpreting the tilt and scatter of the FP} \label{sec:discussion_tilt}

\begin{figure*}
    \centering
    \includegraphics[width=\linewidth]{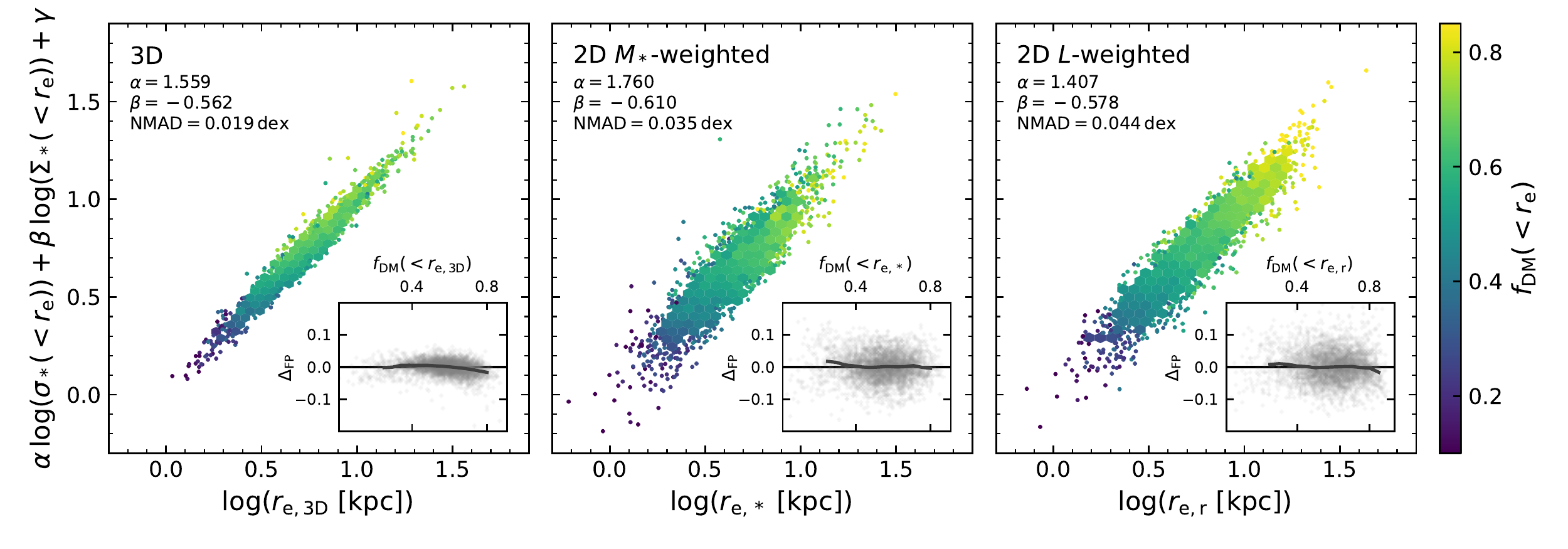}
    \caption{Edge-on view of the stellar mass FP for the intrinsic FP (relying on 3D aperture measurements; left), and the mock observed FPs using measurements weighted by the stellar mass (middle) and the observed light (right). The colour scale shows the dark matter fraction within the effective radius, and insets show the offset from the FP as a function of the same dark matter fraction. Systematic variations in $\fdm$ across the galaxy population set the tilt of the stellar mass FP. The remaining scatter about the FPs anti-correlates only weakly with $\fdm$ (Spearman $\varrho=-0.23$), and is therefore driven by random scatter and measurement uncertainties. }
    \label{fig:fp_edgeon}
\end{figure*}

We set out to measure the low-redshift stellar mass FP, as its tilt and scatter in the zero point trace $\Mdyn/M_*$, and therefore reflect the structural properties and assembly of galaxies. By firstly constructing the total mass FP, we have found that although there is significant variation in the structural properties among the simulated galaxy population, the non-homology of this population has only a weak effect on the tilt of the FP. The resulting deviation from the tilt expected under the assumption of homology is strongest for the population of quiescent galaxies, but still  $<10\%$. In other words, within the effective radius the stellar velocity dispersion, {measured by taking into account both the random and rotational motions of stars (see Section~\ref{sec:vdisp}),} provides a good proxy of the circular velocity (although with an offset of $\approx10\%$). Hence, $\Mdyn\propto\Mtot$, which broadly agrees with observational findings based on dynamical modelling or strongly lensing systems \citep[e.g.,][]{Cappellari2006,Bolton2007,Bolton2008,Li2018}.

Instead, as also suggested by earlier theoretical work \citep[e.g.,][]{Boylan2006,Robertson2006a,Hopkins2008} and more recently specifically for the EAGLE and IllustrisTNG simulations \citep{Ferrero2021}, we have shown that systematic variations in the dark matter fraction as a smooth function of the size and stellar mass drive the strong, observed deviation of the stellar mass FP from the virial plane. Variations in the IMF can affect the inferred dark matter mass fraction, and thereby contribute to the tilt as well, but have a smaller effect. Interestingly, this is true for both the stellar mass FP of the quiescent and star-forming galaxy populations, and leads to a near identical tilt of and scatter about the two simulated FPs: within the parameter space of the mass, velocity dispersion and size, we can therefore regard these two populations as forming a single distribution.

Observationally, it is difficult to constrain $\fdm$, particularly in a way that is independent from the measurement of the FP itself (i.e., through strong lensing), to allow for a robust measurement of potential gradients in $\fdm$ along the FP. Mock observations from the simulations therefore are important to estimate the effects of observational uncertainties and test the role of $\fdm$ in the interpretation of the observed FP. We have found that simply the effects from the random projection of galaxies on the sky, leading to differences in the measured sizes and the velocity dispersion, alter the measured FP by $>5\sigma$, although this can be remedied somewhat by adopting a circularised size rather than a major axis size (or, likely, an alternative correction factor that directly corrects $\sstar$ for the random inclination angle). Moreover, gradients in $M_*/L$ also have a significant effect on the retrieved FP, bringing very good consistency in the $\beta$ parameter (associated with $\Sigma_*$) with the 3D measurements. On the other hand, the $\alpha$ parameter (associated with $\sigma_*$) is biased low and strongly dependent on the sample selection.

The edge-on projections of the intrinsic and mock observed stellar mass FPs are shown in Fig.~\ref{fig:fp_edgeon}, where the colour scale illustrates the correlation with $\fdm$ along the FP. In comparison with observational results, we find that the mock observed EAGLE FP broadly agrees with measurements of the stellar mass FP of quiescent galaxies \citep{HydeBenardi2009,Bernardi2020} as well as the $K$-band luminosity FP \citep{LaBarbera2010,Magoulas2012} when taking into account differences in the sample selection, as these studies have found $\alpha\approx 1.5-1.6$ (and $\alpha\approx1.9$ for the most massive early-type galaxies, of $M_*>10^{11}\,\Msun$). We emphasise that the simulation was not explicitly tuned to reproduce the FP -- only the stellar mass-size relation was used to reject unrealistic subgrid models \citep{Schaye2015} -- and the inferred FP and the drivers of the relation can therefore be considered to be predictive. We note that, unfortunately, the FP of star-forming galaxies has as of yet not been measured explicitly, and we therefore cannot compare those results with observations.

In detail, however, there are some significant differences between the simulated and observed $z\sim0$ stellar mass FP of quiescent galaxies. The values for $\beta$ differ more strongly, with $\beta\approx-0.8$ in observations, whereas the lowest value of $\beta$ measured in the EAGLE simulations is $\beta=-0.66\pm0.04$ for the bottom-heavy IMF model. The top-heavy IMF model on the other hand (with $\beta=-0.37\pm0.16$) is disfavoured, although not ruled out given the large uncertainties on our measurements. { In addition to uncertainties in the IMF affecting both observations and simulations, observations also suffer from other uncertainties in the SED modelling used to obtain stellar mass estimates. \citet{Trcka2020} performed SED modelling on mock observations of galaxies in the EAGLE simulations, finding a constant offset between the inferred and true stellar masses. Importantly, however, \citet{Leja2019a} and \citet{Lower2020} showed that the discrepancy between different stellar mass estimates is often mass-dependent, and arises from uncertainties in the adopted form of the star formation history. Within the mass range considered in this paper, the magnitude of the mass-dependent mass discrepancy found by these authors is comparable to the mass-dependent trend found in Fig.~\ref{fig:imf_mstar} of this work. Although we showed that such a systematic uncertainty on the stellar mass lowers the $\alpha$ and $\beta$ parameters of the measured FP by up to $20\%$, it is not sufficient to bring the observed and simulated FPs into consistency. The physical interpretation of the discrepancy between the observed and simulated FP is therefore discussed further in Section~\ref{sec:TF_Q}. 

The measured scatter also differs by a factor $\approx 2$, with the simulated FP having both a lower observed scatter and intrinsic scatter. }
The difference in the measured scatter can easily be attributed to the fact that the observed stellar mass FP includes additional scatter due to the measurement uncertainties on the stellar masses. 
Yet, the nonzero (intrinsic) scatter in the observed stellar mass FP has been suggested to arise from physical effects, namely variations in $\fdm$ or the IMF through the thickness of the FP \citep{Graves2010III}. The insets in Fig.~\ref{fig:fp_edgeon} show the residuals from the simulated FPs as a function of the dark matter fraction: we find only a weak anti-correlation between $\Delta_{\rm FP}$ and $\fdm$ (Spearman $\varrho=-0.23$) for the 3D FP (left-hand panel), which indicates that galaxies with higher $\Sigma_*$ than predicted from the FP have slightly lower $\fdm$ (since $\Delta\log\Sigma_*\propto \Delta_{\rm FP}$, see Eq.~\ref{eq:delta_fp}). Interestingly, we also find a correlation with the co-rotating fraction $\kappa_{\rm co}$ (Section~\ref{sec:nonhomology}) within the scatter, of $\varrho=0.35$. We do not find any correlations with other bulk galaxy properties, e.g., with mass, velocity dispersion or SFR. This suggests that the scatter in the 3D FP may not be entirely random, but still be partially due to physical effects (e.g., recent mergers). On the other hand, for the mock observed FPs the scatter is predominantly driven by measurement uncertainties.
Therefore, the difference found between the intrinsic scatter of the simulated and observed FP may most easily be explained by an underestimation of the uncertainties in $M_*/L$ from the SED modelling. This includes the systematic uncertainty due to the assumed IMF, which we find to lead to an increase of $\approx10\%$ in the measured scatter in the simulated FPs.

Despite the differences between simulated and observed FP, it is interesting to explore the physical origins of the simulated relation. 
Of course, although we have demonstrated that the variation in $\fdm$ can largely explain the tilt of the FP, with observational uncertainties and biases muddying the picture, the quantity $\fdm$ in itself is merely a consequence of other factors. From Fig.~\ref{fig:sstar_fixedmstar} we see that, at fixed $M_*$, the scatter in $\sigma_*$ is relatively small, but there is strong variation in $\re$. The variation in $\re$ correlates with $\fdm$ (Fig.~\ref{fig:fp_edgeon}), and is in line with the suggestion by \citet{Ferrero2021} that galaxy size is the main differentiating parameter between galaxies. 
The question of which physical mechanisms drive the FP can therefore be recast as: what causes the scatter in $\re$ at fixed $M_*\,$?

This effectively reduces the FP to a 2D scaling relation, namely the stellar mass-size relation, and the question of which galaxy properties show correlations along this plane. Observational studies have shown that galaxy size correlates with colour, age, metallicity and $\alpha$-element enhancement at fixed mass \citep{Franx2008,Scott2017,Barone2020,Barone2022}, and suggested an additional dependence on the structural properties (S\'ersic index, level of rotational support) and environment. For instance, at fixed stellar mass, more compact galaxies have been found to be older and to have higher metallicities, as well as greater $\alpha$-element enhancements. However, these measurements are difficult to interpret, as the stellar population properties are typically luminosity-weighted, global quantities, and therefore are difficult to relate to the overall star formation histories and merger histories of galaxies.

Cosmological simulations may offer valuable insight here, as they allow to trace individual particles within the formation history of a galaxy, and therefore to distinguish between the in-situ and ex-situ growth of galaxies. \citet{Furlong2017} demonstrated that there is a dependence on the sSFR across the stellar mass-size plane for star-forming galaxies in the EAGLE 100\,Mpc simulation, with larger galaxies having higher sSFR at fixed $M_*$, and a correlation with the mass assembly timescale for simulated quiescent galaxies, such that more extended quiescent galaxies assembled later in cosmic time. \citet{Rosito2019b} showed that this also translates to observable measurements, finding trends with not just age and metallicity, but also radial gradients therein, across the dynamical mass-size plane in the EAGLE simulations, and suggest that this in turn correlates with the stellar spin parameter. Therefore, combining these different ideas, investigating the in-situ and ex-situ growth and associated timescales across the stellar mass-size plane and linking these to observable measures may deliver powerful insight into the assembly of galaxies and the physical origins of the tilt of the FP, for simulated, and likely also for observed, populations of galaxies. %\textbf{[something we defer to future work... I started making some plots, but it's a half-hearted attempt]}

\subsection{Reconciling the FP with the TF relation}\label{sec:discussion_TF}

We have found that, at least intrinsically, the simulated star-forming galaxies lie on a total mass FP and stellar mass FP that are approximately the same as the FPs spanned by the quiescent population, and with equally low scatter. On the other hand, star-forming galaxies have been shown to obey the TF relation in both observations and the EAGLE simulations \citep[e.g.,][]{TullyFisher1977,Schaye2015,Ferrero2017}.
These two findings may appear to be contradictory, as the FP is explicitly dependent on surface brightness, yet, extensive literature has shown that the TF relation does not correlate with a third parameter \citep[e.g.,][]{Zwaan1995,Courteau1999,Meyer2008,Lelli2019}. 

\subsubsection{Star-forming galaxies can simultaneously obey the FP and TF relation}

%There is a slight difference between the two relations in the measure of the kinematics that is used: whereas the FP takes {the spatially-integrated velocity dispersion}, which encompasses both disordered motion and rotation, the TF relation uses the inclination-corrected rotational velocity. {Although different in nature, both serve as a proxy for the circular velocity.} 
There is a slight difference between the two relations in the measure of the kinematics that is used: whereas the FP takes {the spatially-integrated velocity dispersion, the TF relation uses the inclination-corrected rotational velocity. Although different in nature, both serve as a proxy for the circular velocity, because the integrated velocity dispersion accounts for both the disordered motion and rotation of the stars (see Section~\ref{sec:vdisp}). 
Furthermore,} there is a difference in the aperture that is considered: rotational velocities are often measured in the outskirts of the disk, whereas the FP probes the effective radius or even smaller radii. The stellar mass or baryonic mass TF relation, described by $M\propto v_{\rm c}^\mu$, has a slope of $\mu\approx 3-4$ depending on the aperture chosen \citep[e.g.,  see][]{Lelli2019}.

If we rewrite the FP in a form that is closer to that of the TF relation, we obtain
\begin{equation}
    M \propto \sigma^{-\alpha/\beta} r_{\rm e}^{(1+2\beta)/\beta}\,.
    \label{eq:mfp_tf}
\end{equation}
Focusing on the 3D measurements of the size, stellar velocity dispersion and the total mass, this results in 
\begin{equation}
\Mtot\propto \sstar^{1.94} \retd^{0.96}\,,
\label{eq:TF_mtot}
\end{equation}
for the $M_*$-selected star-forming population (Table~\ref{tab:FP_3D}). As expected from Section~\ref{sec:nonhomology}, this is very close to the viral relation, which indicates that the star-forming galaxies form a near-homologous sample, and corresponds roughly to the group of galaxies with $T\approx0$, $\epsilon_*\approx0.7$ and $\kappa_{\rm co}\approx0.6$ in Fig.~\ref{fig:sstar_nonhomology}. Although $\sstar$ has a small systematic offset from $v_{\rm c}$ in this parameter range, because of the small impact of non-homology for this sample we can conclude that $\sstar \propto v_{\rm c}(r_{\rm e})$, provided that $\sstar$ does not suffer from projection effects. Hence,
\begin{equation}
    M \propto \sstar^{-\alpha/\beta} r_{\rm e}^{(1+2\beta)/\beta} \propto v_{\rm c}^\mu(r_{\rm e}) r_{\rm e}^\nu \,.
\end{equation}
If we now set $M=M_*$, and continue with the 3D measurements to avoid projection effects on $\sstar$, we obtain (using the bottom rows of Table~\ref{tab:FP_3D})
\begin{equation}
  M_*(<\retd) \propto v_{\rm c}^{2.85}(\retd) \retd^{0.21}\,,
  \label{eq:TF_mstar}
\end{equation}
which is effectively the TF relation, as $\nu\ll \mu\,$. Moreover, this value of $\mu$ corresponds very well to observationally measured values: e.g., using integral field unit (IFU) data from the SAMI Survey, \citet{Bloom2017} found $\mu^{-1} =0.31\pm0.09$ ($\mu\approx3.2$), whereas \citet{Lelli2019} found $\mu=3.06\pm0.08$ for the baryonic TF relation of galaxies in the SPARC dataset using their smallest aperture of $1.3 r_{\rm e}$. We stress that the choice of aperture is critical here, as \citet{Lelli2019} show, using the exact same sample, that the velocity of the flat part of the rotation curve results in a slope of $\mu=3.85\pm0.09$. 

Star-forming galaxies in EAGLE are thus simultaneously compatible with the stellar mass FP and the stellar mass TF relation. We indeed find that the scatter is lower for the FP than the TF relation, finding an orthogonal scatter of $0.0221\pm0.0005\,$dex for the TF relation obtained from an orthogonal linear fit to the stellar masses and circular velocities calculated in Section~\ref{sec:nonhomology} (with a best-fit slope $\mu=3.24\pm0.03$). This is a marginal, although statistically significant, difference of $0.0035\pm0.0006\,$dex with respect to the scatter in the stellar mass FP, and may help to explain why observational studies might not find a correlation with size within the TF relation.

It is unclear, however, to which extent this result can be translated to the observed FP and TF relation. The tilt of the simulated FP differs from observations, and indicates that there may be fundamental discrepancies between the (dynamical) structures of simulated and observed galaxies. As is discussed more extensively in Section~\ref{sec:TF_Q}, we expect this to be of particular importance for quiescent galaxies, but the star-forming galaxies may likely also be affected.

\begin{figure}
    \centering
    \includegraphics[width=\linewidth]{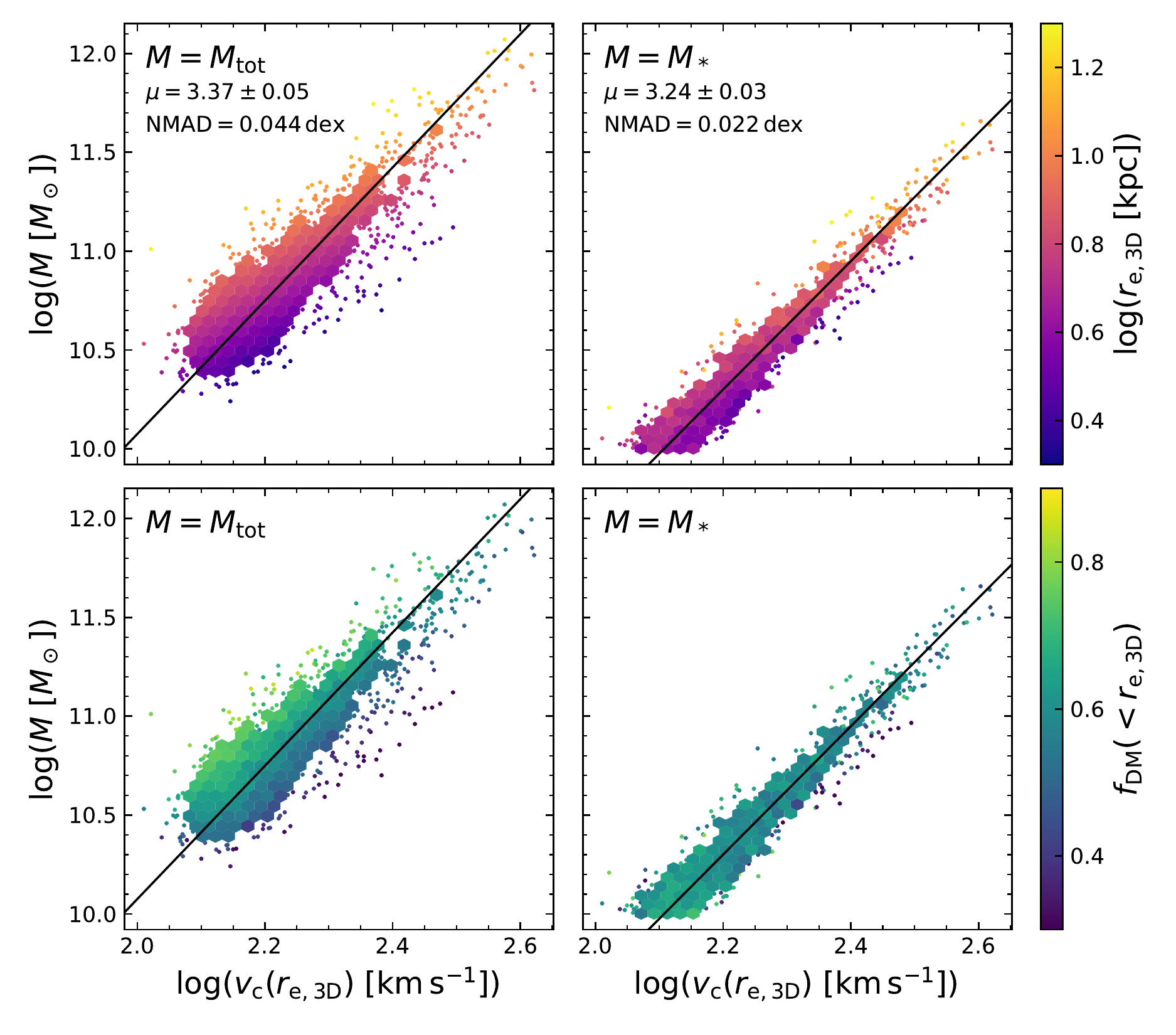}
    \caption{Tully-Fisher relation of star-forming galaxies, using the total mass (left) and stellar mass (right) and circular velocities calculated with Eq.~\ref{eq:vcirc}. Solid lines show the relations obtained with orthogonal distance regression, and the orthogonal scatter is printed in each panel. The offset from the total mass TF relation correlates with the half-mass radius (top), and in turn also with the dark matter fraction (bottom). Because of the correlation between $\retd$ and $\fdm$, the stellar mass TF relation is nearly as tight as the stellar mass FP. }
    \label{fig:TF}
\end{figure}

Nevertheless, we can examine why the TF arises from the FP in the simulations, by dividing Eq.~\ref{eq:TF_mstar} by Eq.~\ref{eq:TF_mtot}:
% \begin{equation}
%     M_* \propto \frac{M_*}{\Mtot} \sstar^{1.94} \retd^{0.96}\,,
% \end{equation}
\begin{equation}
    \frac{M_*}{\Mtot} \propto \sstar^{0.91}\retd^{-0.75}\,,
\end{equation}
which under the assumption of homology, $\sstar \approx v_{\rm c}$ and $\Mtot\approx\Mdyn$, reduces to
\begin{equation}
    \frac{M_*}{\Mtot} \propto \Mtot^{0.46}\retd^{-1.21}\,.
\end{equation}
This relation can be interpreted in terms of the stellar and dark matter density profiles (provided that the gas fractions are low). At fixed $\Mtot$ the stellar-to-total mass ratio decreases strongly with radius, which implies that the dark matter fraction rises rapidly. \citet{Ferrero2021} showed that the half-mass radii of star-forming EAGLE galaxies lie in the dark matter-dominated regions of galaxies, i.e., $\retd>r_{\rm c}$ with the `critical radius' ($r_{\rm c}$) defined as $M_{\rm DM}(<r_{\rm c})=M_*(<r_{\rm c})$. As a result, $M_*/\Mtot$ depends mainly on the mass profile of the dark matter, $M_{\rm DM}(<r)$, which increases monotonically with radius for a NFW profile. The small scatter in the stellar-halo mass relation \citep[$\approx0.15\,$dex for galaxies in the mass range considered here;][]{Matthee2017} implies that galaxies of fixed $M_*$ have similar $M_{\rm DM}$ profiles. Variations in $\retd$ therefore are largely responsible for the slope of the TF relation: %lead to systematic variations in $\fdm$\,: 
Fig.~\ref{fig:TF} shows that although there is no tight TF relation between $\Mtot$ and $\vc$\,, there is a tight relation between $M_*$ and $\vc$ due to the near-perfect correlation between $\retd$ and $\fdm$ at fixed $\vc$\,.

Moreover, there is a weaker dependence on $\Mtot$, such that at fixed size, more massive galaxies are relatively more baryon-dominated within $\retd$. This overall scaling with mass thus reflects the compactness of the stellar mass distribution, and depends on the mass assembly history of galaxies, e.g., through variations in the star formation efficiency within $\retd$ or the merger history. For star-forming galaxies, this possibly reflects the build-up of central bulges in more massive galaxies.

\subsubsection{A TF relation for quiescent galaxies?}\label{sec:TF_Q}

The same reasoning of the previous section can be applied to the quiescent galaxy population. Although the effects of non-homology are stronger for this population, we can obtain a relation similar to Eq.~\ref{eq:TF_mstar} for the quiescent population: $M_*(<\retd) \propto \sstar^{2.68} \retd^{0.23}$.
This form is close to that of the Faber-Jackson relation \citep[FJ;][]{FaberJackson1976}, which is the linear scaling relation between the mass (or luminosity) and velocity dispersion for early-type galaxies.

However, whereas the $r_{\rm e}$ dependence found here is just as weak as for the star-forming galaxies, observational studies have shown that the scatter in the FJ relation correlates significantly with galaxy size, therefore motivating the use of the FP \citep[e.g.,][]{Djorgovski1987,Dressler1987}. This mismatch between the observed and simulated FJ relation suggests that either projection effects on the observed $\sstar$ depend on galaxy size, or, more plausibly, that there is a discrepancy between the simulated and observed FP.

To achieve the weak dependence of $M_*$ on $r_{\rm e}$, requires that the tilt of the FP $\beta \approx -0.5$ with no strong restriction on $\alpha$ (see Eq.~\ref{eq:mfp_tf}). 
Throughout, we have found $\beta\approx-0.56$ with minimal variation, despite significant effects from random inclination angles and $M_*/L$ gradients, and different sample selection effects, and therefore describes a FP that can be easily reconciled with the TF relation.

Yet, observational studies of early-type galaxies at $z\sim0$ have measured a different tilt, with $\beta=-0.84\pm0.02$ or $\beta=-0.776\pm0.019$ for the stellar mass FP \citep{HydeBenardi2009,Bernardi2020}. Similarly, for the luminosity FP, which in principle may differ slightly in the tilt due to $M_*/L$ variations, measurements have consistently resulted in $\beta\approx-0.8$ for early-type galaxies \citep[e.g.,][]{Jorgensen1996,LaBarbera2010,Cappellari2013a}. These measurements all point to a much weaker dependence of $M_*/\Mtot$ on size, i.e., we would expect Fig.~\ref{fig:fdm_var} (showing $\fdm$ in the stellar mass-size plane) to look very different for observed galaxies. This suggests that there are either differences in the dark matter density profiles, or differences in the stellar mass density profiles with respect to the simulated galaxies.

Other theoretical studies of the FP using cosmological simulations have noted a similar systematic discrepancy in the tilt: although these use different sample selections (i.e., a selection of early-type galaxies by the S\'ersic index or kinematic structure), measurements, and fitting methods, \citet{Lu2020} found that the luminosity FP in the IllustrisTNG-100 simulation has a tilt of $\beta=-0.63$ and \citet{Rosito2021} reported $\beta=-0.54$ for the stellar mass FP in the Horizon-AGN simulation. Moreover, \citet{Ferrero2021} showed that the FJ relation of early-type galaxies is systematically offset from the observed relation in both IllustrisTNG and EAGLE. 

As also proposed by the aforementioned studies, the difference between the observed and simulated quiescent galaxy population is likely related to the fact that the sizes of simulated galaxies are systematically too large in comparison with observations, and, correspondingly, the velocity dispersions too low \citep[e.g.,][]{Genel2018,Rodriguez2019,vdSande2019,deGraaff2022}. In addition, the morphological properties also differ from observations, with star-forming galaxies not being flattened enough, and quiescent galaxies not being sufficiently round and having S\'ersic indices that are too low. This implies that the stellar mass distributions diverge from real galaxies, which can be caused by several effects, such as the limited resolution, the gas pressure floor imposed in the simulation, or the details of the star formation and feedback prescriptions in the subgrid models, including the choice of the adopted IMF. 

Interestingly, the three cosmological simulations employ different subgrid models, yet all result in a similarly divergent mass-size relation and FP. The simulations do have similar resolutions, and EAGLE and Illustris-TNG both use a pressure floor with an associated spatial scale of $\approx 1\,$kpc, which likely affects the stellar mass density profiles. An increased resolution for the EAGLE simulations leads to improved S\'ersic indices (i.e., more realistic 1D profiles) and smaller half-mass radii, but similar 3D shapes and velocity dispersions \citep[Appendix~\ref{sec:apdx_highres};][]{Thob2019}. Despite the improvements, the obtained stellar mass FP is similar to that in the simulations at standard resolution. Therefore, either due to the pressure floor or inaccuracies in the subgrid model (e.g., the implementation of the feedback processes, or the IMF), simulated quiescent galaxies do not obtain the correct shapes and dynamical properties.

We suggest that, as a result, the 3D stellar mass distributions are too `puffy', with sizes that are larger than observed or ellipticities that are lower than observed, and thereby containing relatively more dark matter within the effective radius. Even for the EAGLE model that assumes a bottom-heavy IMF, evidence for which has been found in low-redshift early-type galaxies \citep[e.g.,][]{vanDokkum2010,Auger2010}, the inferred tilt deviates by $>3\sigma$ from observations and is likely too dark matter-dominated. In the terminology of the previous section, this would mean that the effective radii of quiescent galaxies are not small enough in comparison with their critical radii. 

On the other hand, \citet{Mukherjee2022} performed a strong lensing analysis to compare the inferred dark matter fractions of massive early-type galaxies in simulations and observations, and found good agreement between the projected measurement of $\fdm(<r_{\rm e})$ from EAGLE and observed lenses, and only a slight discrepancy for the smaller aperture of $\fdm(<r_{\rm e}/2)$. However, their study focuses on the most massive galaxies in EAGLE ($M_*\gtrsim10^{11}\,\Msun$), thereby probing a different regime than considered here. Moreover, as we showed, even though the 1D mass profiles may appear realistic, the 3D structures of the dark matter and stellar mass can still differ and lead to systematic differences in $\sstar$ between observations and simulations. Although relying on projected measurements, the FP is sensitive to the 3D structure, and therefore the stellar mass distribution relative to the dark matter mass distribution.

\begin{figure}
    \centering
    \includegraphics[width=0.8\linewidth]{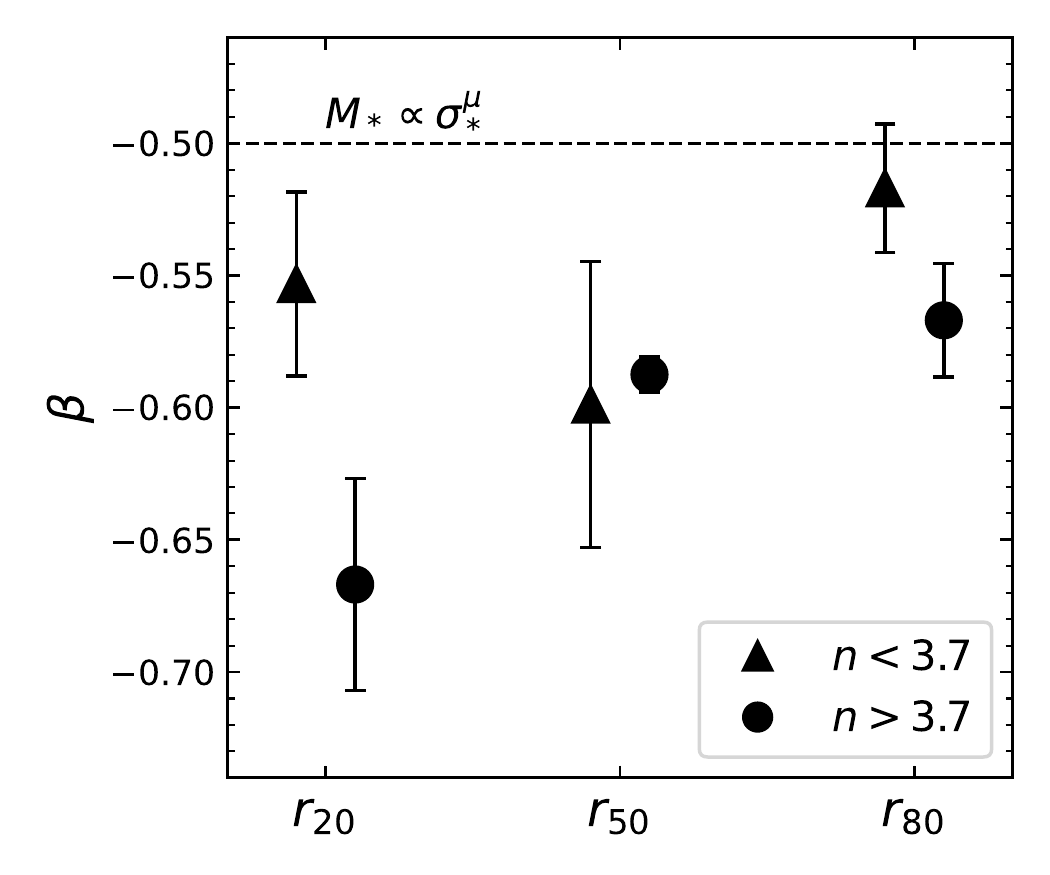}
    \caption{Tilt parameter $\beta$ of the stellar mass FP as a function of the aperture size, for galaxies with low and high S\'ersic indices in the high-resolution EAGLE simulation. The line of $\beta=-0.5$ implies no dependence of the stellar mass on size, i.e., the existence of a perfect TF or FJ relation. There is a weak trend toward this value of $\beta$ for larger aperture sizes, indicating that the dark matter content plays an increasingly important role in the tilt of the stellar mass FP.}
    \label{fig:r20}
\end{figure}

If our assertion, that the greater relative importance of the dark matter in the simulated galaxies affects the inferred FP, is correct, then we would expect to find a dependence of the FP tilt on the chosen aperture. So far, we set this aperture to be half of the enclosed mass or light. 
\citet{Miller2019} proposed the use of the radii enclosing 20\% ($r_{20}$) and 80\% ($r_{80}$) of the light (or stellar mass) distribution instead of the half-light (half-mass) radius, as these sizes are suggested to be more closely linked to the star formation history and halo mass, respectively. 

We use the high-resolution, 25$^3$\,cMpc$^3$ EAGLE simulation to measure the 3D stellar mass $r_{\rm 20}$ and $r_{80}$ radii (where the percentiles are calculated using the stellar mass enclosed within a spherical aperture of radius 100\,kpc), and measure the stellar velocity dispersions within the same apertures. Given the very small number of galaxies with $\rm sSFR<10^{-11}\,yr^{-1}$, we instead divide the sample in two equal-sized subsamples of high and low S\'ersic indices (split at $n=3.7$). Fig.~\ref{fig:r20} shows the measured value of the $\beta$ parameter of the stellar mass FP as a function of the aperture size for the two subsamples. Although a strong conclusion is not possible due to the small sample size and  the limited resolution affecting the measurements for $r_{20}$, there seems to be a trend in the expected direction: larger apertures are more dark matter-dominated, and result in a higher measured value for $\beta$. Moreover, high S\'ersic index galaxies have systematically lower (more negative) $\beta$, in line with the suggestion that galaxy structure is correlated with the dark matter fraction and hence the tilt of the stellar mass FP.

We can therefore conclude that the stellar mass FP offers an important measure of success for the realism of early-type galaxies in cosmological simulations. \citet{Lu2020} similarly proposed the use of the scaling relation between $\Mdyn/L$ and $\Mdyn$\,, however, to obtain a realistic estimate of $L$ requires significant effort in the post-processing of a simulation \citep[e.g.,][]{Trayford2017}. Instead, the stellar mass FP is easily measured, and provides an equally valuable assessment of the 3D stellar mass profile.

Lastly, it is interesting to note that the problem of an inconsistency in the measured tilt occurs mainly at low redshift, as \citet{Lu2020} and \citet{Rosito2021} report no evolution and weak evolution in $\beta$ with redshift, respectively. At the same time, observational work does show evidence for evolution in the tilt of the FP, with values reported in the range $\beta\approx[-0.7,-0.5]$ at $z\approx1$ \citep[e.g.,][]{Saglia2010,Saglia2016,Joergensen2013,Jorgensen2019,Saracco2020,deGraaff2021}. The change in the tilt may be correlated with the observed structural evolution across the same redshift range, as quiescent galaxies become smaller and more disc-like in shape toward higher redshift \citep{Chang2013} and with greater rotational support \citep{Bezanson2018a}. This also appears to be supported by the fact that \citet{Bernardi2020} report a systematically higher value of $\beta$ for low-redshift S0 galaxies than elliptical galaxies.

If the tilt of the FP indeed depends on the average structural properties of the selected galaxy population (e.g., discs versus spheroids), then we would expect star-forming and quiescent galaxies to lie on a single stellar mass FP at higher redshifts \citep[as has been observed;][]{deGraaff2021}, but to span increasingly divergent FPs toward $z\sim0$. The fact that \citet{Bezanson2015} find star-forming galaxies to be consistent with the stellar mass FP of quiescent galaxies at $z\approx0.05$, may then be due to the larger scatter from measurement errors and projection effects for the star-forming population (which affect both the measured velocity dispersion and size), or be caused by the small apertures in which the velocity dispersions were measured, probing only the bulge-like centres of star-forming galaxies at low redshifts. Recently completed large IFU surveys of low-redshift galaxies can shed light on whether the stellar mass FP of star-forming galaxies is truly the same as for quiescent galaxies, and simultaneously offers a direct comparison with the TF relation within the same aperture.

\section{Conclusions}\label{sec:conclusion}

We have used the EAGLE cosmological simulations to measure the tilt and scatter of the stellar mass FP ($\re \propto \sigma^\alpha \Sigma_*^\beta$) for a mass-selected sample of galaxies at $z=0.1$ ($M_*\gtrsim10^{10}\,\Msun$). Using measurements of the total and stellar masses and velocity dispersions within 3D spherical apertures defined by the half-mass radii, we have evaluated the different drivers of the simulated FP. By comparing with measurements of the masses, sizes and stellar velocity dispersions obtained from realistic mock observations, we have quantified the effects of observational uncertainties and the sample selection on the inferred scaling relation.

Our results can be summarised as follows:
\begin{itemize}
    \item We use the measured total masses and 3D stellar velocity dispersions to show that, within the effective radius, the simulated galaxies obey a total mass FP that is very close to the virial relation. The stellar velocity dispersions, {which take into account both the random and streaming motions of the stars,} thus provide a good approximation for the circular velocities (deviating by $\approx 10\%$), with only a weak effect from the galaxy environment (i.e., the classification into central/satellite systems). Therefore, despite significant variation in the structural properties among the simulated galaxy population, the effects of this non-homology on the simulated FP are weak. The velocity dispersion of all (dark matter, stellar and gas) particles deviates more strongly from the circular velocity, due to the dynamically-hot dark matter particles. 
    \item Replacing the total mass by the stellar mass, we find that star-forming and quiescent galaxies span a nearly identical stellar mass FP within the EAGLE simulations, with equally low scatter (0.019\,dex). The stellar mass FP deviates strongly from the virial relation, which is driven by variations in the dark matter fraction within the effective radius ($\fdm$), with negligible impact from variations in the gas content. We show that $\fdm$ is a smooth function of the size and stellar mass, and therefore sets the tilt of the stellar mass FP. We find that the remaining scatter in this FP anti-correlates only very weakly with $\fdm$, and correlates weakly with the degree of rotational support.
    \item For the star-forming galaxies in the simulations, we demonstrate that they are simultaneously compatible with the stellar mass FP and the linear Tully-Fisher relation, provided that both relations are evaluated within the same aperture of the effective radius. The scatter about the TF relation is only slightly higher (0.022\,dex) than the stellar mass FP.  
    \item We create mock observations to show that the projection of galaxies at a random inclination angle along the line of sight affects both the measured sizes and (spatially-integrated) velocity dispersions. These effects can change the inferred tilt of the simulated stellar mass FP by $\approx 10\%$, and increase the scatter by a factor of $\approx 2$. When we use luminosity-weighted measurements instead of $M_*$-weighted measurements, the tilt of the stellar mass FP is changed by a similar amount, but in the opposite direction. The $\alpha$ parameter (associated with $\sigma_*$) in particular is highly sensitive to these changes, and also depends strongly on the sample selection. The scatter about the mock FP is further increased by $\approx 30\%$, which we show is caused by the luminosity weighting of the velocity dispersions.%, and by measurement uncertainties from the noise within our mock observations and the scatter in the inclination angles. 
    \item Systematic uncertainties in the assumed IMF can have a significant effect on the inferred tilt of the stellar mass FP: the parameters of the tilt change by up to $\approx 30\%$ for simulations that employ an observationally-motivated, variable IMF with respect to the stellar mass FP measured for the standard EAGLE model that assumes a universal Chabrier IMF. Nevertheless, we find that regardless of the adopted IMF, variations in $\fdm$, which are themselves correlated with the IMF, are the main driver of the FP.
    \item However, although the tilt and scatter of the measured mock FPs broadly agree with observational results, we find significant differences as well. {Regardless of the resolution of the simulation, the $\beta$ parameter (associated with $\Sigma_*$) differs by $>5\sigma$ from local observations. Systematic uncertainties on the stellar mass, e.g. due to the assumed IMF, may be large, but do not resolve this tension.}
    %Regardless of the adopted IMF or the resolution of the simulation, the $\beta$ parameter (associated with $\Sigma_*$) differs by $>5\sigma$ from local observations. 
    The imposed pressure floor in the simulation, or inaccuracies in the subgrid model likely lead to substantial differences in the 3D stellar mass distributions of the simulated galaxies with respect to local observations. In addition, the standard resolution used in the EAGLE simulations leads to S\'ersic indices that are too low. The effects of non-homology are therefore possibly also weaker for simulated galaxies than for the real Universe.
\end{itemize}

Our work indicates that $\fdm$ is the dominant factor that sets the properties of the FP in the simulations, which in turn is most likely caused by the large variation in $\re$ at fixed $M_*$. We have found that the correlations between the variations in $\fdm$ and $\re$ at fixed $M_*$ naturally give rise to both a FP and TF relation for star-forming galaxies within the aperture of $1\,r_{\rm e}$, although it is unclear to what extent this is the result of the systematic discrepancies in the mass distributions of EAGLE galaxies with respect to observations. For the galaxy population as a whole, it further raises the question of which physical mechanisms may be responsible for the variation in $\re$ at fixed $M_*$, as this may provide valuable insight into the physical origins of dynamical scaling relations such as the FP, as well as the TF relation.

\section*{Acknowledgements}

We acknowledge the Virgo Consortium for making their simulation data available. The EAGLE simulations were performed using the DiRAC-2 facility at Durham, managed by the ICC, and the PRACE facility Curie based in France at TGCC, CEA, Bruy\`eres-le-Ch\^atel.

%%%%%%%%%%%%%%%%%%%%%%%%%%%%%%%%%%%%%%%%%%%%%%%%%%
\section*{Data Availability}
The data underlying this article are available through \url{http://icc.dur.ac.uk/Eagle}. Data products (velocity dispersions, masses) created for this article are available upon reasonable request to the corresponding author.
 
% The inclusion of a Data Availability Statement is a requirement for articles published in MNRAS. Data Availability Statements provide a standardised format for readers to understand the availability of data underlying the research results described in the article. The statement may refer to original data generated in the course of the study or to third-party data analysed in the article. The statement should describe and provide means of access, where possible, by linking to the data or providing the required accession numbers for the relevant databases or DOIs.



\bibliographystyle{mnras}
\bibliography{sims} % if your bibtex file is called example.bib

%\begin{thebibliography}{99}
%\end{thebibliography}

%%%%%%%%%%%%%%%%%%%%%%%%%%%%%%%%%%%%%%%%%%%%%%%%%%

%%%%%%%%%%%%%%%%% APPENDICES %%%%%%%%%%%%%%%%%%%%%

\appendix

\section{High-resolution simulation results}\label{sec:apdx_highres}

In Section~\ref{sec:FP_3D} we investigated the tilt of the FP and showed that there are correlations with the morphological and dynamical properties of galaxies, as well as their mass compositions (particularly $\fdm$). However, the structural properties of galaxies in the EAGLE simulations have been found to deviate from observed galaxies \citep[e.g.,][]{Trayford2017,vdSande2019,deGraaff2022}. \citet{Ludlow2019,Ludlow2021} showed that this is at least in part due to the limited resolution of the simulations, as the 2-body scattering of dark matter and baryonic particles causes a dynamical heating of the baryons, which affects the galaxy size and likely also other structural properties.

These effects can be alleviated by increasing the resolution: the high-resolution $25^3\,$Mpc$^3$ EAGLE simulation (RecalL0025N0752; Section~\ref{sec:eagle}) therefore provides an important test for the robustness of the conclusions drawn from simulations at the standard resolution. Using this simulation, \citet{Thob2019} showed that their measured structural properties are not affected by a change in the resolution. 

\begin{figure}
    \centering
    \includegraphics[width=\linewidth]{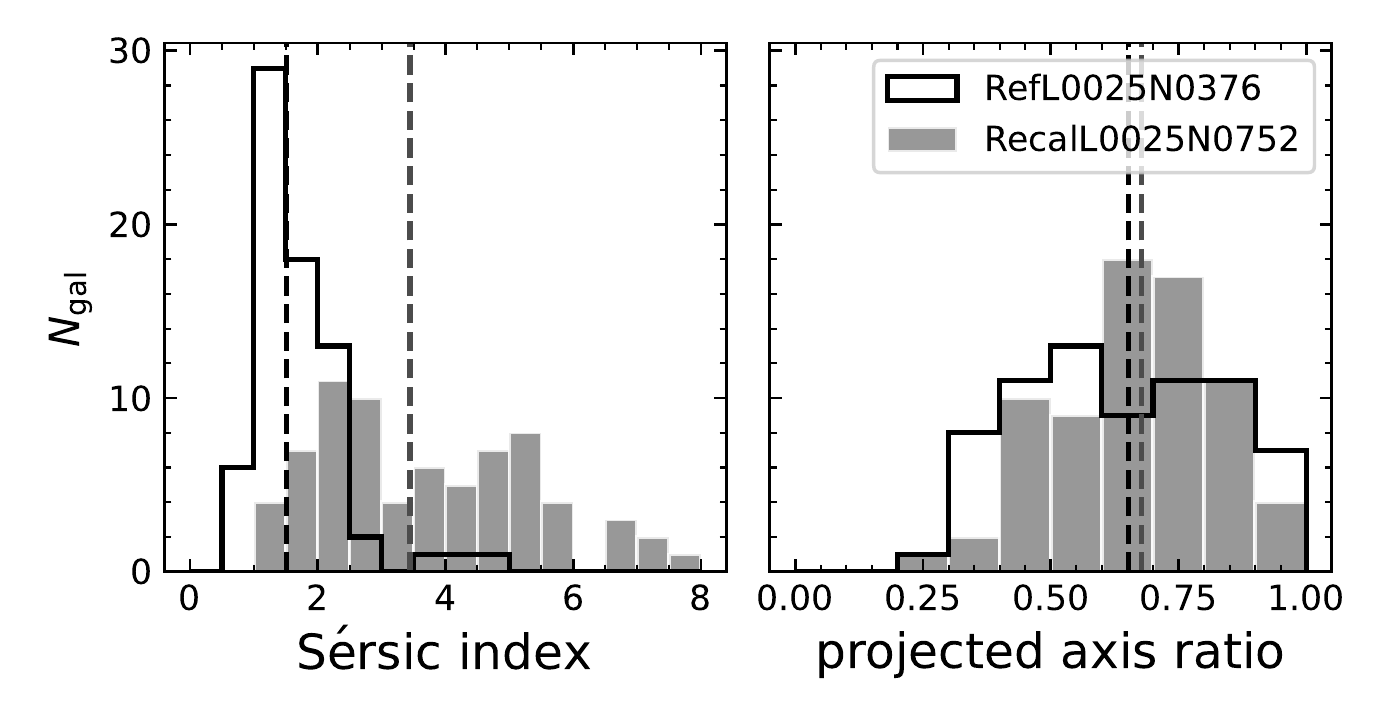}
    \caption{Probability density distribution of the S\'ersic index (left) and axis ratio (right) of the projected stellar mass distribution for EAGLE galaxies of $M_*>10^{10}\,{\rm M_\odot}\,$. The black histogram shows the results for the $25^3\,$Mpc$^3$ simulation run at standard resolution; the grey histogram shows the distribution for galaxies in the high-resolution simulation. The S\'ersic index is strongly dependent on the resolution: the standard resolution does not produce a sufficient number of bulge-like ($n\approx4$) systems in comparison with observations, which is largely solved by the increased resolution. On the other hand, there is little change in the projected axis ratios, with galaxies in the high-resolution simulation being only slightly rounder.}
    \label{fig:sersic_index_comp}
\end{figure}

We perform a similar test for the S\'ersic profile modelling, selecting all (78) galaxies of $M_*>10^{10}\,\rm M_\odot$ (within a spherical aperture of radius 30\,kpc) in the high-resolution simulation. We also select 71 galaxies from the reference model simulation that was run at standard resolution, and has the same volume and initial conditions as the high-resolution simulation (RefL0025N0376). We follow the methodology described in \citet{deGraaff2022} to fit S\'ersic profiles to the projected stellar mass distributions, and show in Fig.~\ref{fig:sersic_index_comp} how the S\'ersic indices and projected axis ratios differ between the RefL0025N0376 and RecalL0025N0752 simulations. The increased resolution has a strong effect on the measured S\'ersic indices, and largely resolves the previously found discrepancy between the RefL0100N1504 simulation and observations in the local Universe.

\begin{figure}
    \centering
    \includegraphics[width=\linewidth]{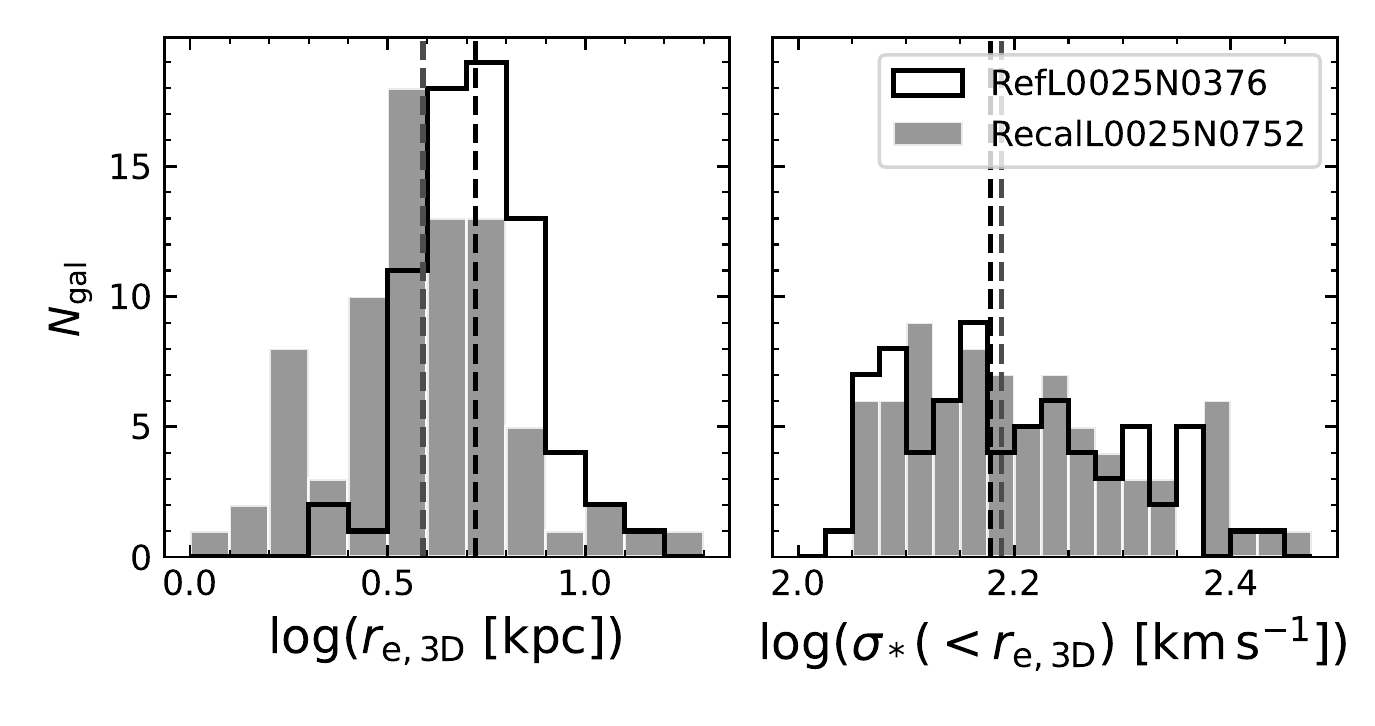}
    \caption{Probability density distribution of the 3D stellar half-mass radius (left) and the stellar velocity dispersion within this radius (right) for EAGLE galaxies of $M_*>10^{10}\,{\rm M_\odot}\,$. The black histogram shows the results for the $25^3\,$Mpc$^3$ simulation run at standard resolution; the grey histogram shows the distribution for galaxies in the high-resolution simulation. The increased resolution leads to more compact galaxies, with higher S\'ersic indices (Fig.~\ref{fig:sersic_index_comp}). However, the velocity dispersions remain unchanged, and are therefore still lower than in local observations.}
    \label{fig:3D_hist_comp}
\end{figure}

On the other hand, the projected axis ratios are largely unchanged, and galaxies in the high-resolution simulation are only slightly rounder. This suggests that the 3D shapes do not depend on the resolution, as also found by \citet{Thob2019}. We therefore also assess the effect of resolution on the half-mass radius and the stellar velocity dispersion within this spherical aperture. The distributions are shown in Fig.~\ref{fig:3D_hist_comp}: in line with the higher S\'ersic indices (indicating more centrally concentrated mass distributions), we find that the half-mass radii are systematically smaller in the high-resolution simulation. However, the velocity dispersions show no dependence on the resolution, and are still smaller than observed \citep[as shown by][]{vdSande2019}.

Next, we assess the convergence of the results obtained in Section~\ref{sec:FP_3D}. We apply the exact same methodology as before, by constructing a sample that is complete in $\Mtot$ and measuring the different mass components within $\retd$\,, as well as $\stot(<\retd)$\,, $\sstar(<\retd)$ and the differences ($\Delta\log\sigma$) with the predicted velocity dispersion of Eq.~\ref{eq:vcirc}. {We present measurements of the tilt for the high-resolution simulations in Table~\ref{tab:FP_highres}, where we also provide measurements of the $25\,$Mpc simulation run at standard resolution for comparison. Due to the small number of galaxies, we do not distinguish between quiescent and star-forming galaxies in the fits. We find that the best-fit parameters and the scatter of the total mass FPs between the two simulations agree to within $<1.8\sigma$, indicating that the increased resolution has at most a modest effect.}

\begin{figure*}
    \centering
    \includegraphics[width=0.7\linewidth]{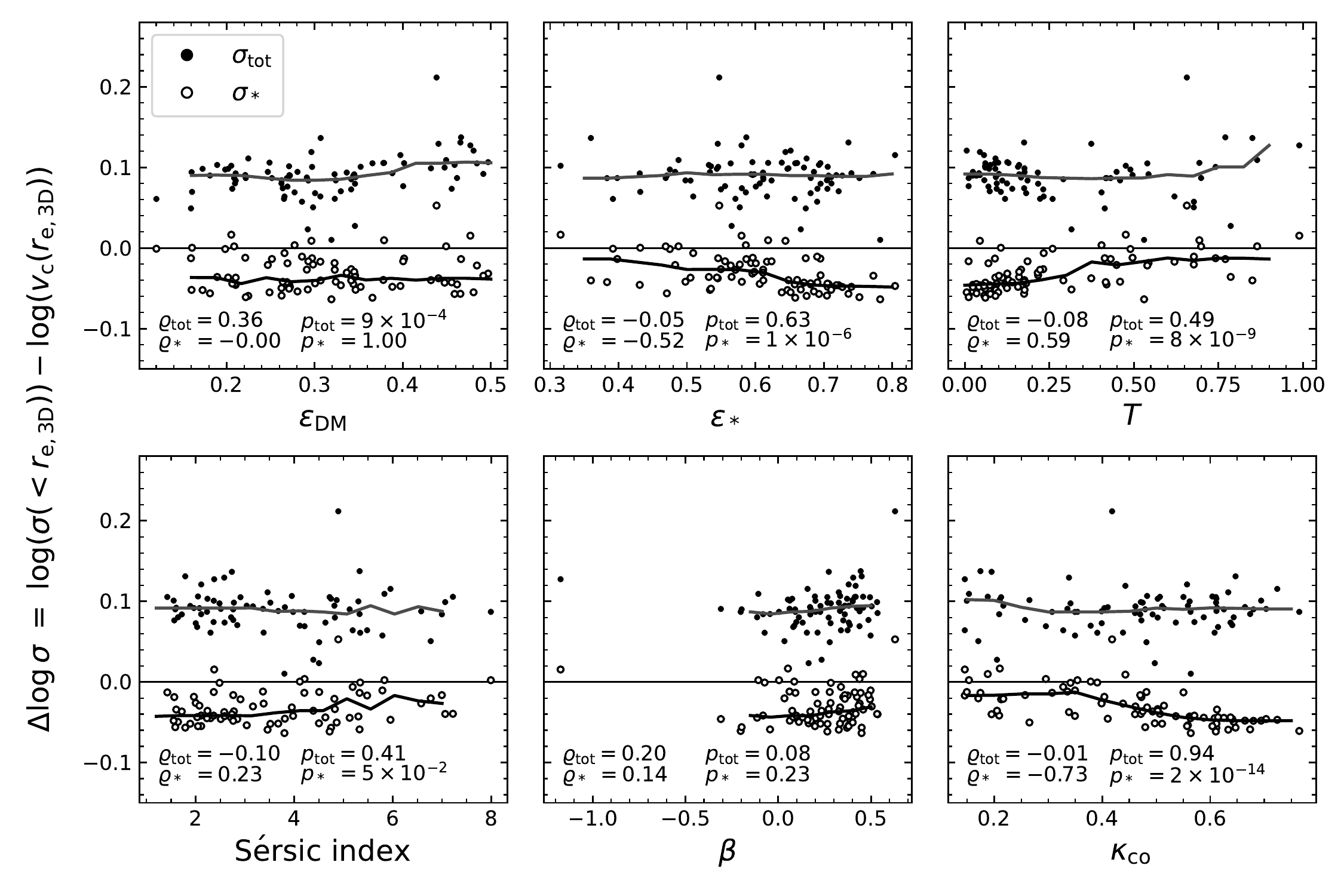}
    \caption{Analogous to Fig.~\ref{fig:stot_nonhomology} and \ref{fig:sstar_nonhomology}, the deviation in the total (filled symbols) and stellar (open symbols) velocity dispersion from the predicted velocity dispersion of Eq.~\ref{eq:vcirc} as a function of different structural properties, for galaxies in the high-resolution RecalL0025N0752 simulation. Lines show the running medians in each panel. Although the number of massive galaxies galaxies in the high-resolution simulation is limited, the measured correlations are similar to those found for the standard resolution. Despite a dependence of the galaxy morphology on the simulation resolution (Fig.~\ref{fig:sersic_index_comp}), the effects of non-homology shown in Section~\ref{sec:nonhomology} are not affected significantly.   }
    \label{fig:nonhomology_highres}
\end{figure*}

\begin{figure}
    \centering
    \includegraphics[width=\linewidth]{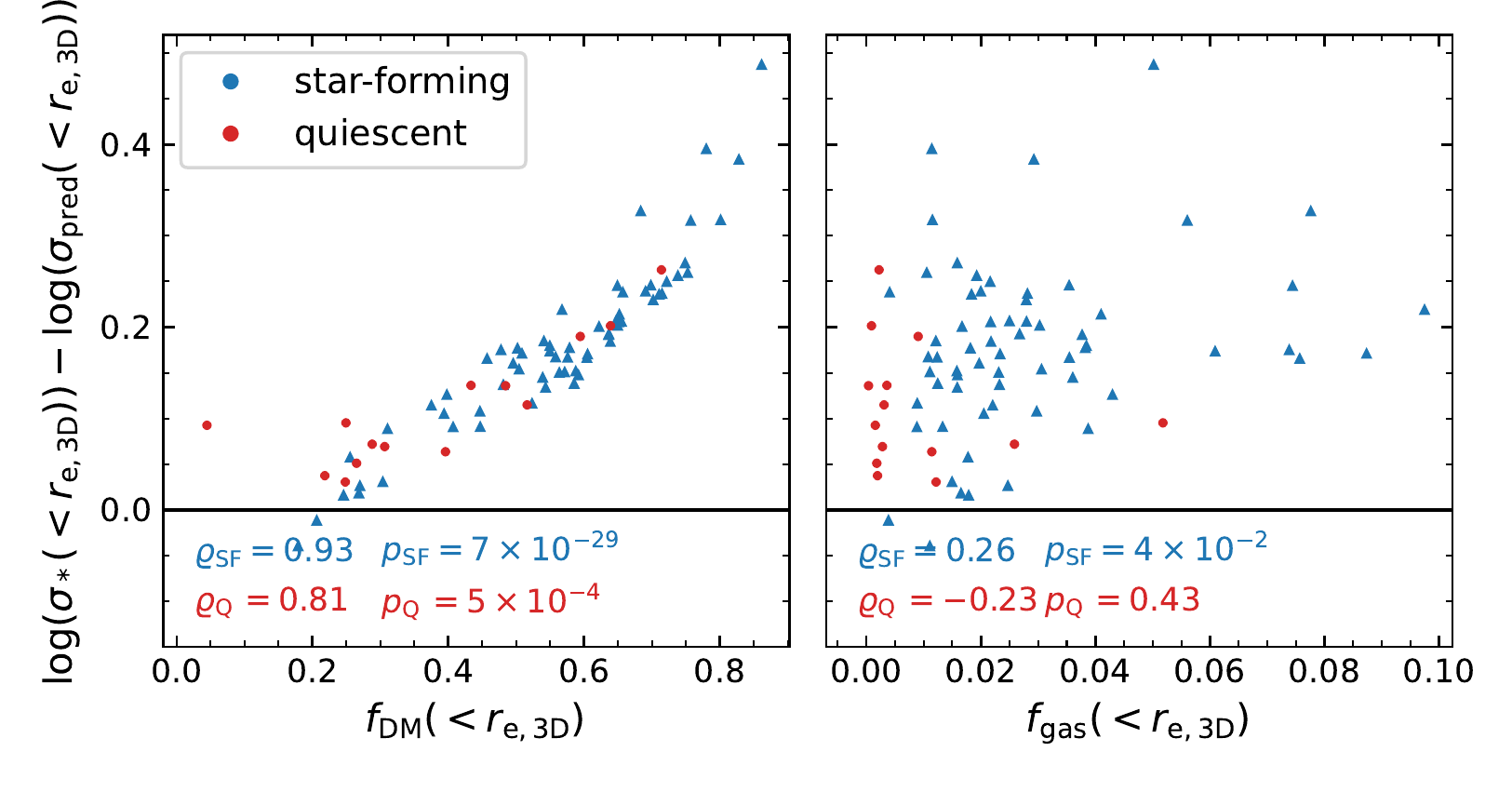}
    \caption{Deviation between the measured stellar velocity dispersion and the dispersion predicted from the stellar mass and half-mass radius (Eq.~\ref{eq:sigma_pred}) versus the dark matter (left) and gas (right) fractions within the half-mass radius, for galaxies in the high-resolution simulation. Unlike Fig.~\ref{fig:sstar_mass_fraction}, there is a weak correlation with the gas fraction for star-forming galaxies ($p$-value $=0.041$). However, the conclusion that the variation in $\fdm$ is the primary driver of the stellar mass FP does not depend on the resolution of the simulation.}
    \label{fig:mass_fractions_highres}
\end{figure}

\begin{table*}
 \caption{Best-fit coefficients of the total mass FP and stellar mass FP, and the orthogonal scatter about the planes, for simulations run at different resolutions. Galaxies for which $\retd<2\,$kpc ($\retd<1\,$kpc) in the RefL0025N0376 (RecalL0025N0752) simulation are excluded from the fits. }
 \label{tab:FP_highres}
 \begin{tabular}{llcccc}
  \hline
  Relation & Simulation & $\alpha$ & $\beta$ & $\gamma$ & NMAD  \\
  \hline
  $\log(\retd) = \alpha \log(\stot(<\retd)) + \beta \log(\Sigma_{\rm tot}) + \gamma$  & RefL0025N0376 & $1.79 \pm 0.09$ & $-0.89 \pm0.05$ & $4.7\pm0.3$ & $0.014\pm0.003$ \\

    & RecalL0025N0752 & $1.72 \pm 0.06$ & $-0.84 \pm0.019$ & $4.39\pm0.17$ & $0.014\pm0.002$ \\

  $\log(\retd) = \alpha \log(\sstar(<\retd)) + \beta \log(\Sigma_{\rm tot}) + \gamma$  & RefL0025N0376  & $1.87 \pm 0.07$ & $-0.99 \pm0.04$ & $5.57\pm0.18$ & $0.011\pm0.002$ \\

   & RecalL0025N0752 & $1.70 \pm 0.07$ & $-0.95 \pm0.03$ & $5.61\pm0.15$ & $0.016\pm0.002$ \\

  $\log(\retd) = \alpha \log(\sstar(<\retd)) + \beta \log(\Sigma_*) + \gamma$  & RefL0025N0376  & $1.44 \pm 0.09$ & $-0.56 \pm0.04$ & $2.48\pm0.15$ & $0.018\pm0.003$ \\

   & RecalL0025N0752 & $1.46 \pm 0.05$ & $-0.59 \pm0.008$ & $2.62\pm0.06$ & $0.010\pm0.002$ \\

  \hline
 \end{tabular}
\end{table*}

Fig.~\ref{fig:nonhomology_highres} further presents the equivalent of Figs.~\ref{fig:stot_nonhomology} and \ref{fig:sstar_nonhomology} for the high-resolution simulation, showing the correlation between $\Delta\log\sigma$ and different structural parameters. We show $\Delta\log\stot$ (filled circles) and $\Delta\log\sstar$ (open circles) within the same figure, with solid lines indicating the running medians. Given the small number of objects (79), we omit the separation into central and satellite galaxies. Compared with Fig.~\ref{fig:stot_nonhomology}, we find that the correlations for $\Delta\log\stot$ are weaker, suggesting that the results found before are partially driven by the effects of the limited resolution (likely particularly that of the dark matter particles).
On the other hand, the results for $\Delta\log\sstar$ are nearly identical to those found in Fig.~\ref{fig:sstar_nonhomology}. Our conclusions on the effects of non-homology on the total mass FP therefore are robust to changes in the resolution. Moreover, the systematic offset between $\stot$ and $\sstar$ remains, which indicates that this is not an effect of the resolution, and gives credence to the interpretation discussed in Section~\ref{sec:nonhomology}.

{Furthermore, we use the $M_*$-selected sample from above to assess the stellar mass FP in the high-resolution simulation, finding excellent agreement with the simulation run at standard resolution (Table~\ref{tab:FP_highres}). }Following Section~\ref{sec:DM_frac}, we measure $\Delta\log\sstar$ using Eq.~\ref{eq:sigma_pred} and examine the relation with the dark matter and gas fraction. Fig.~\ref{fig:mass_fractions_highres} shows star-forming (blue) and quiescent (red) galaxies separately: we find that the main conclusion, that $\fdm$ is the main driver of the stellar mass FP, is unchanged. Unlike the results of Fig.~\ref{fig:sstar_mass_fraction}, we find a weak correlation with the gas fraction for star-forming galaxies, although this is at low statistical significance ($p$-value of 0.041).

In conclusion, we find good convergence between the results obtained with the simulation at standard and high resolution, despite differences in some of the morphological properties between the two different sets of simulations. 

{We note that in this exercise we have not considered possible effects of the increased resolution on the mock observations used in Section~\ref{sec:FP_obs}. By increasing the spatial resolution by a factor of 2, we might expect the distribution of the dust to reach higher densities, leading to stronger attenuation, which in turns alters the $M_*/L$ gradients and hence the obtained size measurements. Such an effect should depend on the inclination of a galaxy, as the effects of dust are most prominent when the galaxy is viewed edge-on. \citet[][Fig. 4]{Trayford2017} used the different 25\,Mpc simulations to compare the ($B$-band) dust attenuation of individual galaxies as a function of the viewing angle. They show that, although a small number of galaxies in the high-resolution simulation are more strongly attenuated, the resolution has no effect on the attenuation for the population average. Therefore, we do not expect our findings of Section~\ref{sec:FP_obs} to depend on the resolution of the dust in the post-processing of the simulation.  }

\section{Stellar mass FP with circularised sizes}\label{sec:apdx_rcirc}

\begin{figure*}
    \centering
    \includegraphics[width=0.92\linewidth]{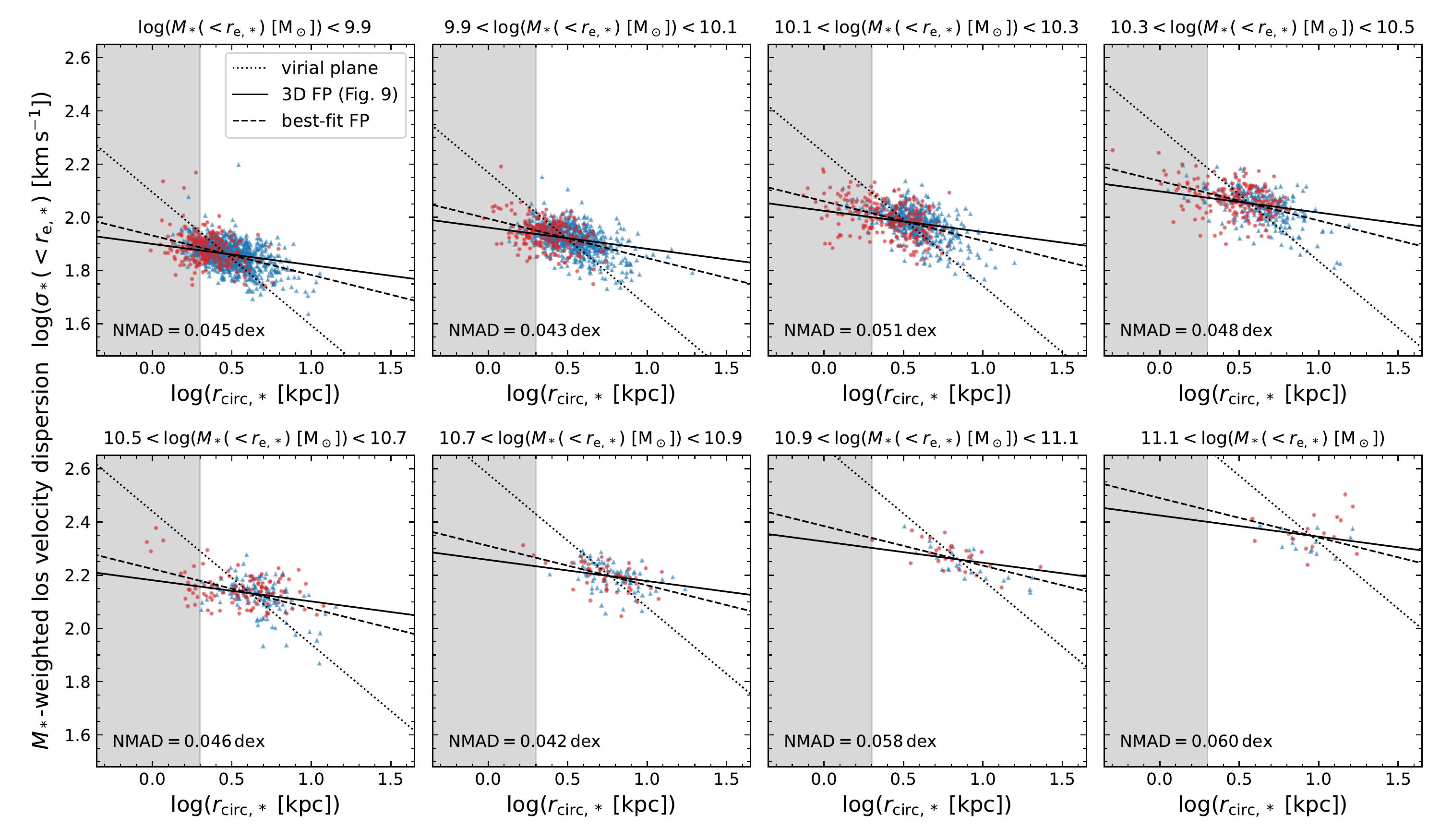}
    \caption{As Fig.~\ref{fig:fp_re_star}, but showing the circularised effective radius rather than the major axis size (obtained from the best-fit S\'ersic profile to the projected stellar mass distributions).}
    \label{fig:fp_rcirc_star}
\end{figure*}

\begin{figure*}
    \centering
    \includegraphics[width=0.92\linewidth]{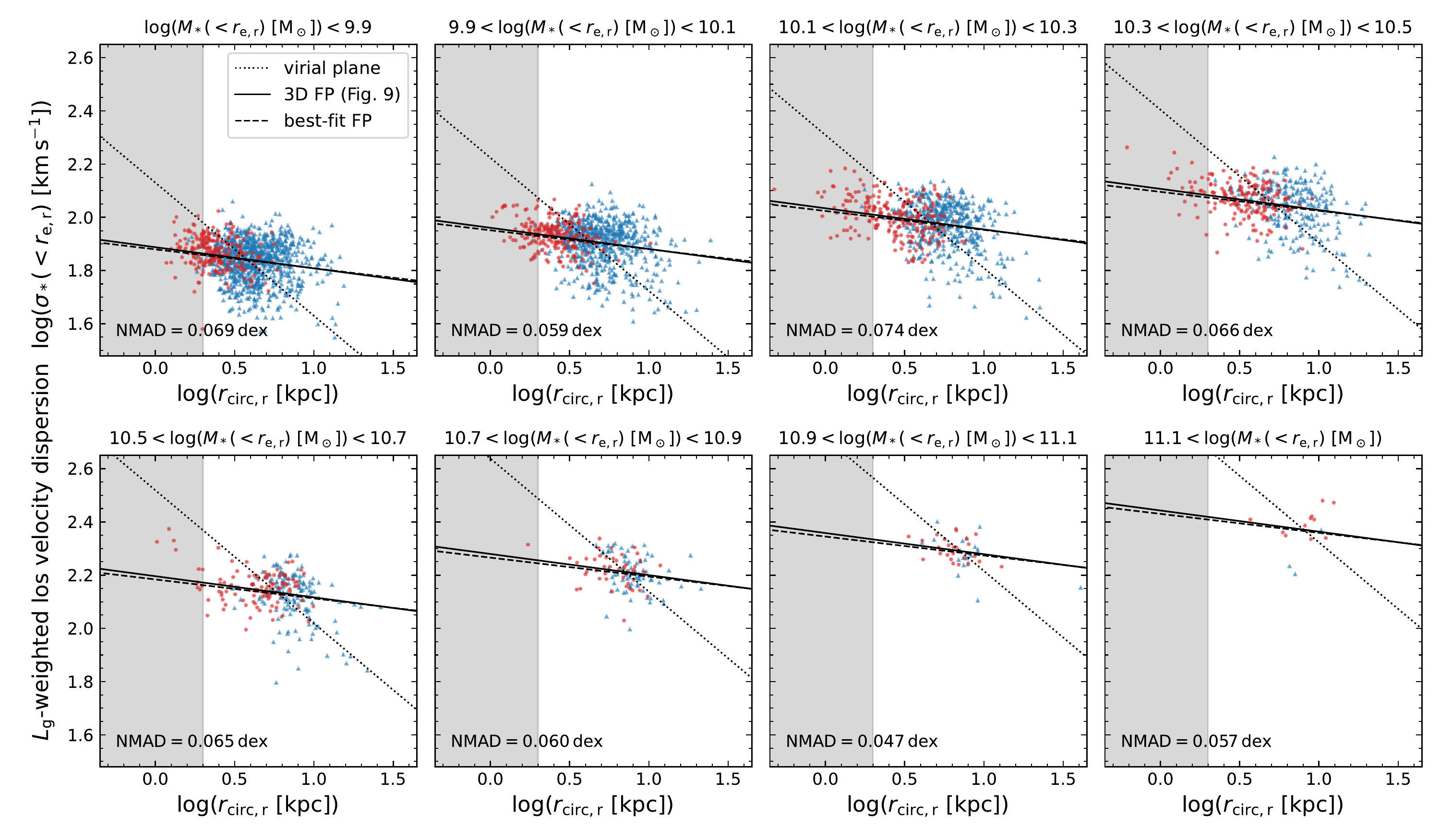}
    \caption{As Fig.~\ref{fig:fp_re_rband}, but showing the circularised effective radius rather than the major axis size (obtained from the best-fit S\'ersic profile to the mock $r$-band images).}
    \label{fig:fp_rcirc_rband}
\end{figure*}

\begin{figure*}
    \centering
    \includegraphics[width=0.92\linewidth]{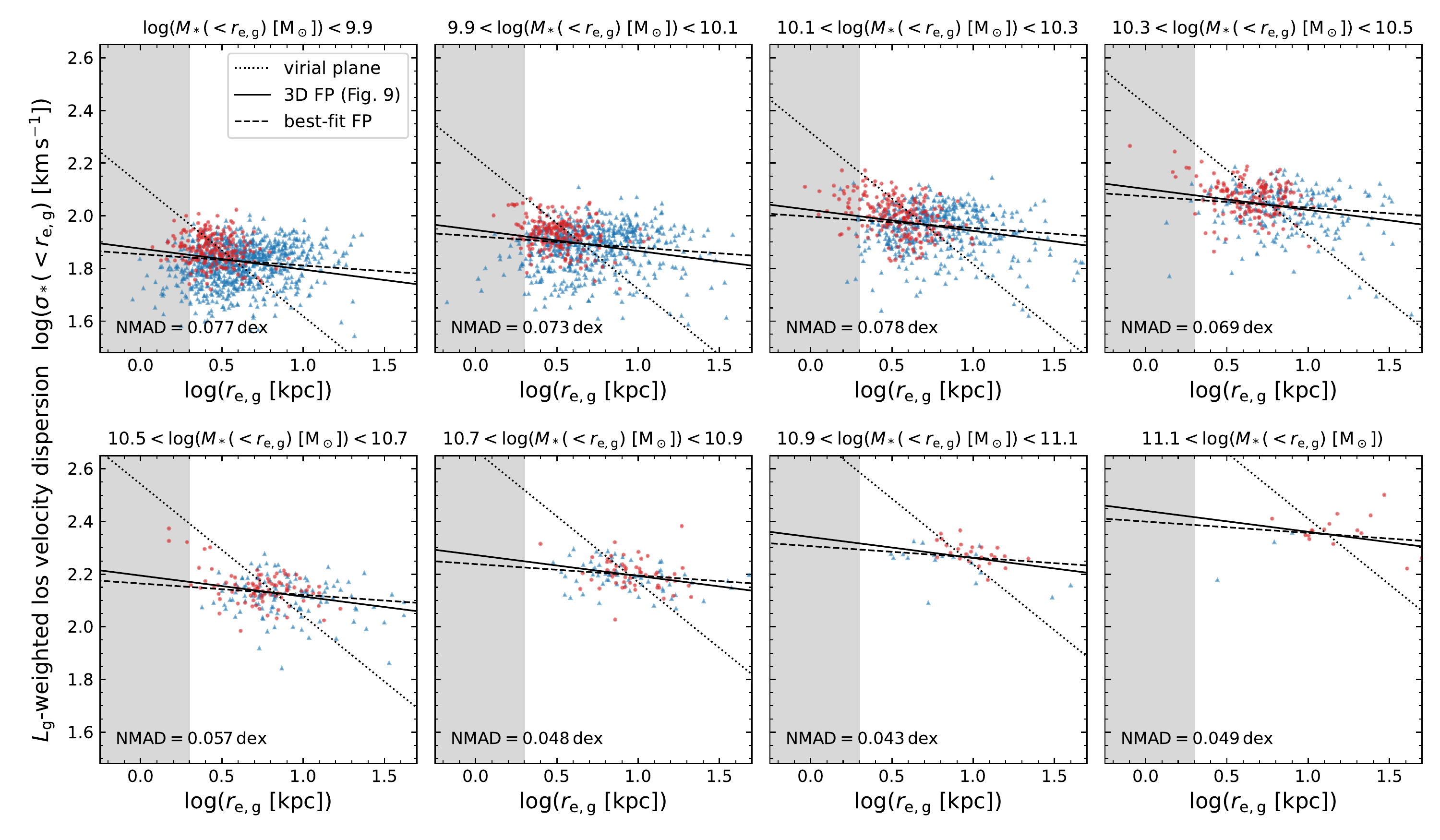}
    \caption{As Fig.~\ref{fig:fp_re_rband}, but using a consistent tracer (the rest-frame $g$-band luminosity) for both the sizes and velocity dispersions: the effective radii are measured from $g$-band images that do not include the effects of dust attenuation using the methodology described in \citet{deGraaff2022}, and the spatially-integrated stellar velocity dispersions are measured within elliptical apertures defined by these unattenuated $g$-band half-light radii. The strong increase in the scatter between Figs.~\ref{fig:fp_re_star} and \ref{fig:fp_re_rband} is therefore not due to an inconsistency in the tracers within Fig.~\ref{fig:fp_re_rband}. Rather, it is the luminosity-weighting itself that causes a significant increase in the scatter.}
    \label{fig:fp_nodust}
\end{figure*}

In Section~\ref{sec:obs_bias} we presented the stellar mass FP, and showed the relation between the projected major axis size and line-of-sight velocity dispersion for different stellar mass bins (Figs.~\ref{fig:fp_re_star} and \ref{fig:fp_re_rband}). However, observational studies often use circularised sizes rather than major axis sizes, which we showed to result in a FP that is in better agreement with the intrinsic stellar mass FP, as the circularised sizes provide an ad hoc correction for the random inclination angles of galaxies.

In Figs.~\ref{fig:fp_rcirc_star} and \ref{fig:fp_rcirc_rband} we show the circularised size as a function of the line-of-sight velocity dispersion, binned by the stellar mass, for the stellar mass-weighted and luminosity-weighted measurements, respectively. These differ from Figs.~\ref{fig:fp_re_star} and \ref{fig:fp_re_rband} by only the measure of size used. The velocity dispersions are unchanged, as these are spatially-integrated measurements within elliptical apertures (see Section~\ref{sec:vdisp}).

As is to be expected, the circularised sizes are smaller than the major axis sizes, with an average offset of $-0.1\,$dex. Most importantly, however, the scatter in $\log r_{\rm circ}$ changes as well: there is a wide spread in the distribution of the projected axis ratios \citep{deGraaff2022}, and the circularised size can differ from the major axis size by $\approx-0.35\,$dex for a galaxy that is projected edge-on ($q\approx0.2$). At the same time, from the top panel of Fig.~\ref{fig:sigma_obs} we can see that, at fixed intrinsic dispersion $\sstar(<\retd)$, the observed line-of-sight velocity dispersion is a factor $\approx2$ greater for edge-on systems in comparison with face-on systems. As $\log\sstar$ is unchanged in the fits of the stellar mass FP and in Figs.~\ref{fig:fp_rcirc_star} and \ref{fig:fp_rcirc_rband}, it is this change in the scatter in $\log r_{\rm circ}$ that counteracts the projection effects on $\log\sstar$ and hence alters the inferred FP.

\section{Effects of luminosity-weighting on the FP}\label{apdx:nodust}

In Section~\ref{sec:obs_bias} we found that the mock observations of the sizes and velocity dispersions introduce significant scatter in the FP. The $M_*$-weighted mock sizes and velocity dispersions shown in Fig.~\ref{fig:fp_re_star} indicate that this is likely due to the random projection of galaxies along the line of sight, as well as the uncertainties on the half-mass radii, as these were measured from mock images with realistic noise and PSF smoothing.

However, in Fig.~\ref{fig:fp_re_rband} we found that the use of luminosity-weighted measurements further increases the scatter by $\approx50\%$ for the less massive galaxies, despite the fact that these measurements were extracted using the exact same methodology. Whereas the measurements in Fig.~\ref{fig:fp_re_star} are both weighted by $M_*\,$, the measurements in Fig.~\ref{fig:fp_re_rband} use slightly different tracers: the sizes were measured from $r$-band images that include the effects of dust attenuation (Section~\ref{sec:size_mass}), but the velocity dispersions were measured using the unattenuated $g$-band luminosities of the stellar particles, which are spatially-integrated measurements within elliptical apertures defined by the $r$-band S\'ersic profiles.

Therefore, we explore whether the inconsistency in the tracer used causes the strong increase in the scatter. We create images of the unattenuated rest-frame $g$-band light, and follow the methodology described in \citet{deGraaff2022} to construct mock images and fit S\'ersic profiles. {We find a small difference between the dust-free $g$-band sizes and the $r$-band sizes with dust attenuation: for star-forming galaxies, the dust-free g-band sizes are on average smaller by 0.08\,dex, with a scatter of 0.12\,dex in $\Delta\log(r_{\rm e,g} / r_{\rm e,r})$. For quiescent galaxies this average difference is only 0.02\,dex with a scatter of 0.04\,dex.} Next, we use these $g$-band S\'ersic profiles to construct elliptical apertures, and hence obtain consistent, spatially-integrated velocity dispersions that are weighted by the $g$-band luminosities of the particles.

We show the resulting relation between the $g$-band half-light radii (major axis sizes) and velocity dispersions in Fig.~\ref{fig:fp_nodust}{, and present the fits to the stellar mass FP in Table~\ref{tab:FP_nodust}. With respect to the results of Table~\ref{tab:FP_obs} the tilt is changed slightly, as the parameters differ by $<10\%$.

Importantly, the scatter is slightly larger than the scatter found in Fig.~\ref{fig:fp_re_rband} that used the inconsistent, luminosity-weighted measurements.} We therefore conclude that it is the luminosity-weighting itself that leads to an increase in the scatter, rather than the difference between the tracers. This is likely caused by the fact that the younger stellar particles (which have low $M_*/L$) are dynamically colder, and the line-of-sight velocity dispersion is therefore more strongly dependent on the inclination angle of the galaxy.

\begin{table*}
 \caption{Best-fit coefficients for the stellar mass FP based on the size measurements without dust attenuation. }
 \label{tab:FP_nodust}
 \begin{tabular}{lcccc}
  \hline
  Sample &$\alpha$ & $\beta$ & $\gamma$ & NMAD  \\
  \hline
  all & $1.40 \pm 0.03$ & $-0.532 \pm0.005$ & $2.49\pm0.04$ & $0.054\pm0.0011$ \\

 quiescent  & $1.53 \pm 0.03$ & $-0.549 \pm0.014$ & $2.32\pm0.12$ & $0.048\pm0.0019$ \\
   
 star-forming  & $1.40 \pm 0.03$ & $-0.505 \pm0.005$ & $2.28\pm0.05$ & $0.0514\pm0.0015$ \\

  \hline
 \end{tabular}
\end{table*}

%%%%%%%%%%%%%%%%%%%%%%%%%%%%%%%%%%%%%%%%%%%%%%%%%%

% Don't change these lines
\bsp	% typesetting comment
\label{lastpage}
\end{document}